\pdfoutput=1
\documentclass[11pt,twoside,a4paper,cmspaper,final,collab]{cms-tdr}
\def\svnVersion{9f5b6e2}\def\svnDate{2023/08/16}\def\cmsCernNoTag{CERN-EP-2022-021}\def\cmsCernDate{\today}\def\cmsMessage{Published in Physical Review D as \href{http://dx.doi.org/10.1103/PhysRevD.108.032013}{\doi{10.1103/PhysRevD.108.032013}.}}
\begin{document}\cmsNoteHeader{HIG-20-007}

\newlength\cmsFigWidth
\ifthenelse{\boolean{cms@external}}{\setlength\cmsFigWidth{0.85\columnwidth}}{\setlength\cmsFigWidth{0.4\textwidth}}
\ifthenelse{\boolean{cms@external}}{\providecommand{\cmsLeft}{upper\xspace}}{\providecommand{\cmsLeft}{left\xspace}}
\ifthenelse{\boolean{cms@external}}{\providecommand{\cmsRight}{lower\xspace}}{\providecommand{\cmsRight}{right\xspace}}
\newlength\cmsTabSkip\setlength\cmsTabSkip{1.6ex}
\newlength\cmsVecKern
\ifthenelse{\boolean{cms@external}}{\setlength\cmsVecKern{0.3em}}{\setlength\cmsVecKern{0pt}}
\newcommand{\mvis}{\ensuremath{m_\text{vis}}\xspace}
\newcommand{\mtautau}{\ensuremath{m_{\PGt\PGt}}\xspace}
\newcommand{\pth}{\ensuremath{\pt^{\PGt\PGt}}\xspace}
\newcommand{\ptvech}{\ensuremath{\ptvec^{\kern\cmsVecKern\PGt\PGt}}\xspace}
\newcommand{\mjj}{\ensuremath{m_{\mathrm{jj}}}\xspace}
\newcommand{\dphi}{\ensuremath{\Delta\phi_{\mathrm{jj}}}\xspace}
\newcommand{\detajj}{\ensuremath{\abs{\Delta \eta_{\mathrm{jj}}}}\xspace}
\newcommand{\NN}{\ensuremath{NN_\text{disc}}\xspace}
\newcommand{\emu}{\ensuremath{\Pe\PGm}\xspace}
\DeclareRobustCommand{\V}{{\HepParticle{V}{}{}}\Xspace}
\DeclareRobustCommand{\f}{{\HepParticle{f}{}{}}\Xspace}
\newcommand{\Hboson}{\ensuremath{\PH}\xspace}
\newcommand{\alphacp}{\ensuremath{\alpha^{\PH\f\f}}\xspace}
\newcommand{\fggh}{\ensuremath{f_{a3}^{\Pg\Pg\PH}}\xspace}
\newcommand{\fCP}{\ensuremath{f_{CP}^{\PH\PQt\PQt}}\xspace}
\newcommand{\fCPff}{\ensuremath{f_{CP}^{\PH\f\f}}\xspace}
\newcommand{\fai}{\ensuremath{f_{ai}}\xspace}
\newcommand{\fathree}{\ensuremath{f_{a3}}\xspace}
\newcommand{\fatwo}{\ensuremath{f_{a2}}\xspace}
\newcommand{\fL}{\ensuremath{f_{\Lambda 1}}\xspace}
\newcommand{\fLzg}{\ensuremath{f_{\Lambda 1}^{\PZ\PGg}}\xspace}

\newcommand{\cgg}{\ensuremath{c_{\Pg\Pg}}\xspace}
\newcommand{\cggtilde}{\ensuremath{\widetilde{c}_{\Pg\Pg}}\xspace}
\newcommand{\kappaf}{\ensuremath{\kappa_\f}\xspace}
\newcommand{\kappaftilde}{\ensuremath{\widetilde{\kappa}_\f}\xspace}
\newcommand{\kappat}{\ensuremath{\kappa_\PQt}\xspace}
\newcommand{\kappattilde}{\ensuremath{\widetilde{\kappa}_\PQt}\xspace}
\newcommand{\kappab}{\ensuremath{\kappa_\PQb}\xspace}
\newcommand{\kappabtilde}{\ensuremath{\widetilde{\kappa}_\PQb}\xspace}

\newcommand{\DBSM}{\ensuremath{\mathcal{D}_{\mathrm{BSM}}}\xspace}
\newcommand{\Dint}{\ensuremath{\mathcal{D}_{\text{int}}}\xspace}
\newcommand{\DVBF}{\ensuremath{\mathcal{D}_\text{2jet}^\mathrm{VBF}}\xspace}
\newcommand{\DCP}{\ensuremath{\mathcal{D}_{CP}^{\mathrm{VBF}}}\xspace}
\newcommand{\Dz}{\ensuremath{\mathcal{D}_{0-}}\xspace}
\newcommand{\Dzh}{\ensuremath{\mathcal{D}_{\mathrm{0h+}}}\xspace}
\newcommand{\DL}{\ensuremath{\mathcal{D}_{\Lambda1}}\xspace}
\newcommand{\DLzg}{\ensuremath{\mathcal{D}_{\Lambda1}^{\PZ\PGg}}\xspace}
\newcommand{\DCPggH}{\ensuremath{\mathcal{D}_{CP}^{\Pg\Pg\PH}}\xspace}
\newcommand{\DzggH}{\ensuremath{\mathcal{D}_{0-}^{\Pg\Pg\PH}}\xspace}
\newcommand{\Dnn}{\ensuremath{\mathcal{D}_\mathrm{NN}}\xspace}
\newcommand{\PP}{\ensuremath{\Pp\Pp}\xspace}

\newcommand{\ggH}{\ensuremath{\Pg\Pg\PH}\xspace}
\newcommand{\ttH}{\ensuremath{\PQt\PAQt\PH}\xspace}
\newcommand{\Hff}{\ensuremath{\PH\f\f}\xspace}
\newcommand{\Hgg}{\ensuremath{\PH\Pg\Pg}\xspace}
\newcommand{\HVV}{\ensuremath{\PH\PV\PV}\xspace}
\newcommand{\Htt}{\ensuremath{\PH\PQt\PQt}\xspace}
\newcommand{\VH}{\ensuremath{\PV\PH}\xspace}
\newcommand{\muggh}{\ensuremath{\mu_{\Pg\Pg\PH}}\xspace}
\newcommand{\muqqh}{\ensuremath{\mu_{\PQq\PQq\PH}}\xspace}

\newcommand{\Wjets} {\PW{}+\text{jets}\xspace}
\newcommand{\FF} {\ensuremath{F_{\text{F}}}\xspace}

\newcommand{\ZTT} {\ensuremath{\PZ/\PGg^{*}\to\PGt\PGt}\xspace}
\newcommand{\HTT} {\ensuremath{\PH\to\PGt\PGt}\xspace}
\newcommand{\Hgamgam} {\ensuremath{\PH\to\PGg\PGg}\xspace}
\newcommand{\Hllll} {\ensuremath{\PH\to4\ell}\xspace}
\newcommand{\ZLL} {\ensuremath{\PZ/\PGg^{*}\to\ell\ell}\xspace}
\newcommand{\ZEE} {\ensuremath{\PZ/\PGg^{*}\to\Pe\Pe}\xspace}
\newcommand{\ZMM} {\ensuremath{\PZ/\PGg^{*}\to\PGm\PGm}\xspace}
\newcommand{\jettotau} {\ensuremath{\text{jet}\to\tauh}\xspace}
\newcommand{\DeltaLL} {\ensuremath{-2\Delta\ln{\mathcal{L}}}\xspace}

\newcommand{\MELA}{\textsc{mela}\xspace}
\providecommand{\JHUGen}{\textsc{JHUGen}\xspace}
\providecommand{\cmsTable}[1]{\resizebox{\textwidth}{!}{#1}}

\cmsNoteHeader{HIG-20-007} 
\title{Constraints on anomalous Higgs boson couplings to vector bosons and fermions from the production of Higgs bosons using the \texorpdfstring{$\PGt\PGt$}{tau tau} final state}

\date{\today}

\abstract{
A study of anomalous couplings of the Higgs boson to vector bosons and fermions is presented. The data were recorded by the CMS experiment at a center-of-mass energy of pp collisions at the LHC of 13\TeV and correspond to an integrated luminosity of~138\fbinv. The study uses Higgs boson candidates produced mainly in gluon fusion or electroweak vector boson fusion at the LHC that subsequently decay to a pair of \PGt leptons. Matrix-element and machine-learning techniques were employed in a search for anomalous interactions. The results are combined with those from the four-lepton and two-photon decay channels to yield the most stringent constraints on anomalous Higgs boson couplings to date. The pure $CP$-odd scenario of the Higgs boson coupling to gluons is excluded at 2.4 standard deviations. The results are consistent with the standard model predictions.
}

\hypersetup{
pdfauthor={CMS Collaboration},
pdftitle={Constraints on anomalous Higgs boson couplings to vector bosons and fermions from the production of Higgs bosons using the tau tau final state},
pdfsubject={CMS},
pdfkeywords={CMS, Higgs boson, anomalous couplings}}

\maketitle 

\section{Introduction}
\label{sec:intro}

The discovery of the Higgs boson (\PH) by the ATLAS and CMS experiments at the
LHC~\cite{Aad:2012tfa, Chatrchyan:2012xdj, Chatrchyan:2013lba}
has opened a new era for particle physics, wherein the characterization of the new boson is of
crucial importance. Studies of the Higgs boson test the standard model (SM) of particle physics and 
probe for new physics. Thus far, the properties of the \PH are found
to be consistent with the SM predictions~\cite{StandardModel67_1, Englert:1964et,Higgs:1964ia,Higgs:1964pj,Guralnik:1964eu,StandardModel67_2,StandardModel67_3}. In particular, nonzero spin assignments
of the \PH have been excluded~\cite{Khachatryan:2014kca,Aad:2015mxa},
and its spin-parity quantum numbers are consistent with
$J^{PC} = 0^{++}$~\cite{
Chatrchyan:2012jja,Chatrchyan:2013mxa,Khachatryan:2014kca,Aad:2015mxa,Khachatryan:2015mma,Khachatryan:2016tnr,Sirunyan:2017tqd,Sirunyan:2019twz, Sirunyan:2019htt,Chatrchyan:2020htt,CMS-HIG-19-009, CMS-HIG-20-006, CMS-HIG-21-013,Aad:2013xqa,Aad:2015mxa,Aad:2016nal,Aaboud:2017oem,Aaboud:2017vzb,Aaboud:2018xdt,Aad:2020mnm,HttAtlas,ATLAS:2021pkb}. 
However, the limited precision of current
studies allows for anomalous couplings of the \PH with two electroweak gauge bosons (\HVV) or gluons (\Hgg). 
Possible $CP$-violating effects in \PH couplings to fermions (\Hff)
have been constrained by the CMS and ATLAS Collaborations in \ttH 
production~\cite{Chatrchyan:2020htt, CMS-HIG-19-009, HttAtlas}, 
and by the CMS Collaboration in the \HTT decay~\cite{CMS-HIG-20-006},
where $CP$-odd couplings may appear at tree level, and are not suppressed by loop effects. 
In the SM, \Hgg is mediated via loops, where the top quark dominates.  Any observed $CP$ violation in the \Hgg
interaction would indicate either a $CP$-odd Higgs coupling to top quarks (\Htt)  or a new effective interaction requiring
new particles. Thus, a study of the \Hgg coupling provides complementary information on the nature
of the \PH and serves as an indirect search for new phenomena. Both the CMS and ATLAS Collaborations have previously
searched for $CP$-violation in the \Hgg coupling, but these constraints are quite weak~\cite{CMS-HIG-19-009,ATLAS:2021pkb}.

In this paper, we report on a search for anomalous effects, including possible signs of $CP$ violation,
in the tensor structure of the \PH interactions with electroweak bosons and gluons in the production of the \PH. 
The analysis is performed in four $\PGt\PGt$ final states: $\Pe\PGm$, $\Pe\tauh$, $\PGm\tauh$, and $\tauh\tauh$, where \Pe, \PGm and \tauh indicate \PGt decays into electrons, muons and hadrons, respectively. 
We follow the formalism used in previous CMS studies of anomalous couplings in Run~1 and Run~2, described in Refs.~\cite{Khachatryan:2014kca,Chatrchyan:2013mxa, Khachatryan:2014kca, Khachatryan:2015mma,Khachatryan:2016tnr,Sirunyan:2017tqd,Sirunyan:2019twz,Sirunyan:2019htt, Chatrchyan:2020htt, CMS-HIG-19-009}.
The two dominant production channels employed in this study are electroweak vector boson fusion (VBF) and gluon fusion (\ggH).
Compared to our previous study in the \HTT channel~\cite{Sirunyan:2019htt}, we have improved the
sensitivity to anomalous effects with multivariate tools, optimization of the final state categorization,
and an increased data sample. The analysis utilizes a matrix element likelihood approach
(\MELA)~\cite{Chatrchyan:2012jja,Gao:2010qx,Bolognesi:2012mm,Anderson:2013afp,Gritsan:2016hjl} and a neural network to optimize the measurement of anomalous couplings using
production and decay kinematic information.
Compared to our similar study in the \Hllll channel~\cite{CMS-HIG-19-009}, where
$\ell$ denotes an electron or muon and both production and decay information is used,
the inclusion of the \HTT channel leads to a substantial improvement in constraints on
anomalous couplings due to a larger sample of VBF and \ggH events reconstructed in association with two jets.
The results obtained from the two decay channels are further combined to form the most stringent constraint on
anomalous couplings. The combined \ggH results are further combined with the \ttH analysis
using the \Hgamgam and \Hllll decays~\cite{Chatrchyan:2020htt, CMS-HIG-19-009} under the assumption of top quark
dominance in \ggH to constrain the \Htt anomalous couplings.

The paper is organized as follows. The phenomenology of anomalous \HVV and \Hff couplings is discussed
in Sec.~\ref{sec:pheno}. The kinematics of the processes studied and the observables utilized in this study
to search for anomalous contributions are described in Sec.~\ref{sec:kinematics}. The CMS detector is described in Sec.~\ref{sec:CMS}.
The data used in this study, the Monte Carlo (MC) simulation, as well as event reconstruction methods are described
in Sec.~\ref{ss:dataAndMC}. The event selection and categorization is documented in Sec.~\ref{sec:Reco}. Methods to
estimate backgrounds are given in Sec.~\ref{sec:background} and the sources of systematic uncertainty are
listed in Sec.~\ref{sec:Systematics}. The analyses of the \Hgg and \HVV
interactions using \HTT decays are presented in Secs.~\ref{sec:ggH} and~\ref{sec:HVV}, respectively. 
The combination of the \HTT results with the \Hllll and \Hgamgam decay channels
is detailed in Sec.~\ref{sec:combination}. Section~\ref{sec:Summary} summarizes the results.
Tabulated results are provided in the {HEPD}ata record for this analysis~\cite{hepdata}.

\section{Phenomenology of anomalous couplings and cross sections}
\label{sec:pheno}
In this study, we follow the formalism used in the measurement of \PH couplings in earlier CMS
analyses~\cite{Khachatryan:2014kca,Chatrchyan:2013mxa, Khachatryan:2014kca, Khachatryan:2015mma,Khachatryan:2016tnr,Sirunyan:2017tqd,Sirunyan:2019twz,Sirunyan:2019htt, Chatrchyan:2020htt, CMS-HIG-19-009}.
The theoretical approach is described
in Refs.~\cite{Plehn:2001nj,Hankele:2006ma,Accomando:2006ga,Hagiwara:2009wt,Gao:2010qx,DeRujula:2010ys,
Bolognesi:2012mm,Ellis:2012xd,Artoisenet:2013puc,Anderson:2013afp,Dolan:2014upa,Greljo:2015sla,Gritsan:2016hjl}.

Interactions of a spin-0 \PH with two spin-1 gauge bosons $\V\V$, such as $\PW\PW$, $\PZ\PZ$, $\PZ\PGg$, $\PGg\PGg$, and $\Pg\Pg$, are parametrized
by a scattering amplitude that includes three tensor structures with expansion of coefficients up
to $(p^2/\Lambda_{1}^2)$:
\begin{linenomath}
\ifthenelse{\boolean{cms@external}}{
\begin{multline}
\mathcal{A}(\HVV) \sim
\left[ a_{1}^{\V\V}
+ \frac{\kappa_1^{\V\V}p_{1}^2 + \kappa_2^{\V\V} p_{2}^{2}}{\left(\Lambda_{1}^{\V\V} \right)^{2}}
\right]
m_{\V1}^2 \epsilon_{\V1}^* \epsilon_{\V2}^*\\
+ a_{2}^{\V\V}  f_{\mu \nu}^{*(1)}f^{*(2)\mu\nu}
+ a_{3}^{\V\V}   f^{*(1)}_{\mu \nu} {\tilde f}^{*(2)\mu\nu},
\label{eq:formfact-fullampl-spin0}
\end{multline}
}{
\begin{equation}
\mathcal{A}(\HVV) \sim
\left[ a_{1}^{\V\V}
+ \frac{\kappa_1^{\V\V}p_{1}^2 + \kappa_2^{\V\V} p_{2}^{2}}{\left(\Lambda_{1}^{\V\V} \right)^{2}}
\right]
m_{\V1}^2 \epsilon_{\V1}^* \epsilon_{\V2}^*
+ a_{2}^{\V\V}  f_{\mu \nu}^{*(1)}f^{*(2)\mu\nu}
+ a_{3}^{\V\V}   f^{*(1)}_{\mu \nu} {\tilde f}^{*(2)\mu\nu},
\label{eq:formfact-fullampl-spin0}
\end{equation}
}
\end{linenomath}
where $p_{i}$, $\epsilon_{Vi}$, and $m_{\V1}$ are the four-momentum, polarization vector,
and pole mass of the gauge boson, indexed by $i=1,2$.
The gauge boson's field strength tensor and the dual field strength tensor are
$f^{(i){\mu \nu}} = \epsilon_{{\V}i}^{\mu}p_{i}^{\nu} - \epsilon_{{\V}i}^\nu p_{i}^{\mu}$
and ${\tilde f}^{(i)}_{\mu \nu} = \frac{1}{2} \epsilon_{\mu\nu\rho\sigma} f^{(i)\rho\sigma}$.
The coupling coefficients $a_{i}^{\V\V}$, which multiply the three tensor structures,
and $\kappa_i^{\V\V}/(\Lambda_{1}^{\V\V})^2$,
which multiply the next term in the $p^2$ expansion for the first tensor structure,
are to be determined from data, where $\Lambda_{1}$ is the scale of beyond the SM (BSM) physics.
The convention $\epsilon_{0123}=+1$ defines the relative sign of the $CP$-odd and $CP$-even 
couplings. The sign in front of the gauge fields in the covariant derivative defines the sign of the photon
field and sets the sign convention of the $\PZ\PGg$ couplings. 
The conventions adopted in this analysis are discussed in Sec.~\ref{ss:dataAndMC}.

In Eq.~(\ref{eq:formfact-fullampl-spin0}), the only nonzero SM contributions at tree level are $a_{1}^{\PW\PW}$
and $a_{1}^{\PZ\PZ}$, which are assumed to be equal under custodial symmetry.
All other ${\PZ\PZ}$ and ${\PW\PW}$ couplings are considered anomalous contributions,
which are either due to BSM physics or small contributions arising in the SM from
loop effects that cannot be detected with the current precision~\cite{Davis:2021tiv}.
Among the anomalous contributions, considerations of symmetry and gauge invariance require
$a_{1}^{\PZ\PGg}=a_{1}^{\PGg\PGg}=a_{1}^{\Pg\Pg}=0$,
$\kappa_1^{\PZ\PZ}=\kappa_2^{\PZ\PZ}$,
$\kappa_1^{\PGg\PGg}=\kappa_2^{\PGg\PGg}=0$,
$\kappa_1^{\Pg\Pg}=\kappa_2^{\Pg\Pg}=0$,
and $\kappa_1^{\PZ\PGg}=0$~\cite{Gritsan:2020pib}. 
For the $\Pg\Pg$ couplings, the only nonzero couplings are $a_{2}^{\Pg\Pg}$ and $a_{3}^{\Pg\Pg}$, which are anomalous contributions due to BSM physics and do not account for interactions mediated by SM particles via loops.
Therefore, in total there are 13 independent parameters that describe the \PH coupling
to the electroweak gauge bosons and two that describe the coupling to gluons.
The $a_{3}^{\V\V}$ couplings are $CP$-odd, and their presence together with any other $CP$-even
couplings would result in $CP$ violation in a given process.

Our earlier measurements~\cite{Khachatryan:2014kca} and a more recent phenomenological study~\cite{Davis:2021tiv} indicated substantially stronger
limits on $a_2^{\PGg\PGg,\PZ\PGg}$ and $a_3^{\PGg\PGg,\PZ\PGg}$ couplings from $\PH\to\PZ\PGg$
and \Hgamgam decays with on-shell photons than from measurements with virtual photons,
so we do not pursue measurements of these parameters in this paper, and they are set to zero when measuring
other anomalous couplings. 

As the event kinematics of the \PH production in ${\PW\PW}$ fusion and in $\PZ\PZ$ fusion are very similar,
it is essentially impossible to distinguish between $a_i^{\PW\PW}$ and $a_i^{\PZ\PZ}$ in the VBF production. 
It is therefore necessary to choose a convention to set the relative size of the $\PH\PW\PW$ and
$\PH\PZ\PZ$ couplings. The results can be reinterpreted for any chosen relationship between the $a_i^{\PW\PW}$
and $a_i^{\PZ\PZ}$ couplings~\cite{Sirunyan:2019twz}.

{\tolerance=1000 
In our measurements, we adopt two approaches to set the relationship between
the $a_i^{\PW\PW}$ and $a_i^{\PZ\PZ}$ couplings. In the first approach (Approach 1) they are analyzed together assuming
${a_i^{\PW\PW}=a_i^{\PZ\PZ}}$ and
${\kappa_i^{\PZ\PZ}/(\Lambda_{1}^{\PZ\PZ})^2}$ = ${\kappa_i^{\PW\PW}/(\Lambda_{1}^{\PW\PW})^2}$.
In the second approach (Approach 2) we reinterpret the results for the $CP$-violating coupling $a_3$ following the procedure
described in Ref.~\cite{Sirunyan:2019twz}. In this reinterpretation we apply additional considerations
of custodial and SU(2)$\times$U(1) symmetries in the relationships of anomalous
couplings~\cite{deFlorian:2016spz,Gritsan:2020pib}.
With $a_3^{\PGg\PGg}$ and $a_3^{\PZ\PGg}$ set to zero, we are left with a simple
relationship between $a_3^{\PW\PW}$ and $a_3^{\PZ\PZ}$, depending on the Weinberg angle $\theta_W$:
\begin{linenomath}
\begin{equation}
a_3^{\PW\PW} = \cos^2\theta_W~a_3^{\PZ\PZ},  
\label{eq:EFT3}
\end{equation}
\end{linenomath}
\par}

It is convenient to measure the effective cross section ratios \fai rather than the
anomalous couplings $a_i$ themselves, as most uncertainties cancel in the ratio.
Moreover, the effective fractions are conveniently bounded between $-1$ and 1, independent of the coupling
convention. The effective fractional cross sections \fai are defined as follows~\cite{CMS-HIG-19-009}:

\ifthenelse{\boolean{cms@external}}
{
  \begin{equation}\begin{aligned}
    \fathree &= \frac{\abs{a_3}^2 \sigma_{3}}
    {\abs{a_1}^2 \sigma_{1}
      + \abs{a_2}^2 \sigma_{2} + \abs{a_3}^2 \sigma_{3} + \abs{\kappa_{1}}^2 {\sigma}_{\Lambda1}
      + \abs{\kappa_{1}^{\PZ\PGg}}^2 {\sigma}_{\Lambda1}^{\PZ\PGg}
    }
    \\ &\times \sgn\left(\frac{a_{3}}{a_{1}}\right), \\
    \fatwo &= \frac{\abs{a_2}^2 \sigma_{2}}
    {\abs{a_1}^2 \sigma_{1}
      + \abs{a_2}^2 \sigma_{2} + \abs{a_3}^2 \sigma_{3} + \abs{\kappa_{1}}^2 {\sigma}_{\Lambda1}
      + \abs{\kappa_{1}^{\PZ\PGg}}^2 {\sigma}_{\Lambda1}^{\PZ\PGg}
    }
    \\ &\times \sgn\left(\frac{a_{2}}{a_{1}}\right), \\
    \fL &= \frac{\abs{\kappa_{1}}^2 {\sigma}_{\Lambda1}}
    {\abs{a_1}^2 \sigma_{1}
      + \abs{a_2}^2 \sigma_{2} + \abs{a_3}^2 \sigma_{3} + \abs{\kappa_{1}}^2 {\sigma}_{\Lambda1}
      + \abs{\kappa_{1}^{\PZ\PGg}}^2 {\sigma}_{\Lambda1}^{\PZ\PGg}
    }
    \\ &\times\sgn\left(\frac{-\kappa_1}{a_{1}}\right), \\
    \fLzg & = \frac{\abs{\kappa_2^{\PZ\PGg}}^2 {\sigma}^{\PZ\PGg}_{\Lambda1}}
    {\abs{a_1}^2 \sigma_{1}
      + \abs{a_2}^2 \sigma_{2} + \abs{a_3}^2 \sigma_{3} + \abs{\kappa_{1}}^2 {\sigma}_{\Lambda1}
      + \abs{\kappa_{1}^{\PZ\PGg}}^2 {\sigma}_{\Lambda1}^{\PZ\PGg}
    }
   \\ &\times \sgn\left(\frac{-\kappa_2^{\PZ\PGg}}{a_{1}}\right),
    \label{eq:fa_definitions}
    \end{aligned}\end{equation}
}{
\begin{equation}\begin{aligned}
\fathree &= \frac{\abs{a_3}^2 \sigma_{3}}
{\abs{a_1}^2 \sigma_{1}
  + \abs{a_2}^2 \sigma_{2} + \abs{a_3}^2 \sigma_{3} + \abs{\kappa_{1}}^2 {\sigma}_{\Lambda1}
  + \abs{\kappa_{1}^{\PZ\PGg}}^2 {\sigma}_{\Lambda1}^{\PZ\PGg}
}
~\sgn\left(\frac{a_{3}}{a_{1}}\right), \\
\fatwo &= \frac{\abs{a_2}^2 \sigma_{2}}
{\abs{a_1}^2 \sigma_{1}
  + \abs{a_2}^2 \sigma_{2} + \abs{a_3}^2 \sigma_{3} + \abs{\kappa_{1}}^2 {\sigma}_{\Lambda1}
  + \abs{\kappa_{1}^{\PZ\PGg}}^2 {\sigma}_{\Lambda1}^{\PZ\PGg}
}
~\sgn\left(\frac{a_{2}}{a_{1}}\right), \\
\fL &= \frac{\abs{\kappa_{1}}^2 {\sigma}_{\Lambda1}}
{\abs{a_1}^2 \sigma_{1}
  + \abs{a_2}^2 \sigma_{2} + \abs{a_3}^2 \sigma_{3} + \abs{\kappa_{1}}^2 {\sigma}_{\Lambda1}
  + \abs{\kappa_{1}^{\PZ\PGg}}^2 {\sigma}_{\Lambda1}^{\PZ\PGg}
}
~\sgn\left(\frac{-\kappa_1}{a_{1}}\right), \\
\fLzg & = \frac{\abs{\kappa_2^{\PZ\PGg}}^2 {\sigma}^{\PZ\PGg}_{\Lambda1}}
{\abs{a_1}^2 \sigma_{1}
  + \abs{a_2}^2 \sigma_{2} + \abs{a_3}^2 \sigma_{3} + \abs{\kappa_{1}}^2 {\sigma}_{\Lambda1}
  + \abs{\kappa_{1}^{\PZ\PGg}}^2 {\sigma}_{\Lambda1}^{\PZ\PGg}
}
~\sgn\left(\frac{-\kappa_2^{\PZ\PGg}}{a_{1}}\right),
\label{eq:fa_definitions}
\end{aligned}\end{equation}
}
where $\sigma_i$ is the cross section for the process corresponding to $a_i=1$ with all other couplings set to zero.
The choice of the sign for the $\kappa_1$ and $\kappa_2^{\PZ\PGg}$ terms follows the convention 
introduced in the prior results~\cite{Khachatryan:2014kca,Sirunyan:2017tqd,Sirunyan:2019twz,CMS-HIG-19-009}. 
The other sign conventions follow the \JHUGen~7.0.2~\cite{Gao:2010qx,Bolognesi:2012mm,Anderson:2013afp,Gritsan:2016hjl} event generator, as discussed in Section~\ref{ss:dataAndMC} and
Ref.~\cite{Davis:2021tiv}.
For consistency with previous CMS measurements in the \Hllll channel~\cite{Khachatryan:2014kca,CMS-HIG-19-009}, 
the $\sigma_i$ coefficients are defined for the $\Pg\Pg\to\PH\to\PV\PV\to2\Pe2\mu$ process. 
The numerical values are given in Table~\ref{tab:ac_crossections} as calculated using the \JHUGen event generator.
It is assumed that the couplings in Eq.~(\ref{eq:formfact-fullampl-spin0}) are constant and real,
and therefore this formulation is equivalent to an effective Lagrangian formalism. 

\begin{table}[ht!]
\centering
\topcaption{
Cross sections for the anomalous contributions ($\sigma_i$) used to define the fractional cross sections~\cite{CMS-HIG-19-009}. The $\sigma_i$ values are defined as the cross section computed with $a_{i}=1$ and all other couplings set to zero. All cross sections are given relative to the SM value ($\sigma_1$). 
In the case of the $\kappa_{1}$ and $\kappa_{2}^{\PZ\PGg}$ couplings, the numerical values $\Lambda_1 = \Lambda_1^{\PZ\PGg} = 100\GeV$ are considered so as to keep all coefficients of similar order of magnitude.
}
\renewcommand{\arraystretch}{1.25}
\begin{scotch}{ccc}
   Coupling  & $\sigma_i$/$\sigma_1$ \\
\hline
$a_3$   & 0.153      \\
$a_2$   & 0.361      \\
$\kappa_{1}$ & 0.682  \\ 
$\kappa_{2}^{\PZ\PGg}$ & 1.746 \\
\end{scotch}
\label{tab:ac_crossections}
\end{table}

The \ggH process is a purely loop-induced process, which in the SM is generated by the top quark, with a smaller 
contribution from the bottom quark~\cite{Hamilton:2015nsa}. This interaction is $CP$-even in the SM. However, a contribution 
of the $CP$-odd interaction in the \PH coupling to fermions is not ruled out, and the search for such 
a $CP$-violating interaction can be performed in \ttH production and \HTT decay. Under the assumption 
that other BSM particles do not contribute to the gluon fusion loop, a $CP$-structure measurement in the \ggH
process is equivalent to the measurement of the $CP$ structure in Yukawa interactions, which can be parametrized 
with the amplitude
\begin{equation}
 \mathcal{A}(\Hff) = - \frac{m_f}{v}
	\bar{\psi}_{\f} \left ( \kappaf  + \mathrm{i} \, \kappaftilde  \PGg_5 \right ) {\psi}_{\f}.
	\label{eq:ampl-spin0-qq}
\end{equation}
The effective fractional cross section for \Hff couplings is defined as~\cite{Chatrchyan:2020htt} 
\begin{equation}
\fCPff = \frac{\abs{\kappaftilde}^2 }{\abs{\kappaf}^2 + \abs{\kappaftilde}^2}
\sgn\left(\frac{\kappaftilde}{\kappaf} \right).
\label{eq:fCP_definitions}
\end{equation}
An equivalent effective mixing angle \alphacp is also used to describe the $CP$-odd contribution to the \PH Yukawa couplings and is defined as
\begin{equation}
\alphacp= \tan^{-1} \left(\frac{\kappaftilde }{ \kappaf} \right),
\label{eq:alpha_defintion}
\end{equation}
where $\abs{\fCPff}= \sin^2\alphacp$. Therefore, with just two contributions to the gluon
fusion loop ($CP$-even and $CP$-odd fermion couplings), the two parameters are equivalent. However, with consideration of multiple 
contributions, as discussed in the case of electroweak \HVV couplings above, multiple fractional contributions have to be 
defined and a single angle is not sufficient. The \ggH loop can be generated by unknown heavy BSM particles, in addition
to the SM fermions, and the effective coupling results in the $CP$-even $a_{2}^{\Pg\Pg}$ and $CP$-odd $a_{3}^{\Pg\Pg}$ couplings,
defined in Eq.~(\ref{eq:formfact-fullampl-spin0}). 
In the effective field theory (EFT) approach~\cite{deFlorian:2016spz}, they correspond to two EFT couplings in the Higgs basis:
\begin{equation}\begin{aligned}
  &  \cgg = -\frac{1}{2\pi\alpS} a_2^{\Pg\Pg},  \label{eq:EFTpar5}  \\
  & \cggtilde = -\frac{1}{2\pi\alpS} a_3^{\Pg\Pg}, 
\end{aligned}\end{equation}
where \alpS is the running strong coupling constant.
Therefore, there are at least four contributions to consider
(\kappat,  \kappattilde, \cgg, \cggtilde), where in the SM we have (\kappat,  \kappattilde, \cgg, \cggtilde) = (1, 0, 0, 0).
The dependence of the \ggH cross section and \PH branching fractions on these parameters is given in Ref.~\cite{Davis:2021tiv}. Under the assumptions that the only SM particles contributing to the loop are the top and bottom quarks and (\kappab, \kappabtilde) = (1, 0), the \ggH cross section relative to the SM expectation is given as  
\ifthenelse{\boolean{cms@external}}
{
\begin{linenomath}
\begin{equation}\begin{aligned}
\muggh = & 1.1068\kappat^2 + 0.0082 - 0.1150\kappat + 2.5717\kappattilde^2 \\
& + 1.0298(12\pi^2 \cgg)^2 + 2.3170(8\pi^2 \cggtilde)^2 \\ 
& + 2.1357(12\pi^2\cgg)\kappat \\
& - 0.1109(12\pi^2 \cgg) 
 + 4.8821(8\pi^2 \cggtilde)\kappattilde.
\label{eq:cgg_crosssection}
\end{aligned}\end{equation}
\end{linenomath}
}{
  \begin{linenomath}
    \begin{equation}\begin{aligned}
    \muggh = & 1.1068\kappat^2 + 0.0082 - 0.1150\kappat + 2.5717\kappattilde^2 + 1.0298(12\pi^2 \cgg)^2 + 2.3170(8\pi^2 \cggtilde)^2 \\ 
    & + 2.1357(12\pi^2\cgg)\kappat - 0.1109(12\pi^2 \cgg) 
     + 4.8821(8\pi^2 \cggtilde)\kappattilde.
    \label{eq:cgg_crosssection}
    \end{aligned}\end{equation}
    \end{linenomath}
}

Within the framework of our analysis, however, it is hard to distinguish between the \kappaf and $a_{2}^{\Pg\Pg}$
contributions, or between \kappaftilde and  $a_{3}^{\Pg\Pg}$. There are small differences in the transverse momentum \pt distributions
of the \PH, and one can also observe effects in the off-shell \PH production~\cite{Gritsan:2020pib}. 
However, the former is too small to have a noticeable effect in this analysis, 
and the latter does not come within the scope of our analysis based on the on-shell production. Therefore, we absorb the SM fermion 
loop contribution, dominated by the heavy top quark, into the overall $a_{2}^{\Pg\Pg}$ and $a_{3}^{\Pg\Pg}$ couplings. The only remaining effective fractional cross section for the \Hgg  interaction is defined as~\cite{CMS-HIG-19-009} 

\begin{equation}\begin{aligned}
\fggh  = \frac{\abs{a_3^{\Pg\Pg}}^2 } {\abs{a_2^{\Pg\Pg}}^2  + \abs{a_3^{\Pg\Pg}}^2 } 
~\sgn\left(\frac{a_{3}^{\Pg\Pg}}{a_{2}^{\Pg\Pg}} \right).
\label{eq:fggH_definitions}
\end{aligned}\end{equation}
Under the assumption that only the top and bottom quarks contribute to gluon fusion with 
$\kappat=\kappab$ and $\kappattilde=\kappabtilde$, the following relationship~\cite{Gritsan:2020pib} holds:
\begin{equation}
\left| \fCP \right| = \left(1 +2.38 \left[  \frac{1}{\left|\fggh \right|} -1  \right]  \right)^{-1}.
\label{eq:fai-relationship-hgg-tth}
\end{equation}

In this paper, we present a search for anomalous \Hgg couplings in the gluon fusion production and
anomalous \HVV couplings in VBF and
associative \PH production with a \PW or \PZ boson (\VH).
In addition, the \Hgg measurement is interpreted in terms of constraints on \Hff couplings
under the assumption of top quark dominance in gluon fusion.
We measure a given anomalous coupling while setting the values of all other anomalous coupling parameters
to zero, with the exception of measuring the $CP$-odd parameters, \fathree and \fggh, as $CP$ violation in VBF and \VH production would modify the same kinematic distributions as those in the \ggH process. Therefore, we treat \fathree as an unconstrained parameter when we measure \fggh, and vice versa. 
The presence of $CP$ violation in the decay of the \PH to a pair of \PGt leptons does not affect the measurements of the production process, and thus we assume the SM kinematics for the \PH decays.

\section{Production and decay kinematics, and discriminants}
\label{sec:kinematics}

\begin{figure*}[!htb]
\centering
\includegraphics[width=0.45\textwidth]{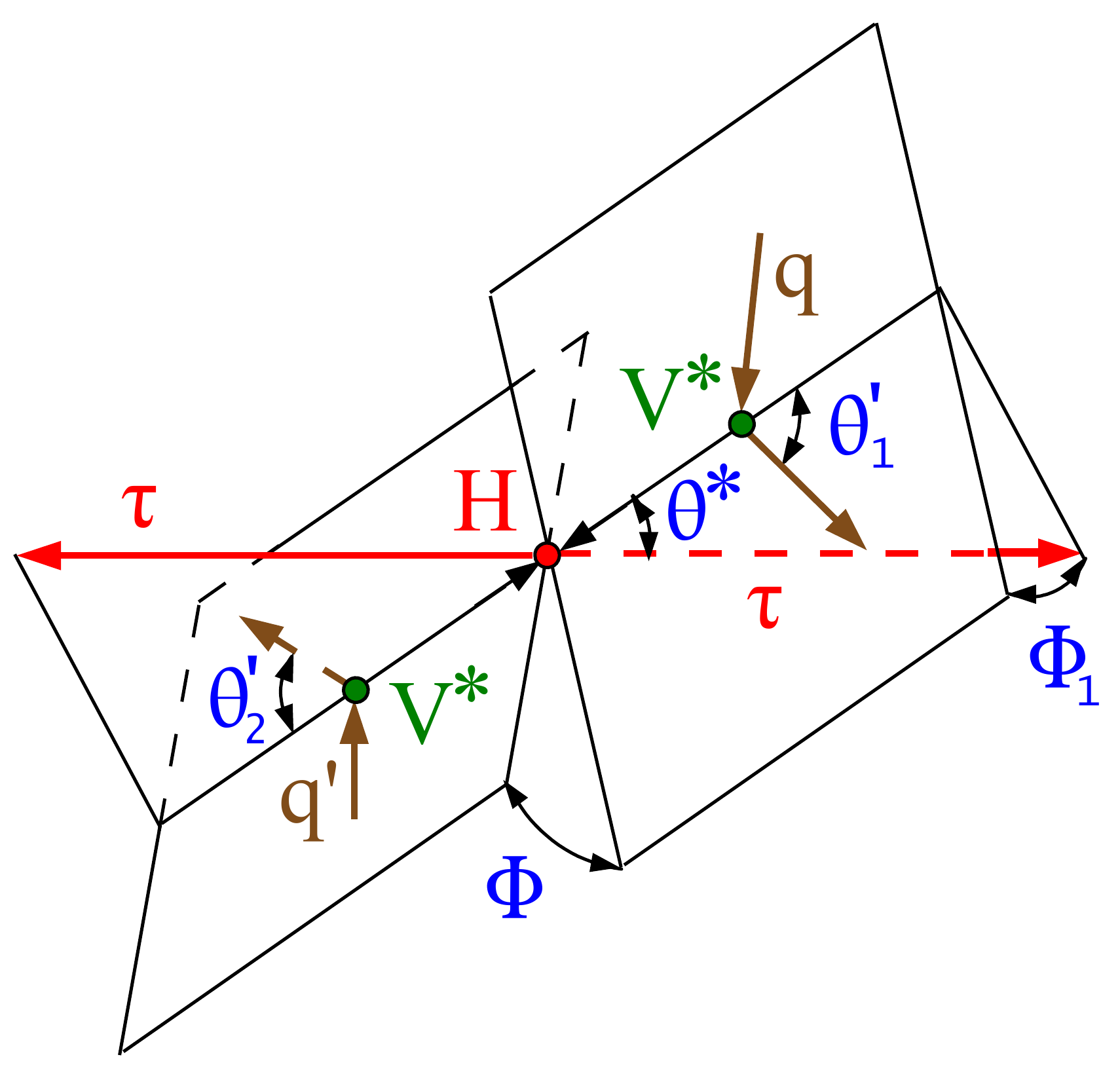}
\includegraphics[width=0.45\textwidth]{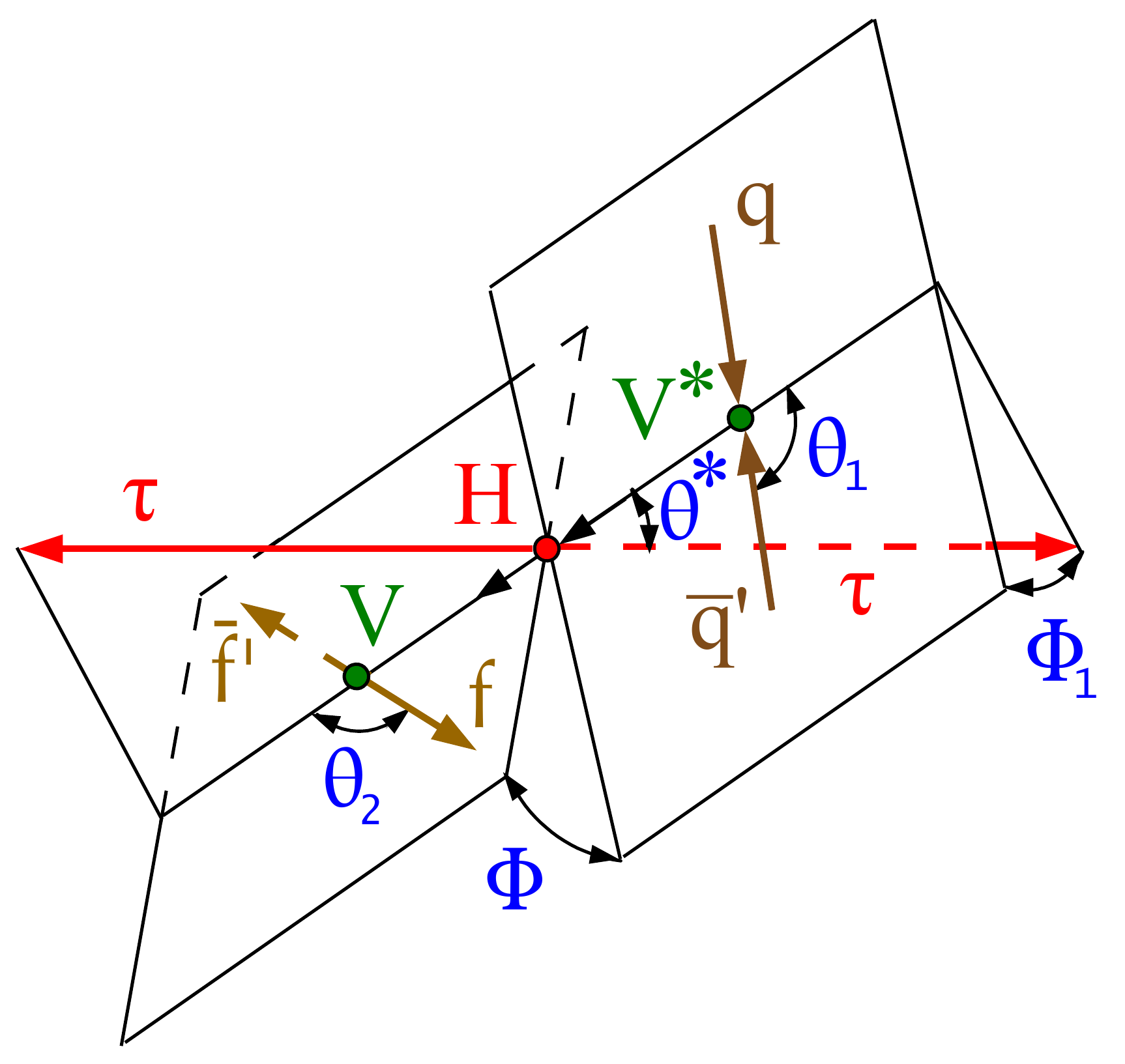}
\caption
{Illustrations of \PH production in VBF ($\PQq{\PQq^\prime}\to \PQq{\PQq^\prime} \PH$) (left) 
and \VH ($\PQq\PAQq^\prime\to \V^*\to \V\PH\to\PQq{\PQq^\prime} \PH$) (right) in the rest frame of the \PH. The decay $\PH\to\PGt\PGt$ is shown without illustrating the further decay chain. 
The incoming partons and fermions in the \PV decay are shown in brown and the intermediate or final-state particles are shown in red and green. 
The angles characterizing kinematic distributions are shown in blue and are defined in the respective rest frames~\cite{Gao:2010qx,Anderson:2013afp}. The illustration for H production via \ggH in association with two jets is identical to the VBF diagram, except with $\PV=\Pg$.
\label{fig:kinematics}}
\end{figure*}

Because exotic nonzero spin assignments of the \PH have been excluded~\cite{Chatrchyan:2012jja,Chatrchyan:2013mxa,Khachatryan:2014kca,Aad:2015mxa,Khachatryan:2015mma,Khachatryan:2016tnr,Sirunyan:2017tqd,Sirunyan:2019twz, Sirunyan:2019htt,Chatrchyan:2020htt,CMS-HIG-19-009, CMS-HIG-20-006, CMS-HIG-21-013,Aad:2013xqa,Aad:2015mxa,Aad:2016nal,Aaboud:2017oem,Aaboud:2017vzb,Aaboud:2018xdt,Aad:2020mnm,HttAtlas,ATLAS:2021pkb},
we focus on the analysis of couplings of a spin-0 \PH.
When combined with the momentum transfer squared of the vector bosons, $p_1^2$ and $p_2^2$, the five angles
in Fig.~\ref{fig:kinematics} provide complete kinematic information for production and decay
of the \PH.

There are four possible and practical ways to access $CP$-violating effects 
(or more generally anomalous \HVV or \Hff couplings) using the reconstructed $\PH\to \PGt\PGt$ events:
\begin{enumerate}
  \item correlation of \PH and two quark jets or leptons in VBF and \VH production;
  \item correlation of \PH and two quark jets in \ggH production;
  \item correlation of \PH and quark jets in \ttH or $\PQt\PH$ production; and
  \item correlation of decay products of two \PGt leptons. 
\end{enumerate}
There are no spin correlations between the production and decay
through a spin-0 object. Therefore, all four of the above processes can be studied independently
and they target different parameters that are independent, even though all of them may be related to
anomalous effects.
This analysis focuses on searching for anomalous effects in the topologies described as the first and second 
items above. We refer to those as the anomalous \HVV and \Hgg
couplings, respectively.

\subsection{Correlation of \texorpdfstring{\PH}{H} and two quark jets or leptons in VBF and \texorpdfstring{\VH}{VH} production}
\label{subsec:vbf}

Kinematic distributions of associated particles in VBF and \VH production 
are sensitive to the quantum numbers and anomalous couplings of the \PH. 
A set of observables could be defined in production, such as 
$\mathbf{\Omega}^\text{assoc}=\{\theta_1^\mathrm{VBF}, \theta_2^\mathrm{VBF}, \theta^{*\mathrm{VBF}}, \Phi^\mathrm{VBF}, \Phi_1^\mathrm{VBF}, p_1^{2,\mathrm{VBF}}, p_2^{2,\mathrm{VBF}} \}$
for the VBF process or
$\mathbf{\Omega}^\text{assoc}=\{\theta_1^{\VH}, \theta_2^{\VH}, \theta^{*\VH}, \Phi^{\VH}, \Phi_1^{\VH}, p_1^{2,\VH}, p_2^{2,\VH} \}$
for the \VH process (as shown in Fig.~\ref{fig:kinematics} and discussed in Ref.~\cite{Anderson:2013afp}).
It is a challenging task to perform an optimal analysis in a multidimensional space of observables. 
The \MELA method introduced earlier~\cite{Chatrchyan:2012xdj,Gao:2010qx,Bolognesi:2012mm,Anderson:2013afp,Gritsan:2016hjl}
is designed to reduce the number of observables to the minimum, while retaining all essential information. Two types
of discriminants are defined for the production process.
One type of discriminant separates the process with anomalous couplings (denoted as generic BSM here) from the SM one:

\begin{equation}
\DBSM = \frac{\mathcal{P}_\mathrm{SM}(\vec\Omega) }{\mathcal{P}_\mathrm{SM}(\vec\Omega) + \mathcal{P}_\mathrm{BSM}(\vec\Omega)},
\label{eq:melaDbsm} 
\end{equation}

where the probability density $\mathcal{P}$ of a certain process (either SM or anomalous signal) is calculated using the~\MELA~\cite{Gao:2010qx,Bolognesi:2012mm,Anderson:2013afp,Gritsan:2016hjl} package, that contains a library of matrix elements for the signal processes from \JHUGen.
The discriminant for each anomalous coupling is listed in Table~\ref{tab:ac_discriminants}. 

\begin{table}[ht!]
\centering
\topcaption{ 
List of discriminants for separating anomalous couplings from the SM contribution. The third column indicates the hypothesis that was assumed for the signal production process when computing the matrix elements that are inputs to the discriminants. For the \DzggH discriminant, the ``$\Pg\Pg\PH$" label indicates that this observable is constructed using matrix elements computed for the $\Pg\Pg\PH$ production process to differentiate it from the equivalent discriminant for the VBF process (\Dz). For the \DLzg for discriminant, the ``$\PZ\PGg$" label is used to indicate that it targets the $\kappa_{2}^{\PZ\PGg}$ anomalous coupling parameter to differentiate it from the \DL discriminant that targets $\kappa_{1}$.
}
\renewcommand{\arraystretch}{1.25}
\begin{scotch}{ccc}
Coupling & Discriminant & Matrix element process \\
\hline
$a_3^{\Pg\Pg}$ & \DzggH & \ggH \\
$a_3$ & \Dz & VBF \\
$a_2$ & \Dzh & VBF \\
$\kappa_{1}$ & \DL & VBF \\ 
$\kappa_{2}^{\PZ\PGg}$ & \DLzg & VBF\\
\end{scotch}
\label{tab:ac_discriminants}
\end{table}

The second type of discriminant isolates the interference contribution:

\begin{equation}
\Dint = \frac{ \mathcal{P}_\mathrm{SM-BSM}^\text{int}(\vec\Omega)} {\mathcal{P}_\mathrm{SM}(\vec\Omega) +\mathcal{P}_\mathrm{BSM}(\vec\Omega)},
\label{eq:melaDint}
\end{equation}

where $\mathcal{P}_\mathrm{SM-BSM}^\text{int}$ is the interference part of the probability distribution for a process with a mixture of the SM and
anomalous contributions. 
This discriminant is called \DCP (\DCPggH) in the $CP$-odd VBF (\ggH) amplitude analysis. The discriminant is this case is a $CP$-odd observable, and a forward-backward asymmetry in its distribution would indicate $CP$ violation.
Probabilities are normalized for the matrix elements to give the same cross sections in the relevant phase space of each process.
 Such normalization leads to a balanced distribution of events in the range between 0 and 1, or between $-1$ and 1, for the \DBSM and \Dint discriminants, respectively. 

The two other observables in Eqs.~(\ref{eq:melaDbsm}) and~(\ref{eq:melaDint})
rely only on signal matrix elements and are well defined. One can apply the Neyman-Pearson lemma to prove that, in the absence of detector smearing,
they become the minimal and complete set of optimal observables~\cite{Anderson:2013afp,Gritsan:2016hjl}
for the measurement of the \fai parameters defined in Sec.~\ref{sec:pheno}.

In application to the $CP$ measurement with the \fathree parameter, the two optimal observables are called
\Dz and \DCP, because $J^P=0^-$ is the BSM hypothesis in this case,
and the interference discriminant is an unambiguously $CP$-sensitive observable. A distinct forward-backward
asymmetry in the \DCP distribution (forward defined as  $\DCP>0$ and backward as
$\DCP<0$) appears only in the presence of $CP$ violation. These observables could be defined
for both VBF and $\V\PH$ processes. However, since the analysis selection is optimized for the VBF process,
the probabilities in the discriminant calculation in Eqs.~(\ref{eq:melaDbsm}) and~(\ref{eq:melaDint}) are defined for the VBF process.

\subsection{Correlation of \texorpdfstring{\PH}{H} and two jets in the production of \texorpdfstring{\PH}{H} via \texorpdfstring{\ggH}{ggH}}
\label{subsec:ggh}

Kinematic distributions of associated particles in \ggH production 
are also sensitive to the quantum numbers and anomalous \Hgg couplings.
The set of observables $\mathbf{\Omega}$ in this topology is identical to the VBF process
(as shown in Fig.~\ref{fig:kinematics} and discussed in Ref.~\cite{Anderson:2013afp}). 

Similar to the VBF and \VH study, we form optimal \DzggH and
\DCPggH observables that are sensitive to $CP$ violation. However,
unlike in the VBF production study, the sensitivity to $CP$ violation using \MELA observables in \ggH
production is comparable to the signed azimuthal difference between two leading jets, defined as~\cite{Klamke:2007cu} 
\begin{linenomath} 
\begin{equation}
\dphi = \phi(\mathrm{j}_{1}) - \phi(\mathrm{j}_{2}),~\text{with}~\eta(\mathrm{j}_{1}) < \eta(\mathrm{j}_{2}).
\end{equation}
\end{linenomath}
The sign is defined by ordering the jets in $\eta$, which ensures that the observable is sensitive to the interference between the $CP$-even and $CP$-odd contributions~\cite{Plehn:2001nj,Hankele:2006ma,Hankele:2006ja}.
Thus, in this study we cross check the results obtained using the \MELA discriminants with a simplified approach that does not rely on multivariate techniques and utilizes \dphi as the $CP$-sensitive observable.
We will refer to these approaches as the ``\MELA method" and the ``\dphi method", respectively.

\section{The CMS detector}
\label{sec:CMS}
The main feature of the CMS detector is a superconducting solenoid of 6\unit{m} internal diameter, that provides
a magnetic field of 3.8\unit{T}. Within the volume of the solenoid, there are a silicon pixel and strip tracker
detectors, as well as a lead tungstate crystal electromagnetic calorimeter (ECAL), and a brass and scintillator
hadron calorimeter; both calorimeters are composed of a barrel and two endcap sections.
Forward calorimeters extend the coverage in pseudorapidity, $\eta$. The CMS muon system is comprised of
gas-ionization chambers embedded in the steel flux-return yoke outside the CMS solenoid.

The CMS data acquisition system employs a two-tiered trigger system~\cite{Khachatryan:2016bia} to select events of
interest. The first level (L1), composed of custom hardware processors, utilizes information from muon detectors
and both calorimeters to select collision events at a rate of about 100\unit{kHz} within a fixed latency below 4\mus.
The second level, also known as the high-level trigger, further reduces the event acceptance rate to about
1\unit{kHz} before data storage by using a full event reconstruction software, optimized for fast processing,
running on a computing farm.

A more detailed description of the CMS detector, together with a definition of the coordinate system used and
the relevant kinematic variables, can be found in Ref.~\cite{Chatrchyan:2008zzk}.

\section{Data and simulation samples}
\label{ss:dataAndMC}
The data samples used in this analysis correspond to integrated luminosities of 36.3, 41.5 and 59.7\fbinv collected in 2016, 2017 and 2018, respectively, for a total of 138\fbinv collected during Run~2 of the CERN LHC at a proton-proton (\PP) center-of-mass collision
energy of 13\TeV~\cite{CMS-LUM-17-003,CMS-PAS-LUM-17-004,CMS-PAS-LUM-18-002}.

MC simulation is used to model signal and background processes in \PP interactions at the LHC
and their reconstruction in the CMS detector. All MC samples are interfaced with the \PYTHIA~\cite{Sjostrand:2014zea}
generator for parton showering, 
where versions 8.212 and 8.226 are used for 2016, and version 8.230 is used for 2017--2018 simulations.
All the MC samples are further processed through a dedicated CMS detector simulation based on the
\GEANTfour program~\cite{Agostinelli2003250}.

Following the formalism discussed in Sec.~\ref{sec:pheno}, the samples with the SM and anomalous \PH couplings
in VBF and \VH production are generated with the~\JHUGen
program at leading order (LO) in quantum chromodynamics (QCD). All the simulated scenarios are reweighted
to model any other set of \PH couplings using the~\MELA package.
The VBF and \VH \JHUGen SM simulations, after parton showering modeling, are explicitly compared with
the next-to-leading order (NLO) QCD SM simulations produced by \POWHEG~2.0~\cite{Nason:2004rx,Frixione:2007vw, Alioli:2010xd,Nason:2009ai, Jezo:2015aia, Granata:2017iod}
and no significant differences are found in kinematic observables. Therefore, the~\JHUGen
simulation is used to describe kinematics of the VBF and \VH processes with anomalous coupling
effects in VBF and \VH processes, with the expected yields scaled to match the SM theoretical
predictions for inclusive cross sections and \HTT branching fraction from Ref.~\cite{deFlorian:2016spz}, and the \POWHEG~2.0 SM prediction of relative event yields in the categorization of events based on associated particles.
 
Anomalous \ggH events are produced with up to two jets at NLO QCD accuracy using~\MGvATNLO~2.6.0~\cite{Alwall:2014hca,Frederix:2012ps,Demartin:2014fia}
and are also studied with \JHUGen at LO. 
The inclusive cross section and \HTT branching fraction are scaled to match the SM theoretical predictions from Ref.~\cite{deFlorian:2016spz}, and the \pt and jet multiplicity distributions are reweighted to match the \POWHEG~\textsc{nnlops} predictions~\cite{Hamilton:2013fea,Hamilton:2015nsa}.
The relationship between the \Hff and \Hgg couplings follows \JHUGen 
with the relative sign of $CP$-odd and $CP$-even coefficients opposite to that assumed in \MGvATNLO~2.6.0.
This choice corresponds to the convention $\epsilon_{0123}=+1$~\cite{Gritsan:2020pib}.
The sign convention of the photon field in \JHUGen is opposite to that in \MGvATNLO,
which leads to the opposite sign of the $\PH\PZ\PGg$ couplings. This sign convention depends on 
the sign in front of the gauge fields in the covariant derivative and this analysis follows
$D_\mu = \partial_\mu -\mathrm{i}e \sigma^i W_\mu^i/(2s_w) + \mathrm{i}e B_\mu/(2 c_w)$
used in \JHUGen~\cite{Davis:2021tiv}.

The \PYTHIA event generator is used to model the \PH decay to \PGt leptons and the decays of the \PGt leptons.
Both scalar and pseudoscalar \HTT decays and their interference have been simulated
to confirm that the observables used in the analysis are not sensitive to anomalous couplings affecting the \HTT decays.
Thus, the default samples are generated with the SM \HTT decay process.

{\tolerance=1000
The \MGvATNLO~\cite{Alwall:2014hca} generator is used to
produce \Wjets and $\PZ\to\Pe\Pe/\PGm\PGm$+jets samples at LO accuracy.
The \MGvATNLO generator is also used for diboson production simulated at NLO, whereas \POWHEG version 2.0 is used for \ttbar~\cite{Alioli:2011as} and
single top quark ($t$-channel) production~\cite{Frederix:2012dh}, and \POWHEG version 1.0 is used for single top quark production in association with a \PW boson~\cite{Re:2010bp}.
\par}
For processes simulated at NLO (LO) in QCD with the \MGvATNLO generator, events characterized by different parton multiplicities from the matrix element calculation are merged via the FxFx~\cite{Frederix:2012ps} (MLM~\cite{Alwall:2007fs}) prescription.

{\tolerance=1000
The \PYTHIA parameters affecting the description of the underlying event are set to the {CUETP8M1}~\cite{Khachatryan:2015pea} tune for 2016 simulations, except for the \ttbar sample where the {CUETP8M2T4}~\cite{CMS:2016kle} tune is used,
and the {CP5}~\cite{TuneCP5} tune for 2017--2018 simulations.         
We use the NNPDF 3.0~\cite{Ball:2014uwa} (3.1~\cite{Ball:2017nwa}) parton distribution functions, PDFs, for 2016 (2017--2018) simulations.
\par}
Simulated events include the contribution from additional \PP interactions within the same or adjacent bunch
crossings (pileup) and are weighted to reproduce the observed pileup distribution in data.

\section{Event selection}
\label{sec:Reco}

The reconstruction of recorded and simulated events relies on the particle-flow (PF)
algorithm~\cite{CMS-PRF-14-001}, which combines the information from the CMS subdetectors to identify and reconstruct muons, electrons, photons, and charged and neutral hadrons emerging from \PP collisions.
Combinations of these PF candidates are used to reconstruct higher-level objects such as jets, \PGt candidates, or
missing transverse momentum, \ptvecmiss.

The primary vertex (PV) is taken to be the vertex corresponding to the hardest scattering in the event, evaluated using tracking information alone, as described in Sec. 9.4.1 of Ref.~\cite{CMS-TDR-15-02}.

Electrons are identified with a multivariate discriminant
combining several quantities describing the track quality, the shape of the energy deposits in the ECAL,
and the compatibility of the measurements from the tracker and the ECAL~\cite{Khachatryan:2015hwa}.
Muons are identified with requirements on the quality of the track reconstruction and on the number of
measurements in the tracker and the muon systems~\cite{Sirunyan:2018muon}.
A relative isolation variable, $I^\ell$, is defined as the total energy deposited in a cone of size of $R<0.3$ (0.4) centered on the electron (muon) direction divided by the \pt of the lepton. The expected contribution to the energy sum from pileup interactions is estimated and subtracted from the total. 
To reject lepton candidates arising from misidentified jet constituents or from hadron decays, we require that $I^\ell<0.15$. 

Hadronic jets are clustered from the reconstructed PF particles using the infrared and collinear safe anti-\kt algorithm~\cite{Cacciari:2008gp, Cacciari:2011ma} with a distance parameter $\DR=\sqrt{\smash[b]{(\Delta\eta)^2+(\Delta\phi)^2}}$ of 0.4. 
Jet momentum is determined as the vector sum of all particle momenta in the jet, and is found from simulation to be, on average, within 5 to 10\% of the true momentum over the whole \pt spectrum and detector acceptance. Pileup interactions can contribute additional tracks and calorimetric energy depositions to the jet momentum. To mitigate this effect, charged particles identified to be originating from pileup vertices are discarded and an offset correction is applied to correct for the remaining contributions. Jet energy corrections are derived from simulation to bring the measured response of jets to that of particle level jets on average. {\it In situ} measurements of the momentum balance in dijet, $\text{photon} + \text{jet}$, $\PZ + \text{jet}$, and multijet events are used to account for any residual differences in the jet energy scale between data and simulation~\cite{Khachatryan:2016kdb}. The jet energy resolution amounts typically to 15\%--20\% at 30\GeV, 10\% at 100\GeV, and 5\% at 1\TeV~\cite{Khachatryan:2016kdb}. Additional selection criteria are applied to each jet to remove jets potentially dominated by anomalous contributions from various subdetector components or reconstruction failures.
In this analysis, jets are required to have $\pt>30\GeV$
and $\abs{\eta}<4.7$, and
to be separated from the reconstructed visible \PGt decay products by a distance parameter of at least 0.5,
where $\phi$ is the azimuthal angle in radians.
Data collected in the most forward region of the ECAL endcaps were affected by large amounts of noise during the 2017
run, which led to disagreements between simulation and data. To mitigate this effect,
jets used in the analysis of the 2017 data are discarded if they have $\pt < 50\GeV$ and $2.650 < \abs\eta < 3.139$.
Hadronic jets that contain b quarks (``{\cPqb} jets'') are identified using
a deep neural network (DNN), called the deep combined secondary vertex algorithm~\cite{Sirunyan:2017ezt}.

Hadronically decaying \PGt leptons are reconstructed with the hadron-plus-strips
algorithm~\cite{Khachatryan:2015dfa, Sirunyan:2018pgf}, which is seeded with anti-\kt jets with $\pt>14\GeV$.
This algorithm reconstructs \tauh candidates based on the
number of tracks and the number of ECAL strips with energy deposits within the
associated $\eta$-$\phi$ plane and reconstructs one-prong,
one-prong+$\PGpz$(s), and three-prong decay modes (where a ``prong" refers to a charged hadron constituent).
For this analysis, a DNN discriminator is used to identify hadronic decays of \PGt leptons~\cite{TAU-20-001}. The input variables to the DNN include variables
related to the \tauh isolation, \tauh lifetime, and other detector-related
variables. These variables serve as input to a DNN, which provides an output
discriminant. The threshold on the output discriminant depends on the \tauh \pt and provides
a \tauh identification (ID) and reconstruction efficiency of about 60\%.
Two other DNNs are used to reject electrons and muons misidentified as \tauh candidates using dedicated criteria
based on the consistency between the measurements in the tracker, the calorimeters, and the muon detectors.

The \ptvecmiss is defined as
the negative vector sum of the \pt of all PF
candidates~\cite{CMS-JME-17-001}. Its magnitude is referred to as \ptmiss.

The invariant mass of the
$\PGt\PGt$ system \mtautau is a key variable for separating \PH candidate events from the background
in this analysis. The \mtautau is reconstructed using the \textsc{FastMTT} algorithm, which is similar
to the \textsc{SVFIT} algorithm~\cite{Bianchini:2014vza} used in previous CMS publications, except that it uses a
simplified mass likelihood function to reduce the computation time.
This algorithm makes use of 
the \ptvecmiss and its uncertainty and the four-vectors of the reconstructed visible \PGt decay products
to calculate an estimate of the mass of the parent boson and
the full four-momenta of the \PH decay products needed to calculate \MELA kinematic
observables discussed in Sec.~\ref{sec:kinematics}.
Compared to the procedure described in Ref.~\cite{Bianchini:2014vza}, the \textsc{FastMTT} algorithm removes the contributions of the leptonic and hadronic \PGt decay matrix elements to the likelihood function, and assumes that the neutrinos are collinear to the visible \PGt leptons.  
This gives a similar \mtautau resolution as the \textsc{SVFIT} algorithm, but the computation time is reduced by two orders of magnitude. 

\subsection{Event categorization}
\label{subsec:Selection}

Selected events are classified according to four decay channels,
$\Pe\PGm$, $\Pe\tauh$, $\PGm\tauh$, and $\tauh\tauh$, where \Pe and \PGm indicate \PGt decays into electrons and muons, respectively.
The resulting event samples are made mutually exclusive by
discarding events that have additional loosely identified
and isolated electrons or muons. In cases where multiple pairs can be formed due to the presence of additional \tauh candidates, we select the pair that includes the \tauh candidate(s) with the largest value of the \tauh ID discriminant.  

The largest irreducible source of background is Drell-Yan production of
\ZTT, while the dominant background sources with jets misidentified as leptons are QCD multijet and
\Wjets.
Other contributing background sources are \ttbar, single top quark, \ZEE, \ZMM, and diboson production.

The two \PGt lepton candidates assigned to the \PH decay are required to have opposite charges.
Events are selected online using a combination of single-lepton, lepton+\tauh, double-\tauh, and electron+muon triggers.
The trigger requirements, geometrical acceptances, and \pt
criteria are summarized in Table~\ref{tab:inclusive_selection}.
The \pt thresholds in the selections are optimized to increase the sensitivity to
the \HTT signal, while also satisfying the trigger requirements. The $\eta$
selections are driven by reconstruction and trigger requirements.

\begin{table*}[htbp]
\centering
\topcaption{Kinematic selection requirements for the four di-$\PGt$ decay channels.
The trigger requirement is defined by a combination of trigger candidates with \pt 
over a given threshold, indicated inside parentheses in \GeVns. The pseudorapidity 
thresholds come from trigger and object reconstruction constraints. The $\pt$ thresholds 
for the lepton selection are driven by the trigger requirements, 
except for the $\tauh$ candidate in the $\PGm\tauh$ 
and $\Pe\tauh$ channels, and the sub-leading lepton in the $\Pe\PGm$ channel, where they have been 
optimized to increase the analysis sensitivity.
\label{tab:inclusive_selection}
}
\begin{scotch}{llllll}
  Channel     &  Trigger     & Year                 & \multicolumn{3}{c}{Selection criteria}                        \\ 
              &  requirement &                      & $\pt$ (\GeVns)  & $\eta$                    &    Isolation      \\
\hline  
$\tauh\tauh$  &  $\tauh (35)\,\&\,\tauh (35)$ & 2016       & $\pt^{\tauh}>40$    & $\abs{\eta^{\tauh}}<2.1$  &   DNN $\tauh$ ID  \\
              &  $\tauh (40)\,\&\,\tauh (40)$ & 2017, 2018 &                     &                           &                   \\
\hline

$\mu\tauh$    &  $\PGm(22)$ & 2016                         & $\pt^\PGm>\pt^{\text{trigger}}+1\GeV$       &  $\abs{\eta^\PGm}<2.1$    &   $I^{\PGm}<0.15$ \\
              &  $\PGm(19)\,\&\,\tauh (21)$ & 2016         & $\pt^{\tauh}>30$    &  $\abs{\eta^{\tauh}}<2.3$ &   DNN $\tauh$ ID  \\
              &  $\PGm(24)$ & 2017, 2018                   &                     &                           &                   \\
              &  $\PGm(20)\,\&\,\tauh (27)$ & 2017, 2018   &                     &                           &                   \\
\hline

$\Pe\tauh$    &  $\Pe(25)$ & 2016                          & $\pt^\Pe>\pt^{\text{trigger}}+1\GeV$        &  $\abs{\eta^\Pe}<2.1$     &   $I^{\Pe}<0.15$  \\
              &  $\Pe(27)$ & 2017                          & $\pt^{\tauh}>30$    &  $\abs{\eta^{\tauh}}<2.3$ &   DNN $\tauh$ ID  \\
              &  $\Pe(32)$ & 2018                          &                     &                           &                   \\
              &  $\Pe(24)\,\&\,\tauh (30)$ & 2017, 2018    &                     &                           &                   \\
\hline
$\Pe\PGm$     & $\Pe(12)\,\&\,\PGm (23)$ & all years       & $\pt^{\Pe}>15$ , $\pt^{\PGm}>24$      & $\abs{\eta^\Pe}<2.4$      & $I^{\Pe}<0.15$    \\
              & $\Pe(23)\,\&\,\PGm (8)$ & all years       & $\pt^{\PGm}>15$ , $\pt^{\Pe}>24$    & $\abs{\eta^\PGm}<2.4$     & $I^{\PGm}<0.15$   \\
\end{scotch}
\end{table*}

In the $\ell\tauh$ channels, the large \Wjets background
is reduced by requiring the transverse mass, \mT, to be less than 50\GeV.
The \mT is defined as follows,
\begin{linenomath}
\begin{equation}
\mT \equiv \sqrt{\smash[b]{2 \pt^\ell \ptmiss [1-\cos(\Delta\phi)]}},
\end{equation}
\end{linenomath}
where $\pt^\ell$ is the transverse momentum of the electron or muon
and $\Delta\phi$ is the azimuthal angle between the lepton momentum and \ptvecmiss.

In the $\Pe\PGm$ and $\ell\tauh$ channels, events with b jets are vetoed to reduce the background from \ttbar production. 
In the $\Pe\PGm$ channel, this background is further mitigated by requiring
$p_\zeta = p_\zeta^{\text{miss}} - 0.85 \, p_\zeta^{\text{vis}} > -35$\GeV,
where $p_\zeta^{\text{miss}}$ is the component of \ptvecmiss along the bisector of
the \pt of the two leptons and $p_\zeta^{\text{vis}}$
is the sum of the components of the lepton \pt along
the same direction~\cite{Jang:2006dt}. 

Event categories are designed to increase the sensitivity to the signal by isolating regions with
large signal-to-background ratios, and to provide sensitivity to the \Hgg and
\HVV parameters. They follow closely the selection in Ref.~\cite{Sirunyan:2017khh}:

\begin{itemize}

\item {0-jet category}: This category targets \PH events produced via
  \ggH. Events containing no jets with $\pt>30$\GeV are selected.

\item {VBF category}: This category targets \PH events produced via the VBF process
and \ggH in association with two jets.
Events are selected with at least two jets with $\pt>30$\GeV.
In the $\Pe\PGm$ and $\ell\tauh$ channels the invariant mass of the two leading jets \mjj is required to be larger than 300\GeV.
In the $\tauh\tauh$ channel we require the separation \detajj between the two leading jets to be greater than 2.5, and the transverse component of the vector sum of the \ptvecmiss and the \ptvec of the visible decay products of the \PGt leptons, defined as \ptvech, to have a magnitude (\pth) greater than 100\GeV.

\item {Boosted category}: This category contains all the events that do not
enter one of the previous categories, namely events with one jet and events
with several jets that fail the requirements of the VBF category. It targets events with an
\PH produced in \ggH and recoiling against an initial-state radiation jet.

\end{itemize}

\section{Background estimation}
\label{sec:background}
\begin{figure*}[htbp]
  \centering
    \includegraphics[width=0.45\textwidth]{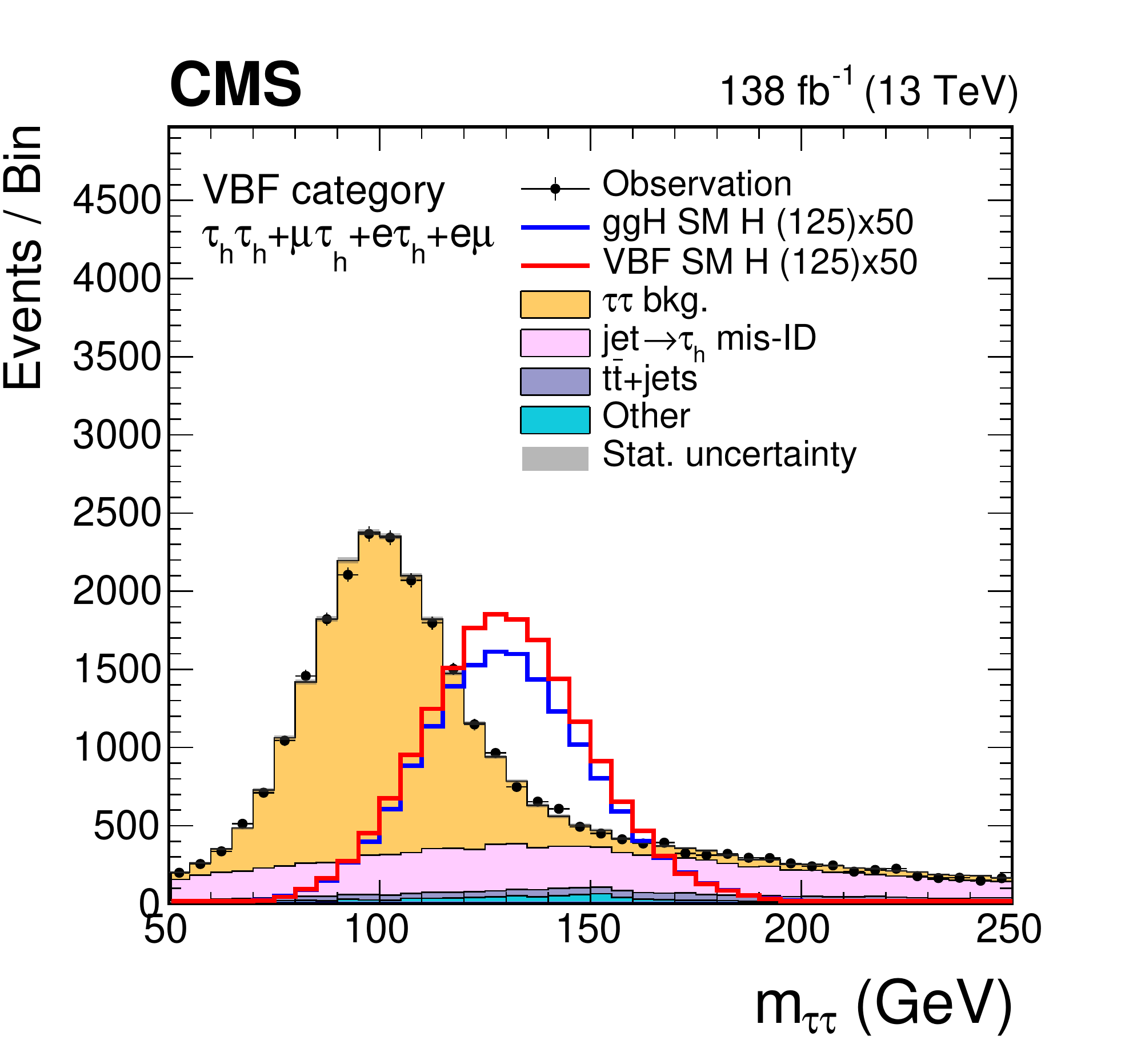}
    \includegraphics[width=0.45\textwidth]{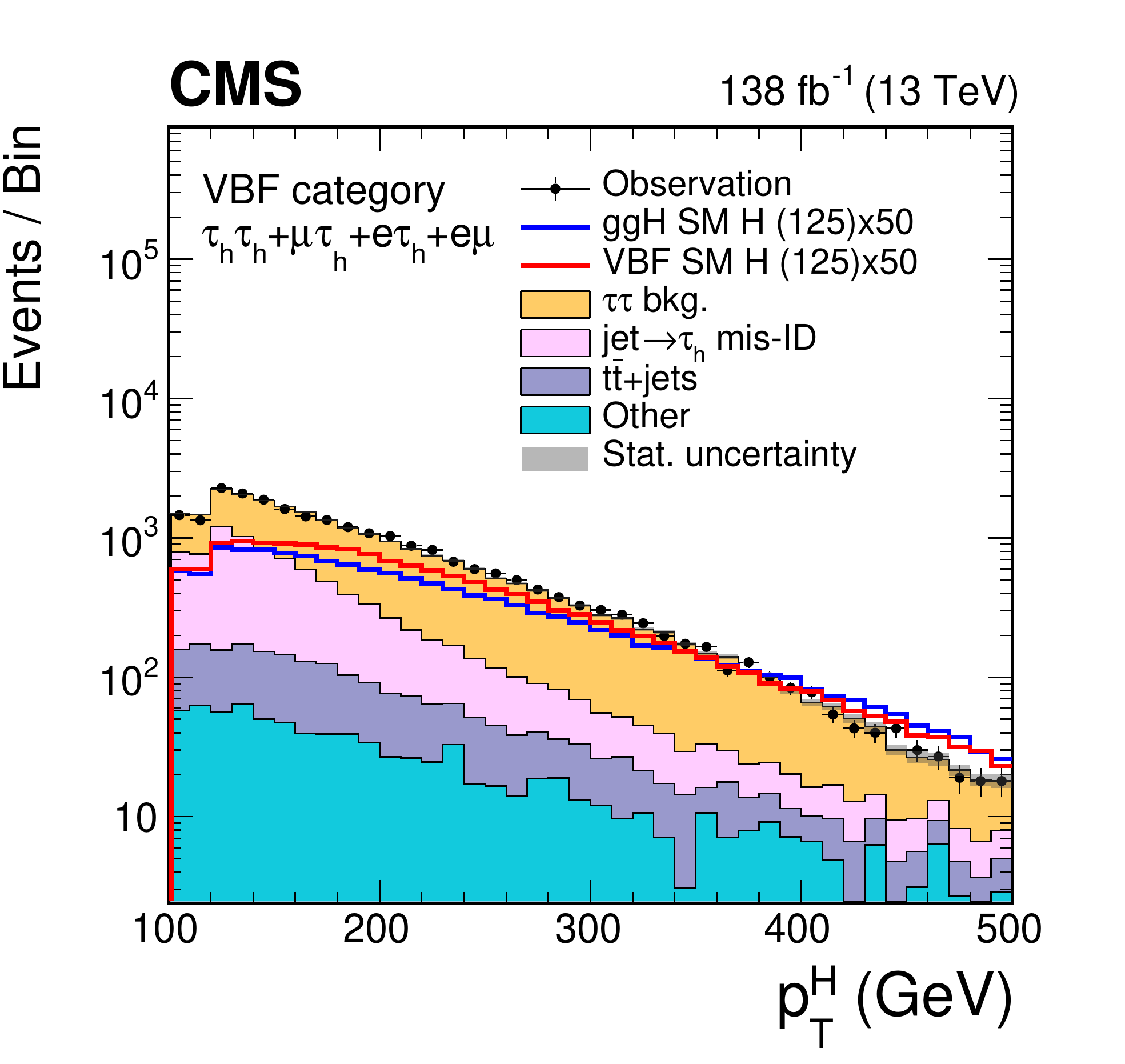}
    \caption{The \mtautau (left) and \pt (right) distributions for \PH candidate di-\PGt lepton pairs in the VBF category.
      All events selected in the $\Pe\PGm$, $\Pe\tauh$, $\PGm\tauh$, and $\tauh\tauh$ final states are included.
      The yields of the \PH processes are scaled to match 50 times the SM predictions.
      Only statistical uncertainties are shown.}
    \label{fig:mttAndHpt}
\end{figure*}

In this section we describe the background processes to the \HTT signal
and methods to estimate their contributions. The major background is from Drell-Yan
production, where the \PZ boson decays to a pair of \PGt leptons, followed by backgrounds
from jets misidentified as \PGt lepton candidates. Whenever possible we rely on data to estimate
background contributions. The data, signal, and background predictions in the VBF
category are illustrated in Fig.~\ref{fig:mttAndHpt}.

\subsection{Backgrounds due to \texorpdfstring{$\PGt\PGt$}{tautau} events}

The Drell--Yan \ZTT process is the dominant background to the \HTT signal as both processes
share the same final state and have very similar kinematic properties. Additionally, several other process such as \ttbar can produce a $\PGt\PGt$ final state.
Given the dominance of these backgrounds in data, we rely on an embedding
method~\cite{Sirunyan:2019Embedded} to simulate them:
in a dedicated control region, \ZMM candidate events are selected from data;
calorimeter deposits and tracks produced by the pair of muons in the event are removed;
the muons in the event are replaced with simulated \PGt leptons with the same kinematic properties as the removed muons; and
the \PYTHIA generator is used to model the decay of the \PGt leptons in the same way as a typical \ZTT decay.

The embedding process is performed separately for each decay channel and results in greatly improved statistical accuracy
compared to that of a typical MC simulation. Using data to describe effects such as pileup and detector noise results
in a much more reliable description of \ptmiss and jet-related variables, which in turn reduces systematic uncertainties arising from, \eg,
jet energy corrections.

\subsection{Background due to jet misidentification}
One of the major backgrounds to the \HTT signal in the $\Pe\tauh$, $\PGm\tauh$, and $\tauh\tauh$
channels originates from events where a jet is misidentified
as a \tauh candidate. 
We refer to such backgrounds as \jettotau processes.
The background processes include QCD multijet production,
\Wjets, \ttbar, and diboson processes. 
As the \jettotau misidentification rates are typically $\sim$ 0.1--1\%,
the use of MC simulation to model this background is undesirable due to the statistical limitation and systematic
uncertainty associated with the correct modeling of the detector response to jets identified as \PGt leptons.
Thus, we rely on data to model this background, using the ``fake factors'' (\FF) method~\cite{Sirunyan:2018qio}.

To estimate the \jettotau background in the signal region, we define a sideband region enriched in \jettotau events. The sideband region is defined by selecting \tauh candidates that fail the nominal requirement on the \tauh DNN ID discriminator but pass a looser requirement.
This looser requirement will be referred to in the following as ``relaxed’'.
To estimate the \jettotau background in the signal region, the events in the sideband region are scaled by extrapolation factors known as the \FF, which are defined as the ratio between the nominal and relaxed \jettotau misidentification rates.
In the $\tauh\tauh$ channel, there are two \tauh candidates and it is possible that either one or both candidates originate from a \jettotau.
However, the dominant process producing \jettotau in this channel is multijet QCD, which results in both \tauh candidates originating from \jettotau misidentifications.
In this case, we define the sideband region by requiring only the leading \tauh to meet the relaxed ID requirement.

The \FF are determined separately for each channel, and for each of the dominant processes contributing to the \jettotau background. 
For the $\tauh\tauh$ channel, the \jettotau background originates almost entirely from QCD multijet events, and therefore \FF are derived only for this process. 
For the $\Pe\tauh$ and $\PGm\tauh$ channels, separate \FF are derived for QCD multijet, \Wjets, and \ttbar processes.
The QCD and \Wjets \FF are measured in dedicated control regions enriched with the events from the given process. 
The QCD control region is defined by inverting the opposite-sign charge requirement on the di-\PGt candidate pair.
The \Wjets control region is defined by selecting events with $\mT>70\GeV$. 
Obtaining a \ttbar control region with high purity is not possible, and the \FF are therefore measured in simulation for this subdominant process.
For all control regions, we subtract the contributions of events with \tauh candidates arising from genuine hadronic \PGt decays or from misidentified electrons or muons using simulations. 
We parametrize the \FF as a function of the \tauh \pt. 
As events in this study are categorized primarily by the number of jets in the event, the \FF are measured in jet multiplicity
bins: no jets, one jet, and two or more jets. 

The \FF for each of the individual processes are then weighted into the overall \FF to account for their relative contributions to the events in the relaxed identification region.
For this purpose, simulated events are used to determine the expected contributions of \Wjets and \ttbar events, and the QCD contribution is estimated by subtracting all simulated non-QCD processes from the data. 
For the $\Pe\tauh$ and $\PGm\tauh$ channels, the \FF measured for the \Wjets process are weighted to also account for all other subdominant
\jettotau processes (all processes except multijet QCD, \Wjets, and \ttbar). 
For the $\tauh\tauh$ channel, the multijet QCD \FF account also for all other subdominant processes where the leading \tauh candidate is a misidentified jet.
The events in which the subleading \tauh candidate is a misidentified jet and the leading \tauh candidate is a genuine \PGt lepton are modeled via simulation; these events constitute only a small fraction
($\mathcal{O}$(2\%)) of the total misidentified jet background in this channel.

Finally, the \FF are further corrected to accommodate residual differences observed when applying the measured \FF to events in control regions.
Such corrections are needed to account for: differences in the \jettotau misidentification rates in control and signal regions arising from, for example, slight differences in the jet flavor compositions, the choices of functional forms for parametrizing the \pt dependence, the finite binning of the parametrizing variables, and the omission of dependencies on kinematic or topological variables, such as $\eta$ for which \FF values are averaged out.   
Sub-leading dependencies of the \FF on $\pt^{\ell}$ and \mT for the $\ell\tauh$ channels, or the \pt of the subleading \tauh candidate and the mass of the visible $\PGt\PGt$ decay products for the $\tauh\tauh$ channel, enter via these corrections.

In the $\Pe\PGm$ channel, one of the minor backgrounds stems from the multijet QCD process, where at least one jet is misidentified as an electron or a muon candidate. The majority of these events involve $\text{b}\bar{\text{b}}$ production, with electron and muon candidates produced in semileptonic decays of heavy-flavor quarks.  
This background is estimated from a sideband region using events with an electron and a muon with same-sign (SS) electric charges. 
Scale factors are then applied to extrapolate from this sideband region to the signal region, where the electron and the muon have opposite-sign (OS) electric charges. 
These so-called OS/SS scale factors are derived from a control region where the muons pass a relaxed isolation requirement but fail the signal region isolation criteria. 
The dependence of the OS/SS factors on the \DR~between the two leptons and jet multiplicity in the event is taken into account. A correction is also applied to account for any bias introduced by the inversion of the isolation requirement on the muon candidate.
Finally, we subtract contributions from known SM processes using embedded and MC simulation samples.

\subsection{Other backgrounds processes}
The remaining backgrounds include Drell--Yan processes, where the \PZ boson decays to a pair of electrons or muons and one or more of the final state leptons is misidentified as the \PGt lepton, as well as \ttbar, single top quark and multiboson
production with fewer than two genuine \PGt leptons and additional electrons or muons that are misidentified as leptonically or hadronically decaying \PGt leptons. 
These backgrounds are small and we rely on MC simulation to estimate their contribution to the signal region. To avoid double-counting of
backgrounds arising from jet misidentification, we remove the events with the generator-level quark or gluon
matched to the reconstructed \PGt lepton candidate in the final state. Similarly, Drell--Yan MC events as well as any
other MC simulation events with two genuine \PGt leptons are discarded to avoid overlap
with the embedded samples.

\subsection{Corrections to simulated data}

To improve the agreement between the signal and background processes modeled with simulations and the data, the following corrections are applied to the simulated events (including $\PGt\PGt$-embedded events for corrections pertaining to the simulated \PGt leptons):

\begin{itemize}
\item The pileup distribution in simulation is reweighted in order to match the pileup in data.

\item The electrons and muons channels are corrected to account for their trigger efficiencies, reconstruction, identification, and isolation requirements.
The channels containing \tauh candidates are corrected for their trigger efficiencies, reconstruction, and identification requirements.
Corrections are also applied to events including $\Pe\to\tauh$ and $\PGm\to\tauh$ candidates to account for differences in the misidentification probabilities.

\item For the $\Pe\PGm$ and $\ell\tauh$ channels, corrections are applied to account for differences in the number of events passing the b jet veto, as a result of variations in the probabilities for jets to be tagged as {\it b} jets.

\item The \tauh energy scales are corrected per decay mode to match the energy scale in data using the \ZTT visible mass peak. Separate corrections are derived for \tauh candidates that originate from genuine hadronically decaying \PGt leptons and those that originate from electron and muon misidentifications.
The electron energy scale is adjusted in data and simulation using the \PZ boson mass peak, and the resolution of the simulated electrons is also adjusted to match the data.

\item Jet energy scale corrections are applied to both data and simulated events, and the energy resolution of simulated jets is adjusted to match the resolution in data. For the Drell--Yan, \Wjets, and \PH events estimated from simulation, corrections are applied to the \ptmiss based on the vector difference of the measured \ptmiss and the total \pt of the neutrinos from the \PZ, \PW, or \PH decay products ("recoil corrections").

\item The \PZ boson mass and \pt spectra in simulation are corrected to better match the data. To this purpose the \PZ mass and \pt are measured in data and simulation in dimuon events. The observed differences between the data and simulations are taken as event weights that are subsequently applied to the simulated \ZLL events.  
The size of the corrections are typically less than 20\%.
For \ttbar events, the top quark \pt spectra are also reweighted to match the \pt spectra in data. The procedure used to derive these corrections is described in Ref.~\cite{Khachatryan:2015oqa}. The sizes of the corrections are less than 20\%.

\item During the 2016--2017 data-taking, a gradual shift in the timing of the inputs of the ECAL L1 trigger in the region at $\abs{\eta} > 2.0$ caused a specific trigger inefficiency~\cite{CMS:2020cmk}. For events containing an electron (a jet) with \pt larger than $\approx$50\GeV ($\approx$100\GeV), in the region $2.5 < \abs{\eta} < 3.0$ the efficiency loss is $\approx$10--20\%, depending on \pt, $\eta$, and time. Correction factors were computed from data and applied to the acceptance evaluated by simulation to account for this inefficiency.
This results in a small decrease in the estimated signal and background yields, \eg, the inclusive SM VBF yields are reduced by about 2\%-3\%.

\end{itemize}

\section{Systematic uncertainties}
\label{sec:Systematics}
A variety of systematic uncertainties are taken into account in the analysis. 
The uncertainty model consists of normalization uncertainties that only scale the yield of a distribution while leaving its shape unchanged, and shape uncertainties that also alter the shapes of the distributions.
The leading systematic uncertainty sources result from the jet energy scales and resolutions, and the statistical uncertainties in the background predictions. All systematic uncertainties are implemented in the form of nuisance parameters in the likelihood, which can be further constrained by the fit to the data. The uncertainties considered in this analysis are summarized in Table~\ref{tab:unc} and detailed below.

\begin{table*}[ht!]
\topcaption{Sources of systematic uncertainties.}
\label{tab:unc}
\centering
\begin{scotch}{lc}
Source                   & Uncertainty \\
\hline
$\tauh$ ID                    & \pt/decay-mode dependent (3\%--10\%) \\
$\tauh$ separation from $\Pe/\PGm$       & 3\% \\
$\Pe\to\tauh$ ID              & $\eta$ dependent (9\%--40\%) \\
$\PGm\to\tauh$ ID             & $\eta$ dependent (10\%--70)\% \\
$\Pe$ ID      & 2\% \\
$\PGm$ ID                     & 1\% \\
b jet veto                    & 0--10\% \\
Integrated luminosity                    & 1.6\% \\
Trigger                       & 2\% for $\Pe$/$\PGm$, \pt/decay-mode dep. for $\tauh$ [$\mathcal{O}$(10\%)] \\
$\ttbar$ cross section        & 4.2\% \\
Diboson cross section         & 5\% \\
Single top quark cross section      & 5\% \\
Drell-Yan cross section       & 2\% \\
L1 trigger timing (2016 and 2017)   & Event-dependent (0.2\%--15\%) \\
$\mathcal{B}(\HTT)$ & 2.1\% \\
$\tauh$ energy scale          & Decay-mode dependent (0.2\%--1.2\%) \\
$\Pe\to\tauh$ energy scale    & Decay-mode dependent (1--7\%)  \\
$\PGm\to\tauh$ energy scale   & 1\%  \\
Electron energy scale         & \pt/$\eta$ dependent ($<$ 1.25\%) \\
Muon energy scale             & $\eta$ dependent 0.4--2.7\% \\
Jet energy scale              & \pt/$\eta$ dependent (0.5\%--14\%)\\
Jet energy resolution         & $\eta$ dependent (2\%--95\%) \\
$\ptmiss$ unclustered energy scale & Event-dependent (0\%--20\%) \\
$\ptmiss$ recoil corrections  & 0.3--5.8\% \\
Jet~$\to\tauh$ misidentification          & Event-dependent [$\mathcal{O}(10\%)$] \\
QCD multijet in the $\Pe\PGm$ channel          & Event-dependent [$\mathcal{O}(20\%)$] \\
Embedded yield                & 4\% \\
$\ttbar$ in embedded          & 10\% \\
Signal theoretical uncertainty      & Event-dependent (up to 25\%) \\ 
Top quark \pt reweighting            & \pt dependent (0\%--21\%) \\
Drell--Yan \pt and mass reweighting             & \pt/mass dependent (0\%--11\%) \\
\end{scotch}
\end{table*}

The integrated luminosities of the 2016, 2017, and 2018 data-taking periods are individually known with uncertainties in the 1.2--2.5\% range~\cite{CMS-LUM-17-003,CMS-PAS-LUM-17-004,CMS-PAS-LUM-18-002}, while the total Run~2 (2016--2018) integrated luminosity has an uncertainty of 1.6\%, the improvement in precision reflecting the (uncorrelated) time evolution of some systematic effects.
The uncertainty in the L1 ECAL trigger timing correction factors described in Sec.~\ref{sec:background} ranges between 0.2--15\%.

The uncertainties in the (electron) muon reconstruction, identification, and isolation efficiencies amount to (2) 1\%. 
The electron and muon triggers contribute an additional 2\% uncertainty in the yield of simulated processes.
The uncertainty in the electron energy scale depends on \pt and $\eta$ and is typically less than 1\%.
The muon energy scale uncertainty varies between 0.4 and 2.7\% depending on $\eta$.

The \tauh reconstruction and identification efficiency is measured in three \pt bins (30--35, 35--40, $>40$\GeV) or four \tauh decay mode bins and its uncertainty is dominated by the statistical component.
The uncertainty is taken to be uncorrelated for the individual measurements and is within the 3\%--10\% range.
In addition, a yield uncertainty of 3\% is taken into account for genuine \tauh candidates due to the discrimination against electrons and muons.
An additional uncertainty is applied for the \tauh reconstruction in the embedded samples to account for differences in the charged hadrons and $\pi^{0}$ reconstruction efficiencies, which ranges from 0.8 to 3\%.
The uncertainty in the \tauh trigger efficiencies depends on the \pt and decay mode of the \tauh candidates, and is therefore treated as a shape uncertainty. The magnitude of this uncertainty is typically $\mathcal{O}$(10\%). 
For electrons and muons misidentified as \tauh candidates, an uncertainty derived in bins of \pt, $\eta$, and decay mode of the misidentified \tauh candidate is applied and amounts to between 9\%--40\% and 10\%--70\%, respectively.
The uncertainty in the \tauh energy scale ranges from 0.2 to 1.1\% depending on the decay mode.
For electrons and muons misidentified as \tauh candidates, the uncertainty in the energy scale amounts to 1\%--6.5\% for electrons and 1\% for muons.

Uncertainties in the jet energy scale come from different sources and with partial correlations. 
These sources typically affect different regions of the detector and their magnitude depends on the jet \pt and $\eta$. The collective magnitude of these uncertainties per jet typically ranges from 0.5 to 14\%. 
Uncertainties in the jet energy resolution are also taken into account and range from 2 to 95\% per jet depending on $\eta$. The jet energy scale and resolution uncertainties create migrations between categories defined on the basis of the jet multiplicity or \mjj, and affect the shapes of the \dphi and \MELA discriminants. 
For all MC samples without recoil corrections applied, the uncertainties in the jet energy scale and resolution are also propagated to the \ptvecmiss.

For simulated events that have recoil corrections, the uncertainties in the resolution and response of the \ptvecmiss are derived as part of the estimate
of the recoil corrections and range from 0.3 to 5.8\%. 
Other processes suffer from uncertainties in the energy measurement for the
energy depositions in the calorimeter, not associated with jets and photon candidates, so-called unclustered energy scale uncertainties. The magnitudes of
these uncertainties depend on the \pt, $\eta$, and types of the unclustered
PF candidates. The overall sizes of these uncertainties are typically less than 20\%. 

The yield uncertainty related to discarding events with a {\it b}-tagged jet varies up to 10\% for backgrounds with heavy-flavor jets, whereas for backgrounds with mostly gluon and light-flavor jets it is less than 1\%.

For background with \jettotau misidentifications, the uncertainties in the measured \FF are propagated to the background predictions as shape uncertainties. 
This includes statistical uncertainties in the fitted functions as well as systematic uncertainties coming from residual differences observed in control regions. Altogether the uncertainties on the \FF are $\mathcal{O}$(10\%). 
Similarly, for the multijet QCD estimation in the $\Pe\PGm$ channel uncertainties in the OS/SS extrapolation factors are taken into account. Altogether the uncertainties amount to $\mathcal{O}$(20\%). 

Uncertainties related to the embedding method are taken into account in addition to those pertaining to the simulated \PGt lepton decay products described previously. 
Embedded samples include all events with two \PGt lepton candidates, essentially Drell--Yan events, but also contain small fractions of diboson and $\ttbar$ events. 
A shape uncertainty is applied to take into account the contamination from these non Drell--Yan events, which amounts to 10\% of the $\ttbar$ and diboson contribution to embedded samples, as estimated from simulation.
Data events with muons are selected with a muon trigger before embedding the simulated \PGt leptons. 
The uncertainty in this trigger requirement amounts to 4\%.

Uncertainties in the \ttbar, Drell--Yan, diboson, and single top quark production cross sections amount to 4.2, 2.0, 5.0, and 5.0\%,
respectively. This includes uncertainties due to missing higher-order corrections, the PDFs, and \alpS. For the \ttbar cross section the uncertainty in the top quark mass is also included.
Uncertainties due to the reweighting of the top \pt and Drell--Yan \pt and mass spectra are also included.   
For the \ttbar samples, the size of the correction is taken as the uncertainty, while for the Drell--Yan samples, the correction is varied by 10\%. 

The theoretical uncertainties in the \PH production cross sections and \HTT branching fraction follow the recommendations in Ref.~\cite{deFlorian:2016spz}. 
The uncertainty in the branching fraction of the \PH to \PGt leptons includes a 1.7\% uncertainty due to missing higher-order corrections, a 1\% parametric uncertainty in the quark masses, and a 0.62\% parametric uncertainty in \alpS.
The inclusive uncertainty related to the PDFs amounts to 3.2, 2.1, 1.8, and 1.3\%, respectively, for the \ggH, VBF, $\PW\PH$, and $\PZ\PH$ 
production modes. 
Acceptance uncertainties for the \ggH signal due to renormalization and factorization scale variations are applied following the uncertainty schemes proposed in Ref.~\cite{deFlorian:2016spz}. The sizes of these uncertainties are typically smaller than 25\%.
Acceptance uncertainties for the VBF signal due to renormalization and factorization scale variations are applied as yield uncertainties. The sizes of the uncertainties are typically smaller than 5\%. 

Uncertainties arising from the limited sample size of the simulated events, or data control regions, are taken into account using the``Barlow--Beeston" method~\cite{BarlowBeeston,Conway:2011in}. They are considered for all bins of the distributions used to extract the results.

\section{Analysis of \texorpdfstring{\ggH}{ggH} production}
\label{sec:ggH}

In this section we first review the analysis methods employed to extract the \Hgg anomalous coupling parameters. 
In Sec.~\ref{ssec:MVA-ggH} we describe the \MELA method from which we obtain our most stringent expected limits on the anomalous coupling parameters. 
The \dphi method, which is used to cross check the results obtained by the \MELA method, is briefly described in Sec.~\ref{ssec:cut_based_ggH}. 
The results obtained are then presented in Sec.~\ref{ssec:ggH_Results}.

\subsection{The \MELA method}
\label{ssec:MVA-ggH}

For events entering the VBF category, a combination of simple neural networks and
\MELA discriminants is used. The former provide optimal separation of the dominant backgrounds for a given
channel from the \PH production, while the latter offer powerful handles to distinguish
different signal hypotheses.

A feed-forward network containing two hidden layers is used in each channel. As dominant backgrounds
vary by channel, the observables used in the neural network training change and therefore the architecture
of the network is modified for each channel. The number of nodes per layer is kept to
a minimum to reduce the complexity of the neural network without compromising its performance.
As sensitivity to the \Hgg anomalous coupling is maximal for events with kinematics similar to those of
VBF production, we use VBF signal events as the signal process for all neural networks. This provides
the added benefit that the same network can be used in the analysis of both \ggH and VBF production
processes.

The simplest neural network is employed in the $\Pe\tauh$ and $\PGm\tauh$ channels where the
background is dominated by the \ZTT production. Thus, a simple binary classifier
is trained to distinguish VBF production from the \ZTT process. We use all
seven \MELA input variables ($\mathbf{\Omega}^\text{assoc}$) defined for the VBF process in Sec.~\ref{sec:kinematics}, \mtautau, \mjj, and \pth as input features for the network.

In the $\tauh\tauh$ and \emu channels, there are two background processes that have significant event yields in the VBF category. For the $\tauh\tauh$ channel, the two background processes are \ZTT and \jettotau. While for the \emu channel the backgrounds are \ZTT and \ttbar. We thus train multiclass neural networks to divide events into three classes: two background classes and one signal class targeting the VBF \PH production process. The same input features used for the binary $\ell\tauh$ networks are also used for $\tauh\tauh$ and \emu multiclass neural networks. However, in the \emu channel two additional features, the jet multiplicity and $p_\zeta$, are included to improve the rejection of \ttbar events.

For the binary classifiers we use the neural network output scores as discriminating variables, whereas for the multiclass networks we use the output scores for the VBF signal classes. We will refer to these discriminants collectively as \Dnn. 

Three \MELA discriminants are used in the \Hgg analysis.
In order to separate \ggH production from VBF production,
\DVBF as defined
in Eq.~(\ref{eq:d2j-vbf}) is used. The discriminant \DzggH, defined in Eq.~(\ref{eq:d0-ggh}), is used to
separate \ggH produced with the SM couplings from \ggH produced with a pure pseudoscalar
coupling. Lastly, \DCPggH, defined in Eq.~(\ref{eq:dcp-ggh}), provides sensitivity to the
interference between the $CP$-odd and $CP$-even contributions:
\begin{equation}
    \DVBF = \frac{\mathcal{P}_\mathrm{SM}^{\ggH} + \mathcal{P}_\mathrm{0-}^{\ggH}}{\mathcal{P}_\mathrm{SM}^{\ggH} + \mathcal{P}_\mathrm{0-}^{\ggH} + {\mathcal{P}_\mathrm{SM}^\mathrm{VBF}}},
    \label{eq:d2j-vbf}
\end{equation}

\begin{equation}
    \DzggH = \frac{\mathcal{P}_\mathrm{SM}^{\ggH}}{\mathcal{P}_\mathrm{SM}^{\ggH} + \mathcal{P}_\mathrm{0-}^{\ggH}},
    \label{eq:d0-ggh}
\end{equation}

\begin{equation}
    \DCPggH = \frac{\mathcal{P}_\mathrm{SM-0-}^{\ggH}}{\mathcal{P}_\mathrm{SM}^{\ggH} + \mathcal{P}_\mathrm{0-}^{\ggH}}.
    \label{eq:dcp-ggh}
\end{equation}
The results of the analysis are extracted with a global maximum likelihood fit based on signal-sensitive observables.
We summarize the observables utilized in the analysis of the \ggH production in Table~\ref{tab:vbf_obs}.
As the same set of observables is used in the \HVV study, except for the superscripts of anomalous coupling
specific \MELA discriminators, we define them in Table~\ref{tab:vbf_obs} as well.

\begin{table*}[ht!]
\centering
\topcaption{List of observables used in the \MELA method. }
\label{tab:vbf_obs}
\begin{scotch}{lll}
Category & Observable & Goal \\
\hline
0-jet & \mtautau                           & Separate \PH signal from backgrounds \\
Boosted & \pth, \mtautau & Separate \PH signal from backgrounds and BSM from SM \HVV \\
VBF & \Dnn & Separate VBF-like \PH signal from backgrounds  \\
VBF & \DVBF & Separate \ggH from VBF \PH production  \\
VBF & \DzggH (\Dz) & Separate BSM from SM \Hgg (\HVV) \\
VBF & \DCPggH (\DCP) & Sensitive to the interference between the $CP$-even and $CP$-odd \\
&    & contributions to the \Hgg (\HVV) coupling \\
\end{scotch}
\end{table*}

We use four observables in total to construct fitted distributions in the VBF category: \DzggH, \DCPggH, \Dnn, and \DVBF.
The selected events are binned in multidimensional histograms (templates) of these observables.
The binning of these templates has been optimized to ensure sufficient statistical populations of all bins, to retain kinematic information, and for memory usage and speed of computer calculations.

The inclusion of the \DCPggH observable is intended to bring sensitivity to the sign of the interference between the $CP$-even and $CP$-odd contributions, which would manifest as an asymmetry between the number of events detected with positive and negative values of \DCPggH.
We therefore include two bins in this discriminant, $\DCPggH < 0$ and $\DCPggH\ge 0$. 
It should be noted that \DCPggH is symmetric about $\DCPggH=0$ in the absence of $CP$-violation, and this symmetry is thus enforced for the background and $CP$-conserving signal templates to reduce the influence of statistical fluctuations. 
In practice, this procedure involves selecting pairs of bins that have \DCPggH values of opposite sign, but otherwise identical bin boundary definitions, and setting the predicted event yields of both bins to the average value of the pair.       
For the remaining three observables we allocate more bins to those that have a stronger influence on the expected sensitivity.
For the $\ell\tauh$ channels, we use 10, 8, and 4 equally sized bins for \DzggH, \Dnn, and \DVBF, respectively.
For the \emu channel, which is the least sensitive channel in this analysis, we respectively use 3, 2, and 4 bins for these observables.
In all cases, neighboring bins are merged such that the background prediction has no bins with statistical uncertainty larger than 50\% to prevent cases where bins have very low statistical populations.
For the $\tauh\tauh$ channel, it was not possible to define a suitable set of equally spaced bins that fulfilled the optimization criteria. Therefore, we employ variable bin widths for the \DzggH, \Dnn, and \DVBF observables, and select bin boundaries that optimize the expected sensitivity, while minimizing the total number of bins in the templates.

We use two observables to construct our templates in the boosted category, \mtautau and \pth, and one observable in the 0-jet category, \mtautau. 
There are no dedicated \MELA observables sensitive to anomalous couplings in these channels, as the events either have fewer than the two jets needed to construct the observables, or do not display significant separation between different signal scenarios to justify their inclusion.
However, the \pth observable used in the boosted category has some sensitivity to anomalous \HVV couplings, as the BSM VBF events generally have larger \pth. Similarly, the relative yields of the signal events across categories also has some sensitivity to anomalous \HVV couplings, as the signal acceptance in each category will vary depending on the \pt of the \PH.     
Despite not bringing any sensitivity to anomalous \Hgg couplings, the 0-jet and boosted categories are included in the fit nonetheless to constrain backgrounds and to provide sensitivity to the inclusive \ggH cross section. The chosen binning in these categories is thus similar to what was employed in previous CMS measurements~\cite{Sirunyan:2017khh}.

\begin{figure*}[htbp]
  \centering
    \includegraphics[width=0.45\textwidth]{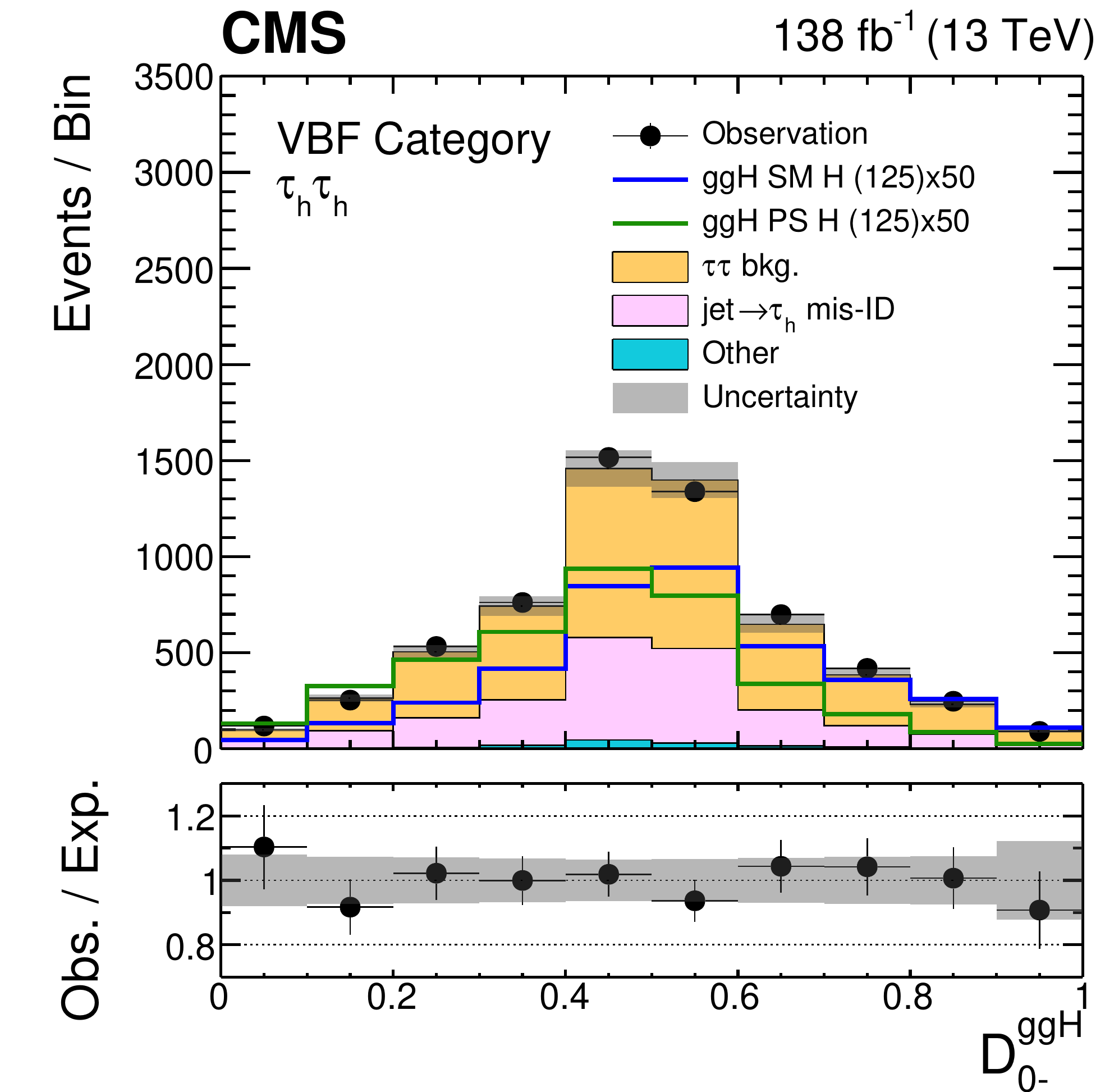}
    \includegraphics[width=0.45\textwidth]{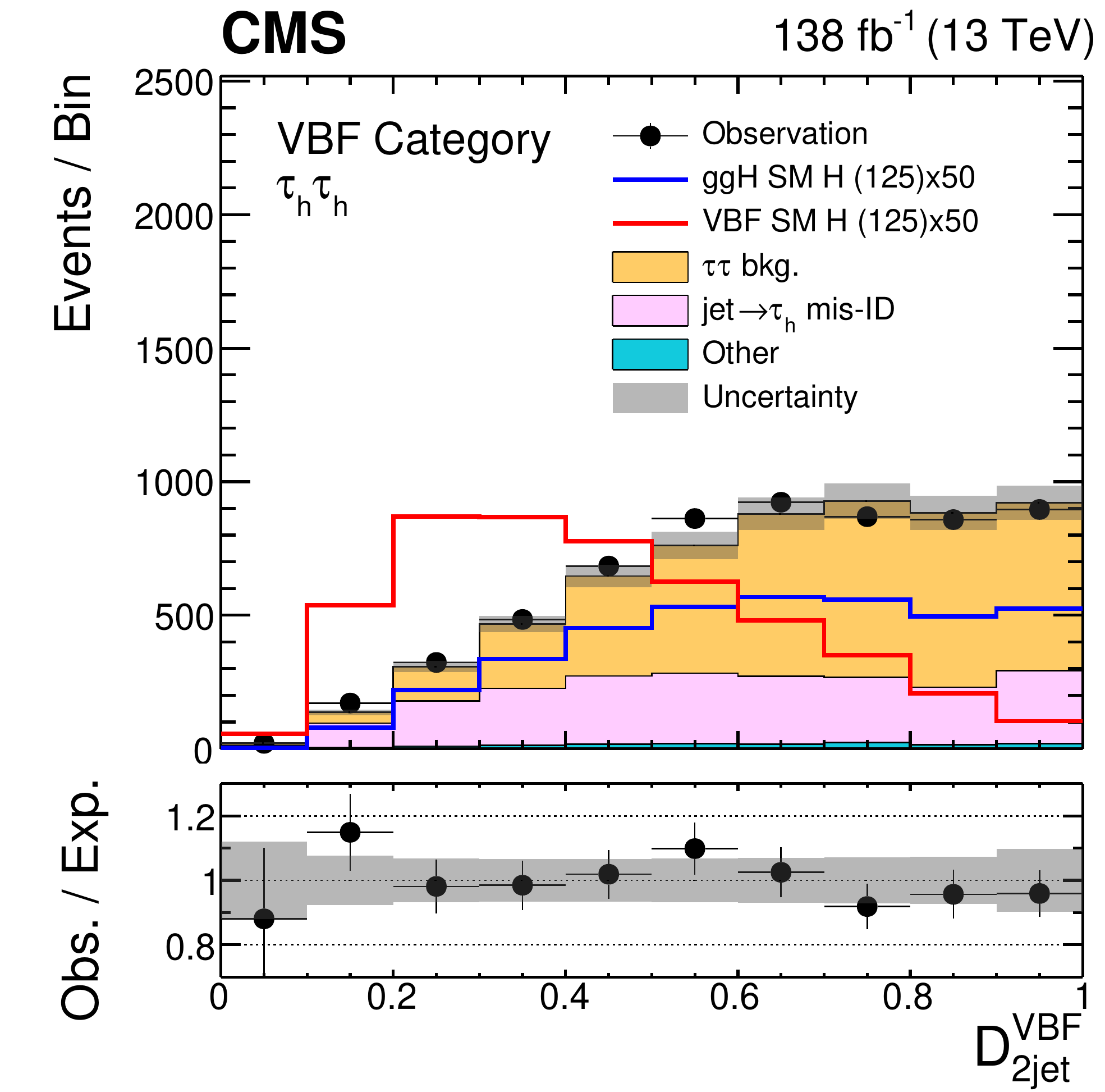}
    \includegraphics[width=0.45\textwidth]{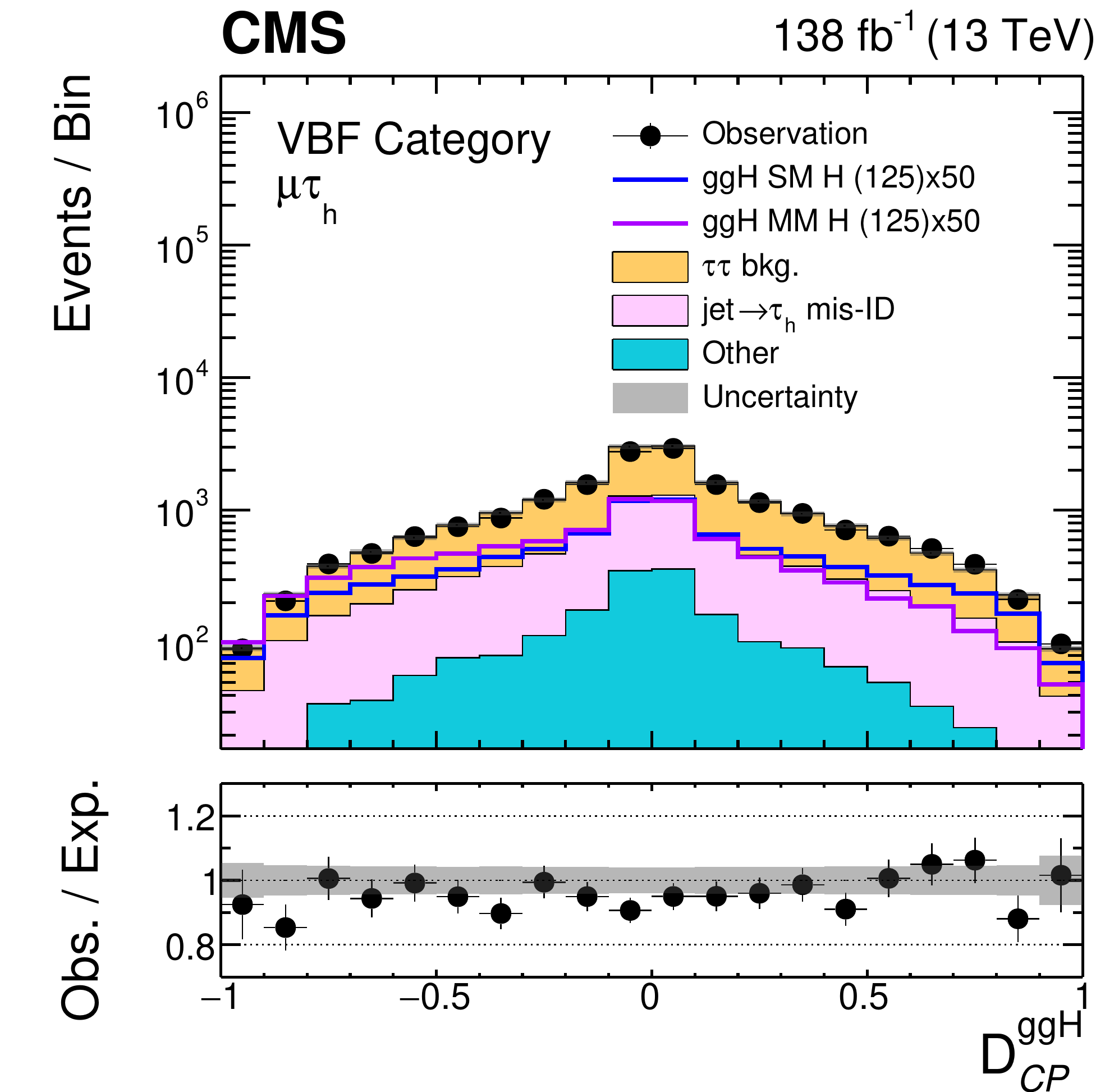}    
    \includegraphics[width=0.45\textwidth]{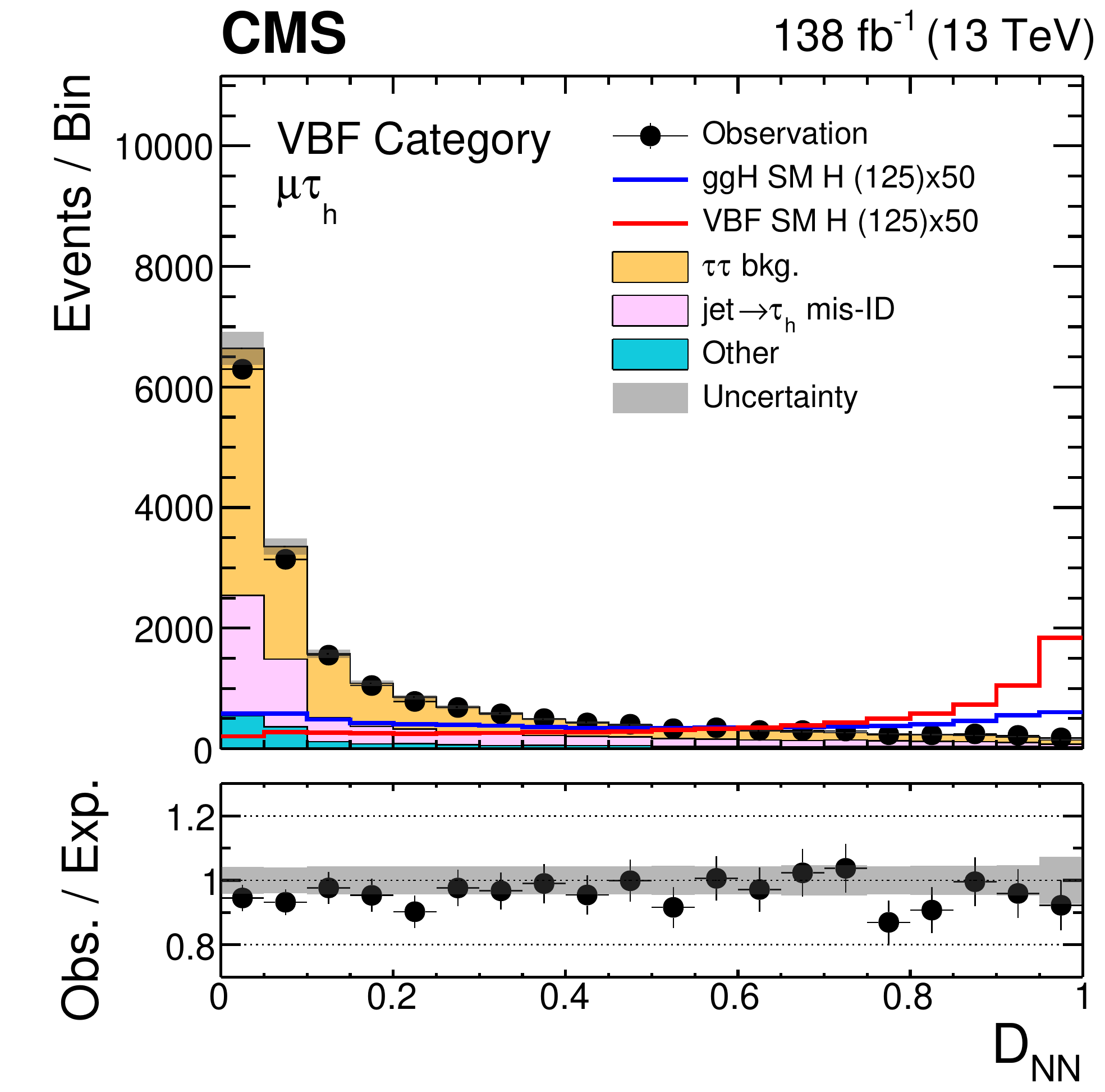}
    \caption{Examples of data and signal and background predictions for \MELA and neural network discriminants in the $\tauh\tauh$ and $\PGm\tauh$ channels. Events passing the selections outlined in Sec.~\ref{sec:Reco} and allocated to the VBF category are included.
      The yields of the \PH processes are scaled to match 50 times the SM predictions.
      The uncertainty band includes statistical uncertainties and systematic uncertainties that affect the normalization of the background distribution.
      The expectation in the ratio panel is the sum of the estimated backgrounds and the SM \PH signal. For the \DzggH discriminant the distribution expected for a pseudoscalar \PH hypothesis (labeled "PS" in the legend) is overlaid to be compared to the SM signal. Similarly, for the \DCPggH discriminant the distribution for a $CP$-violating scenario with the maximum-mixing between $CP$-even and $CP$-odd couplings (labeled "MM" in the legend) is shown. 
    }
    \label{fig:tt_mt_ggH}
\end{figure*}

Example distributions of the observables in the most sensitive $\tauh\tauh$ and $\PGm\tauh$ channels are given
in Fig.~\ref{fig:tt_mt_ggH}.

\subsection{The \texorpdfstring{\dphi}{Delta phi} method}
\label{ssec:cut_based_ggH}
This cross check method is based on the strategy proposed in Ref.~\cite{Klamke:2007cu}. The \dphi variable, defined in Sec.~\ref{subsec:ggh}, provides sensitivity to the $CP$ properties of the \Hgg vertex. The VBF-like ggH events are targeted as they have been shown to be most sensitive to the \Hgg anomalous couplings~\cite{Klamke:2007cu}.

The event selection and categorization follows closely those described in Sec.~\ref{subsec:Selection}. The only notable differences are in the definitions of the VBF signal categories, which are therefore described below.

The VBF category definition described in Sec.~\ref{subsec:Selection} is adopted for the $\Pe\PGm$ and $\ell\tauh$ channels. For the $\tauh\tauh$ channel, the VBF category selections defined previously for the $\ell\tauh$ channels are utilized.
The selected VBF-like events are then further subdivided into four categories based on \mjj and \pth 
to enhance the separation between different $CP$ scenarios and to provide additional differentiation between the signal and backgrounds.
The four categories are defined as follows: 
events with $\mjj < 500\GeV$ and $\pth < 150\GeV$ (low-\mjj category);
events with $\mjj < 500\GeV$ and $\pth \ge 150\GeV$ (low-\mjj boosted category);
events with $\mjj \ge 500\GeV$ and $\pth < 150\GeV$ (high-\mjj category); and
events with $\mjj \ge 500\GeV$ and $\pth \ge 150\GeV$ (high-\mjj boosted category).

We summarize the observables utilized in Table~\ref{tab:vbf_obs_cb}.
In this case we use two-dimensional (2D) templates in the VBF categories to extract the results. These templates are constructed using the \dphi and \mtautau observables. 
We use 12 equally-spaced bins for \dphi.
Variable bin widths are used for the \mtautau observable, where the bin boundaries are selected to capture the peaking structures of the signal distributions close to $\mtautau\sim 125\GeV$. 
The initial choice of the \mtautau bin boundaries is the same for all channels and categories but we apply an additional merging of neighboring bins in cases where the statistical fluctuations in the signal or background templates are excessive. 

\begin{table*}
\centering
\topcaption{List of observables used in the \dphi  method. }
\label{tab:vbf_obs_cb}
\begin{scotch}{lll}
Category & Observable & Goal \\
\hline
0-jet & \mtautau                           & Separate \PH signal from backgrounds \\
Boosted & \pth, \mtautau & Separate \PH signal from backgrounds \\
VBF & \mtautau & Separate \PH signal from backgrounds  \\
VBF & \dphi & Differentiate between $CP$-even, $CP$-odd, and mixed $CP$ scenarios \\
\end{scotch}
\end{table*}

\subsection{Results of the \texorpdfstring{\ggH}{ggH} analysis}
\label{ssec:ggH_Results}

The results are extracted by performing a binned maximum likelihood fit to the data combining all categories for the different channels and data-taking years. 
The likelihood function is defined as a product of conditional probabilities over all bins $i$:
\ifthenelse{\boolean{cms@external}}
{  \begin{multline}
    \mathcal{L}(\text{data} | \muggh, \muqqh, \vec{f}, \theta) =\\
     \prod_{i}\mathrm{Poisson}(n_i|s_{i}(\muggh, \muqqh, \vec{f}, \theta)+b_{i}(\theta))\cdot p(\tilde{\theta} | \theta),
    \label{eq:likelihoodDensity}
  \end{multline}
}{
  \begin{equation}
    \mathcal{L}(\text{data} | \muggh, \muqqh, \vec{f}, \theta) = \prod_{i}\mathrm{Poisson}(n_i|s_{i}(\muggh, \muqqh, \vec{f}, \theta)+b_{i}(\theta))\cdot p(\tilde{\theta} | \theta),
    \label{eq:likelihoodDensity}
  \end{equation}

}
where $n_{i}$ is the observed number of data events in each bin. 
The signal and background expectations are given by $s_{i}$ and $b_{i}$ respectively, which are functions of $\theta$, that represents the full set of nuisance parameters corresponding to the systematic uncertainties, and the parameters that modify the \PH signal processes: \muggh, \muqqh, and $\vec{f}$. 
The parameters \muggh and \muqqh are the \PH signal strength modifiers that respectively modify the \ggH and VBF+\VH cross sections with respect to the SM values. 
The $\vec{f}$ term represents the set of anomalous coupling parameters that modify the distributions of the \ggH and/or VBF+\VH signals. 
In the case of \Hgg anomalous coupling measurements, $\vec{f}=(\fggh, \fathree)$.
Finally, the $p(\tilde{\theta}|\theta)$ term represents the full set of probability density functions of the uncertainties in the nominal values of the nuisance parameters $\tilde{\theta}$. The systematic uncertainties that affect only the normalizations of the signal and background processes are assigned log-normal external constraints, whereas the shape altering systematic uncertainties are assigned Gaussian external constraints. 
The negative log-likelihood is defined as
  \begin{equation}
    \DeltaLL=-2\Delta\ln{\frac{\mathcal{L}(\text{data} | \muggh, \muqqh, \vec{f}, \theta)}{\mathcal{L}(\text{data} | \hat{\mu}_{\Pg\Pg\PH}, \hat{\mu}_{\PQq\PQq\PH}, \vec{\hat{f}}, \hat{\theta})}},
  \end{equation}
with $\hat{\mu}_{\Pg\Pg\PH}$, $\hat{\mu}_{\PQq\PQq\PH}$, $\vec{\hat{f}}$, and $\hat{\theta}$ as the best fit values of the signal modifiers and nuisance parameters. 
The 68 and 95\% confidence level (\CL) intervals are identified when $\DeltaLL = $ 1.00 and 3.84, respectively, for which exact coverage is derived using the asymptotic approximation~\cite{Cowan:2010js}.  

\begin{figure*}[!h]
\centering
\includegraphics[width=\textwidth]{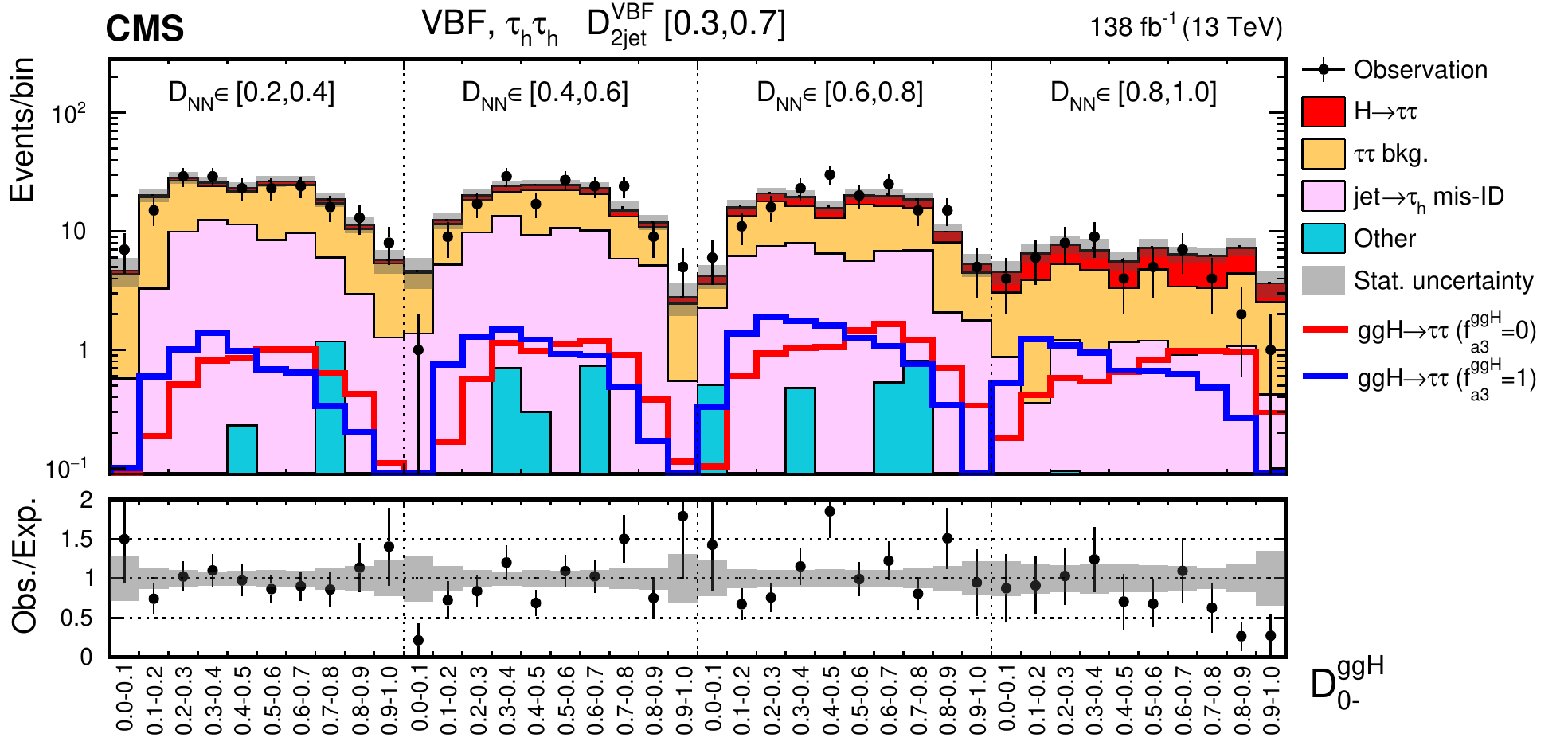} 
\caption{
  The observed and predicted 2D distribution of ($\DzggH$, $\Dnn$) before the fit to data in the most sensitive VBF category region with $0.3 < \DVBF < 0.7$ in the $\tauh\tauh$ channel. The total \PH signal, including VBF, \ggH, and \VH processes, is shown stacked on top of the background in the solid red histogram. The \ggH signal for the $CP$-even ($CP$-odd) scenario is also shown overlaid by the red (blue) line. Only the statistical uncertainties are included in the uncertainty band.}
\label{fig:plot_3d_mt_tt_ggH}
\end{figure*}

\begin{figure*}[!h]
\centering
\includegraphics[width=\textwidth]{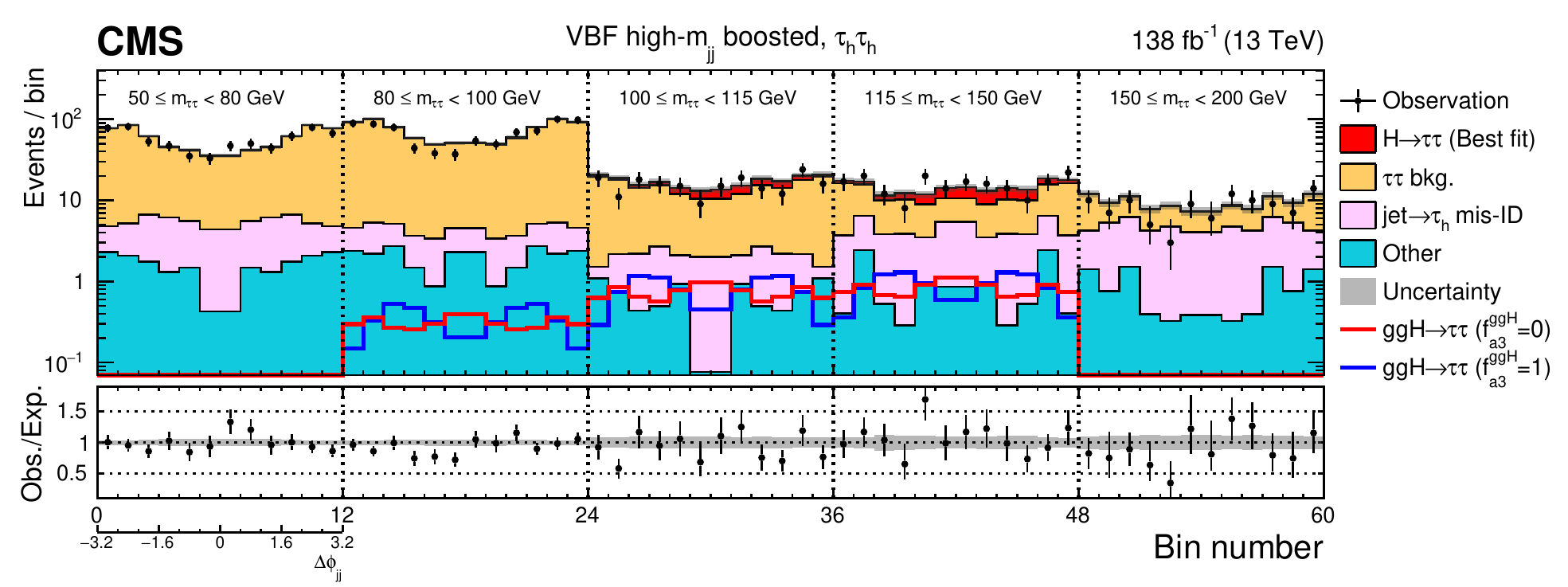}
\caption{ 
Observed and predicted 2D distributions after the fit to data in the VBF high-\mjj boosted category in the $\tauh\tauh$ channel. The total \PH signal, including VBF, \ggH, and \VH processes, is shown stacked on top of the background in the solid red histogram. The \ggH signal for the $CP$-even ($CP$-odd) scenario is also shown overlaid by the red (blue) line. The uncertainty band accounts for all sources of systematic uncertainty in the signal and background predictions. The expectation in the ratio panel is the sum of the estimated backgrounds and the best fit signal.}
\label{fig:plot_2d_mt_tt_vbf} \end{figure*}

\begin{figure*}[!htb]
\centering
\includegraphics[width=0.45\textwidth]{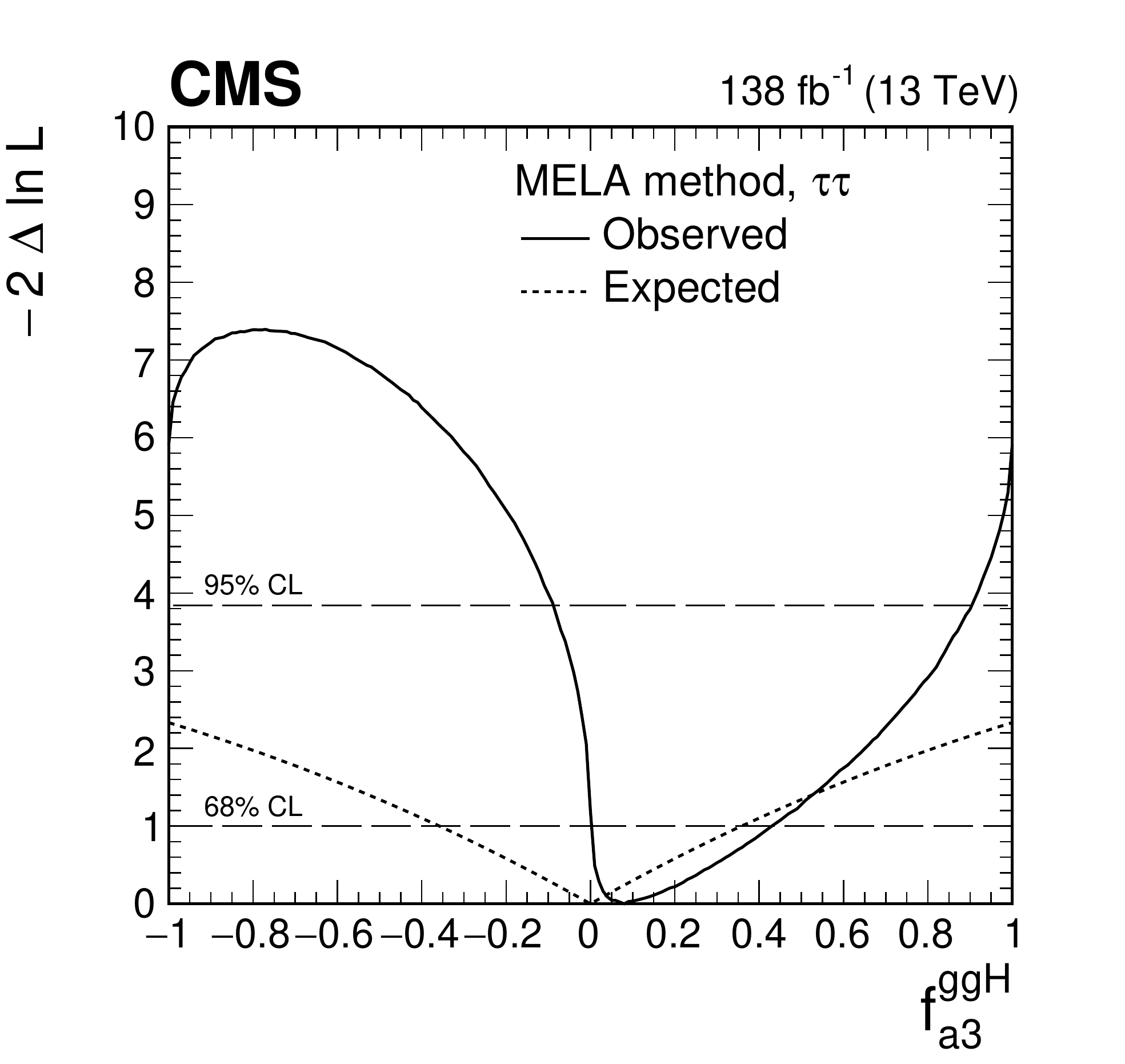}
\includegraphics[width=0.45\textwidth]{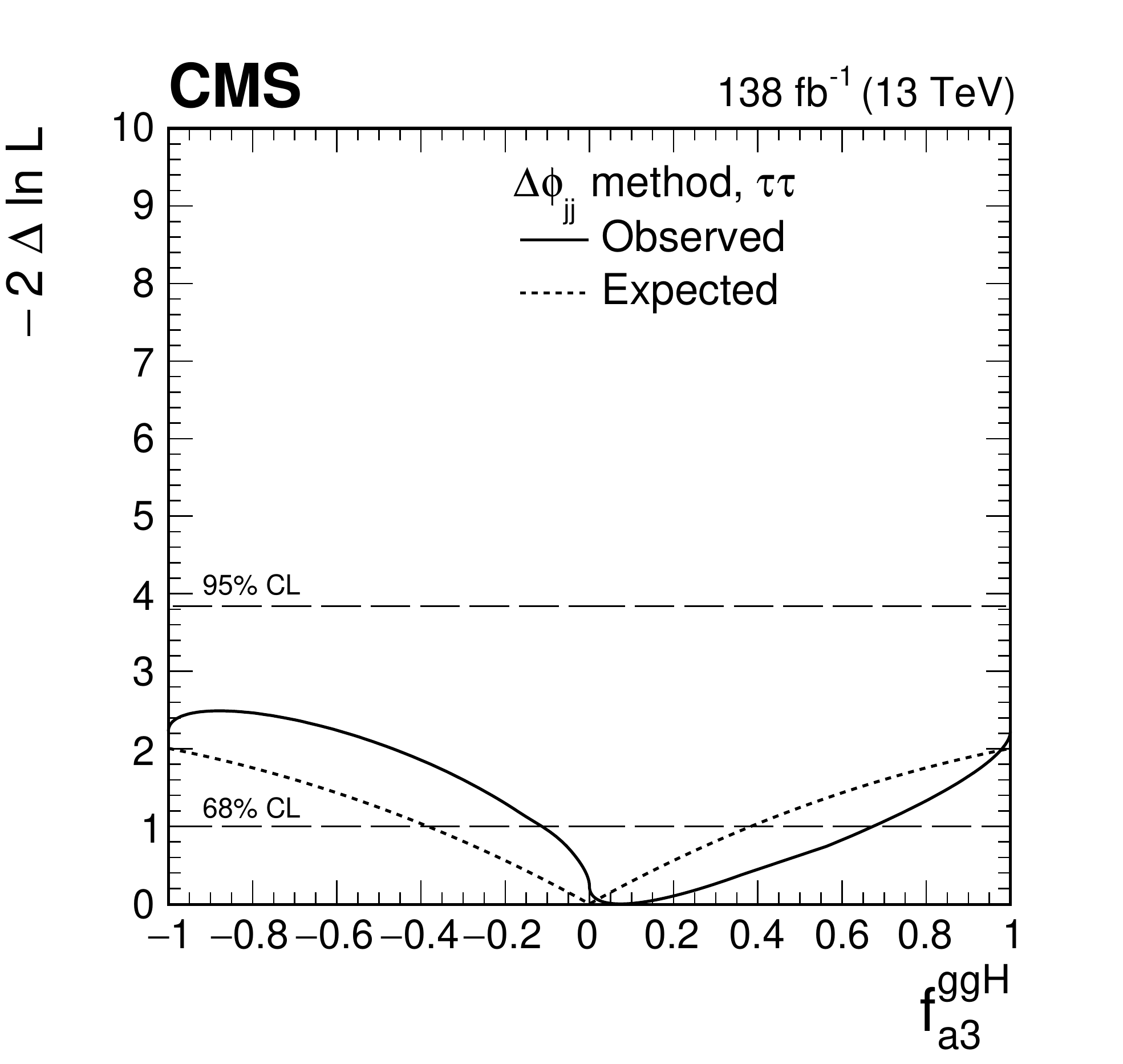}
\caption
{
Observed (solid) and expected (dashed) likelihood scans of
\fggh obtained with the \MELA method (left) and the \dphi method used as a cross check (right).
\label{fig:fa3_ggH}
}
\end{figure*}

\begin{figure*}[!htb]
\centering
\includegraphics[width=0.45\textwidth]{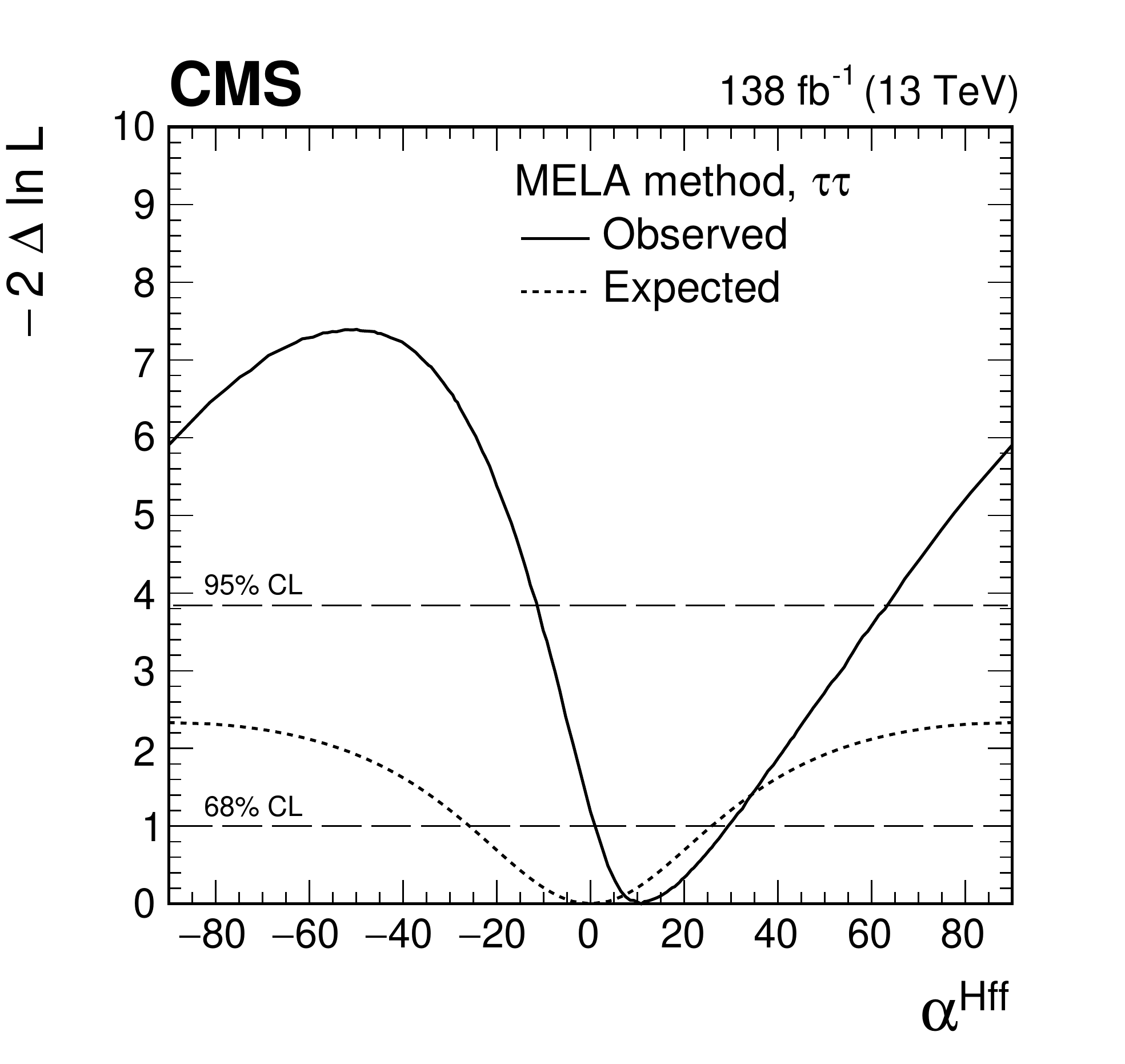}
 \includegraphics[width=0.45\textwidth]{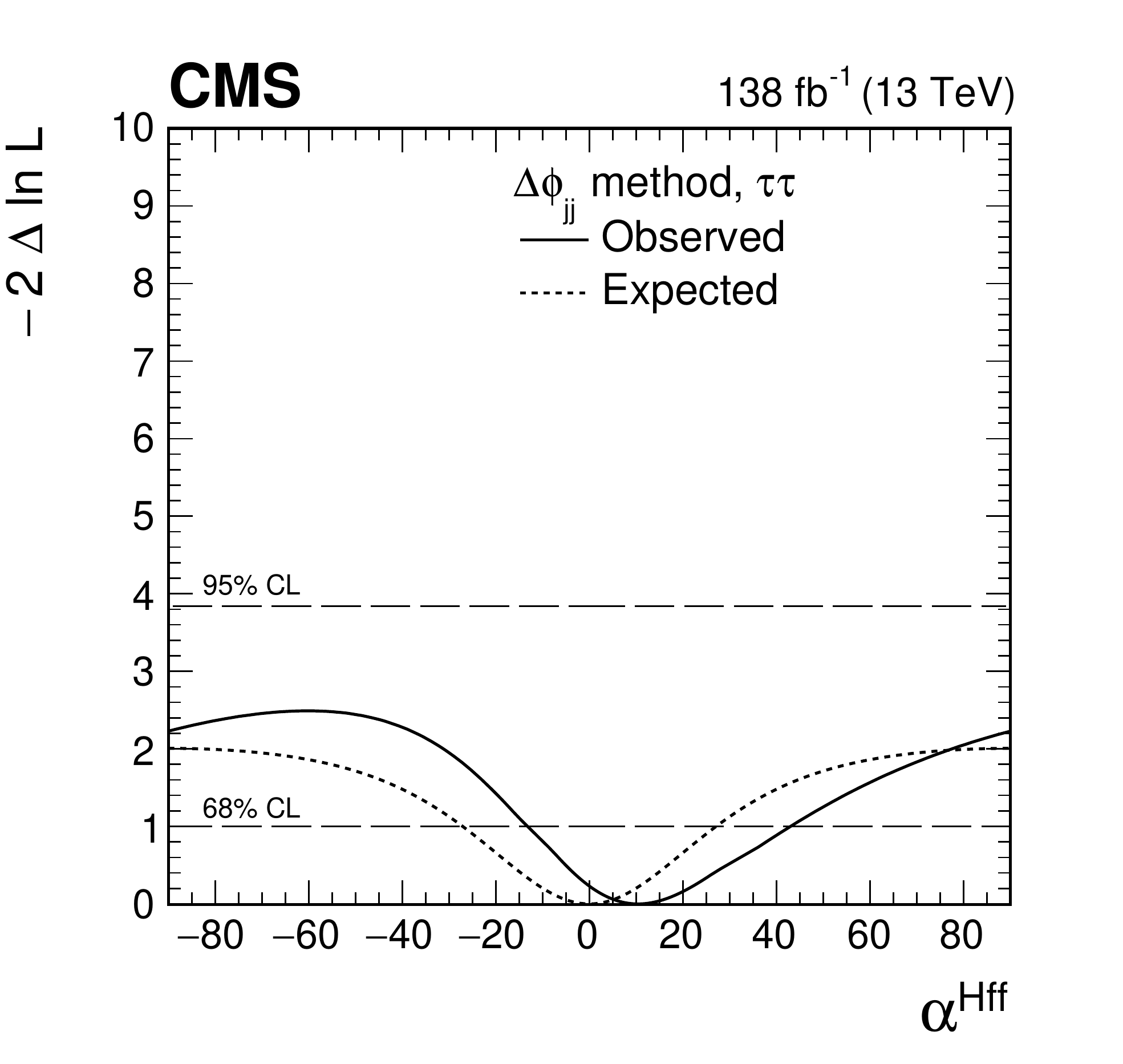}
\caption
{
Observed (solid) and expected (dashed) likelihood scans of
\alphacp (in degrees) obtained with the \MELA method (left) and the \dphi method used as a cross check (right).
\label{fig:alpha_Hff}
}
\end{figure*}

The measurements of \fggh, or equivalently \fCP or \alphacp according to Eq.~(\ref{eq:fai-relationship-hgg-tth}), is performed using the two methods based on \MELA and \dphi.
An example of a pre-fit distribution for the \MELA method is given in Fig.~\ref{fig:plot_3d_mt_tt_ggH} for one of the most sensitive signal categories.
Figure~\ref{fig:plot_2d_mt_tt_vbf} shows the postfit distribution
in the VBF high-\mjj boosted category in the $\tauh\tauh$ channel, which is
the most sensitive category used to extract the results using the \dphi method.
The results of the likelihood scans are shown in Figs.~\ref{fig:fa3_ggH}--\ref{fig:alpha_Hff} and listed in Table~\ref{tab:summary_spin0_ggH}.

\begin{table*}[htb!]
\centering
\topcaption{
Allowed 68\% (central values with uncertainties) and 95\%~\CL (in square brackets)
intervals on anomalous \Hgg coupling parameters using the \HTT decay. The use of ``\NA" indicates cases where no exclusion at the 95\%~\CL was found. As indicated in the Table, the results are presented for the \MELA method, as well as the \dphi method for comparison. The final results of this study are from the \MELA method.
The \alphacp results are derived from \fggh following Eqs.~(\ref{eq:fCP_definitions}),~(\ref{eq:alpha_defintion}), and~(\ref{eq:fai-relationship-hgg-tth}).
}
\renewcommand{\arraystretch}{1.25}
\begin{scotch}{cccccc}
 Parameter              & Method                     &  \multicolumn{2}{c}{Observed} &  \multicolumn{2}{c}{Expected}    \\
                        &                             & 68\%~\CL &  95\%~\CL& 68\%~\CL &  95\%~\CL \\
\hline
\fggh & \MELA & $0.08^{+0.35}_{-0.08}$ & $[-0.09,0.90]$    &   $0.00\pm0.36$ & \NA  \\
\fggh & \dphi & $0.07^{+0.59}_{-0.19}$ & \NA   &   $0.00\pm0.39$ & \NA  \\
\alphacp & \MELA & ${(11^{+18}_{-10})}^{\circ}$ & $[-11,63]$     &   $(0\pm26)^{\circ}$ & \NA  \\
\alphacp & \dphi & ${(10^{+32}_{-24})}^{\circ}$ & \NA  &   $(0\pm 27)^{\circ}$ & \NA \\
\end{scotch}
\label{tab:summary_spin0_ggH}
\end{table*}

For the \MELA method, the maximum value of the \DeltaLL is noticeably larger for the observed scan (occurring at $\fggh\approx -0.8$) compared to the expected.
We checked and excluded a possibility that the discrepancy originated from artificial effects of our analysis procedure.
We used pseudoexperiments to estimate the probability of obtaining a maximum value of the \DeltaLL greater than or equal to the maximum \DeltaLL of the observed scan.
This probability was determined to be 33\%.
We thus conclude that the observed results are affected by statistical fluctuations but compatible with the signal plus background model.

The use of the \MELA method is shown to improve the expected uncertainty in \fggh (\alphacp) by 8 (4)\%.
This improvement owes partly to the use of neural networks that differentiate the \PH signal from the backgrounds more effectively, and partly to the inclusion of matrix-element discriminants that improve the separation between $CP$-even, $CP$-odd, and mixed $CP$ scenarios.
In the case of the \ggH
production, the SM process is generated by the quark loop represented by the
$a_2^{\Pg\Pg}$ term in Eq.~(\ref{eq:formfact-fullampl-spin0}). This makes it harder
to distinguish the anomalous $a_3^{\Pg\Pg}$ contribution from the SM, as both
are generated by dimension-six operators and many of their kinematic features
are similar. Most of the sensitivity to $CP$-odd couplings is primarily in the
azimuthal correlation of two jets, which explains why the full multivariate
\MELA treatment of kinematic information does not bring as much
additional information as in the case of the VBF production, described in
the following section, where the SM process is generated by the tree-level
coupling $g_1^{\V\V}$ in Eq.~(\ref{eq:formfact-fullampl-spin0}),
which is a dimension-four operator. This leads to kinematic differences
between the anomalous contributions and the SM in multiple observables
and a much larger gain from the full multivariate \MELA treatment.

We also cross-checked the results neglecting $CP$-odd contributions to the VBF and \VH processes (\fathree fixed to zero).
This was found to have only a minor effect on the best fit value and uncertainty of \fggh (\alphacp).
Therefore, we present results only for the more general case where \fathree is unconstrained.

\section{Analysis of VBF production}
\label{sec:HVV}
As \MELA-based observables offer superior sensitivity and discrimination among different possible anomalous
\HVV couplings in the VBF and \VH productions compared to single kinematic observables, such as
\dphi, the VBF study is conducted with \MELA-based observables only. As the sensitivity to \VH
anomalous effects is small with the present dataset, the analysis is optimized for the VBF process. However, we allow the anomalous couplings to modify the \VH kinematics in the fit to data. 

We employ the same neural networks to separate VBF-like signal and background processes as in the \ggH analysis, 
while \MELA discriminants offer optimal separation between different signal hypotheses
(shown in Table~\ref{tab:vbf_obs}).
We use \DVBF, defined in Eq.~(\ref{eq:d2j-vbf}), to separate SM \ggH
production from the SM VBF production. Several \MELA discriminants, as defined in Eq.~(\ref{eq:d0-vbf}), are constructed to
optimally separate the SM hypothesis from the potential anomalous coupling in the VBF production: 
\begin{linenomath}
\begin{equation}
    \begin{aligned}
        \Dz &= \frac{\mathcal{P}_\mathrm{SM}^\mathrm{VBF}}{\mathcal{P}_\mathrm{SM}^\mathrm{VBF} + \mathcal{P}_\mathrm{0-}^\mathrm{VBF}},
        & \Dzh = \frac{\mathcal{P}_\mathrm{SM}^\mathrm{VBF}}{\mathcal{P}_\mathrm{SM}^\mathrm{VBF} + \mathcal{P}_\mathrm{a2}}, \\
        \DL &= \frac{\mathcal{P}_\mathrm{SM}^\mathrm{VBF}}{\mathcal{P}_\mathrm{SM}^\mathrm{VBF} + \mathcal{P}_{\Lambda1}},
        & \DLzg = \frac{\mathcal{P}_\mathrm{SM}^\mathrm{VBF}}{\mathcal{P}_\mathrm{SM}^\mathrm{VBF} + \mathcal{P}_{\Lambda1}^{\PZ\PGg}}. \\
    \end{aligned}
    \label{eq:d0-vbf}
\end{equation}
\end{linenomath}

As for the \Hgg analysis, we also define a pure $CP$-odd \MELA discriminant \DCP in Eq.~(\ref{eq:dcp-vbf}), that is sensitive to interference effects between the SM and pseudoscalar \PH
contributions to directly probe for $CP$-violation in the \HVV vertex:
\begin{linenomath}
\begin{equation}
    \DCP = \frac{\mathcal{P}_\mathrm{SM-0-}^\mathrm{VBF}}{\mathcal{P}_\mathrm{SM}^\mathrm{VBF} + \mathcal{P}_\mathrm{0-}^\mathrm{VBF}}.
    \label{eq:dcp-vbf}
\end{equation}
\end{linenomath}

The results of the VBF analysis are extracted with a global maximum likelihood fit based on 4D or 3D, 2D, and 1D distributions built in each of the VBF, boosted, and 0-jet categories, respectively. 
The templates constructed for the boosted and 0-jet categories are identical to those described for the \ggH analysis in Sec.~\ref{ssec:MVA-ggH}, although we note that in this case, in contrast to the former case,
the chosen observables do provide some differentiation between anomalous coupling scenarios.
 
Depending on the anomalous coupling parameter being measured, we use three or four observables in total to construct the distributions in the VBF category.
In all cases we include the \Dnn and \DVBF observables. 
We additionally include \Dz, \Dzh, \DL, or \DLzg, when we measure \fathree, \fatwo, \fL, or \fLzg, respectively; which we will collectively refer to as \DBSM in the following. 
The fourth observable, which is included only for the measurement of the $CP$-odd parameter \fathree, is \DCP.
The selected events are binned into templates constructed using these observables.

The binning of the templates has been optimized following the criteria outlined in Sec.~\ref{ssec:MVA-ggH}.
For the $\ell\tauh$  and $\tauh\tauh$ channels, we use 10, 8, and 4 equally sized bins for \DBSM, \Dnn, and \DVBF, respectively.
For the \emu channel, we respectively use 3, 2, and 4 bins for these observables.
In all cases, neighboring bins are merged such that the background prediction has no bins with statistical uncertainty larger than 50\%.
For the measurement of the \fathree parameter we include two bins in the \DCP discriminant, $\DCP < 0$ and $\DCP \ge 0$,
to bring sensitivity to the sign of the interference between the $CP$-even and $CP$-odd contributions.
The expected symmetry between these bins is enforced for the background and $CP$-conserving signal templates to reduce the influence of statistical fluctuations.

\subsection{Results of the \texorpdfstring{\HVV}{HVV} analysis}
\label{ssec:HVV_Results}

The four \fai parameters describing anomalous \HVV couplings, as defined in
Eqs.~(\ref{eq:formfact-fullampl-spin0}) and~(\ref{eq:fa_definitions}), are tested against the data according to the
likelihood function defined in Eq.~(\ref{eq:likelihoodDensity}), following the same approach as that utilized
in the analysis of the \Hgg vertex.

\begin{figure*}[!h] 
\centering
  \includegraphics[width=\textwidth]{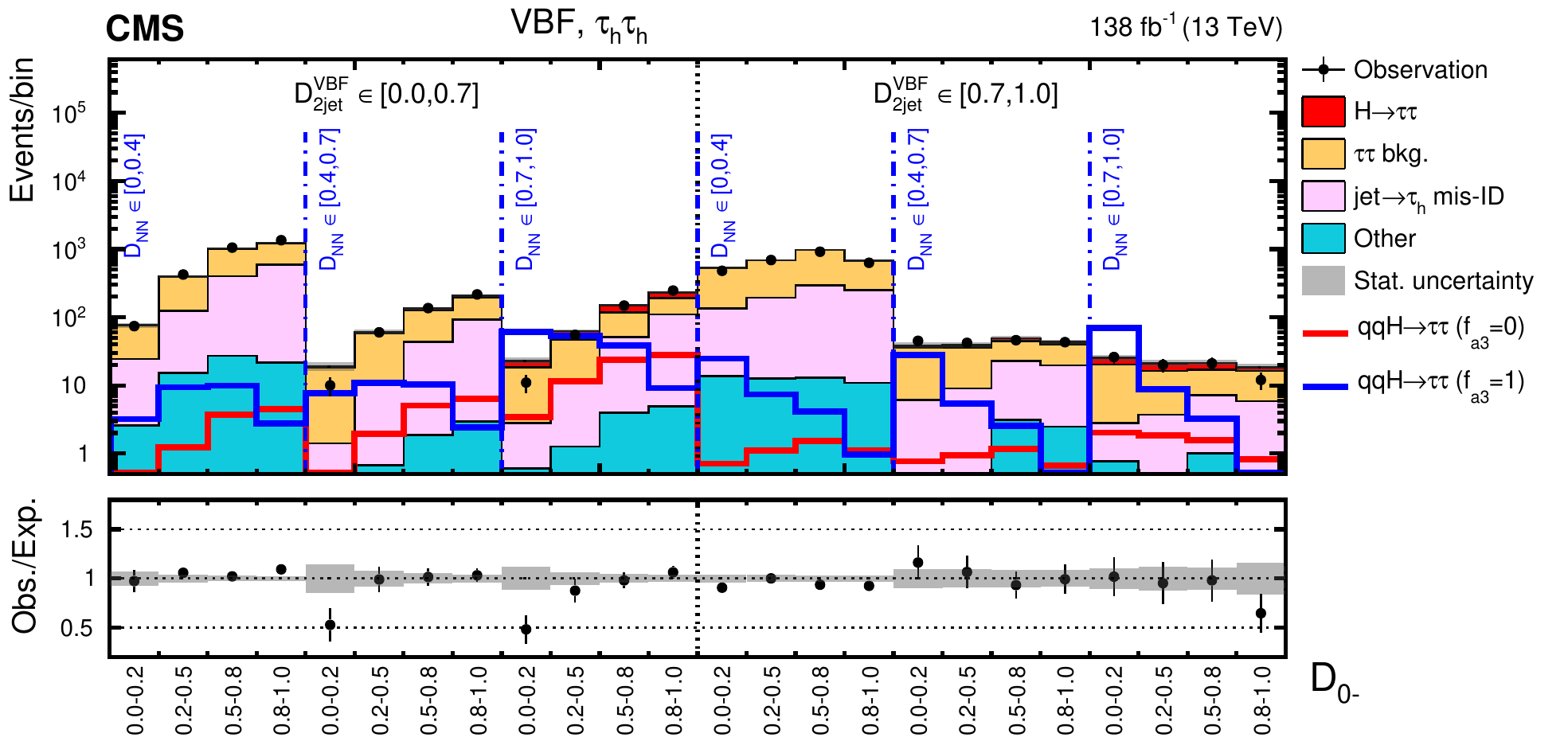}
  \caption{The observed and predicted 3D distribution of (\Dz, \Dnn, \DVBF) before the fit to data in the $\tauh\tauh$ channel for the most sensitive VBF category. The total \PH signal, including VBF, \ggH, and \VH processes, is shown stacked on top of the background in the solid red histogram. The VBF+\VH signal for the $CP$-even ($CP$-odd) scenario is also shown overlaid by the red (blue) line.
    Only the statistical uncertainties are included in the uncertainty band.} 
\label{fig:plot_3d_mt_tt_HVV} \end{figure*}

\begin{figure*}[!htb]
\centering
\includegraphics[width=0.45\textwidth]{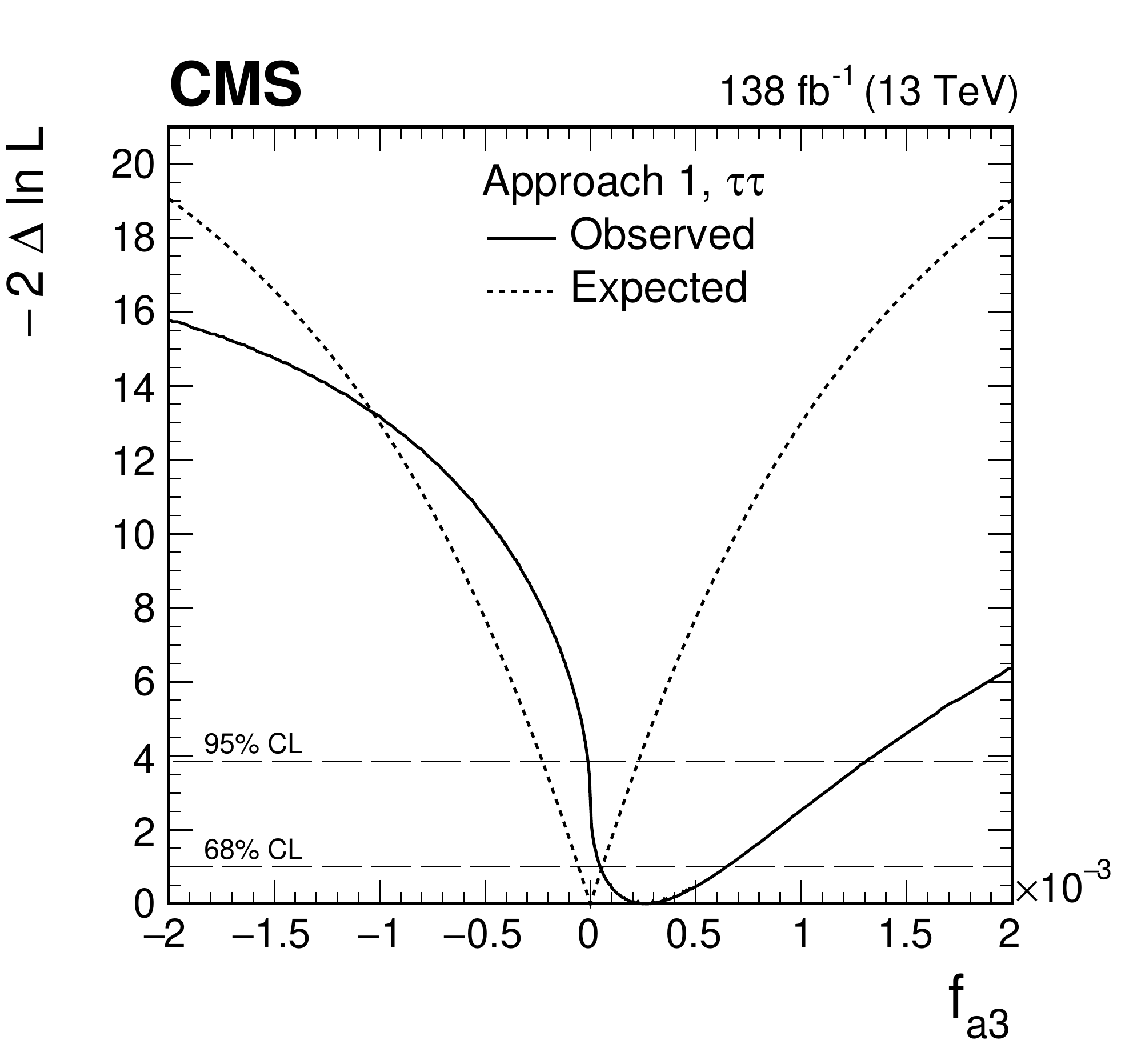}
\includegraphics[width=0.45\textwidth]{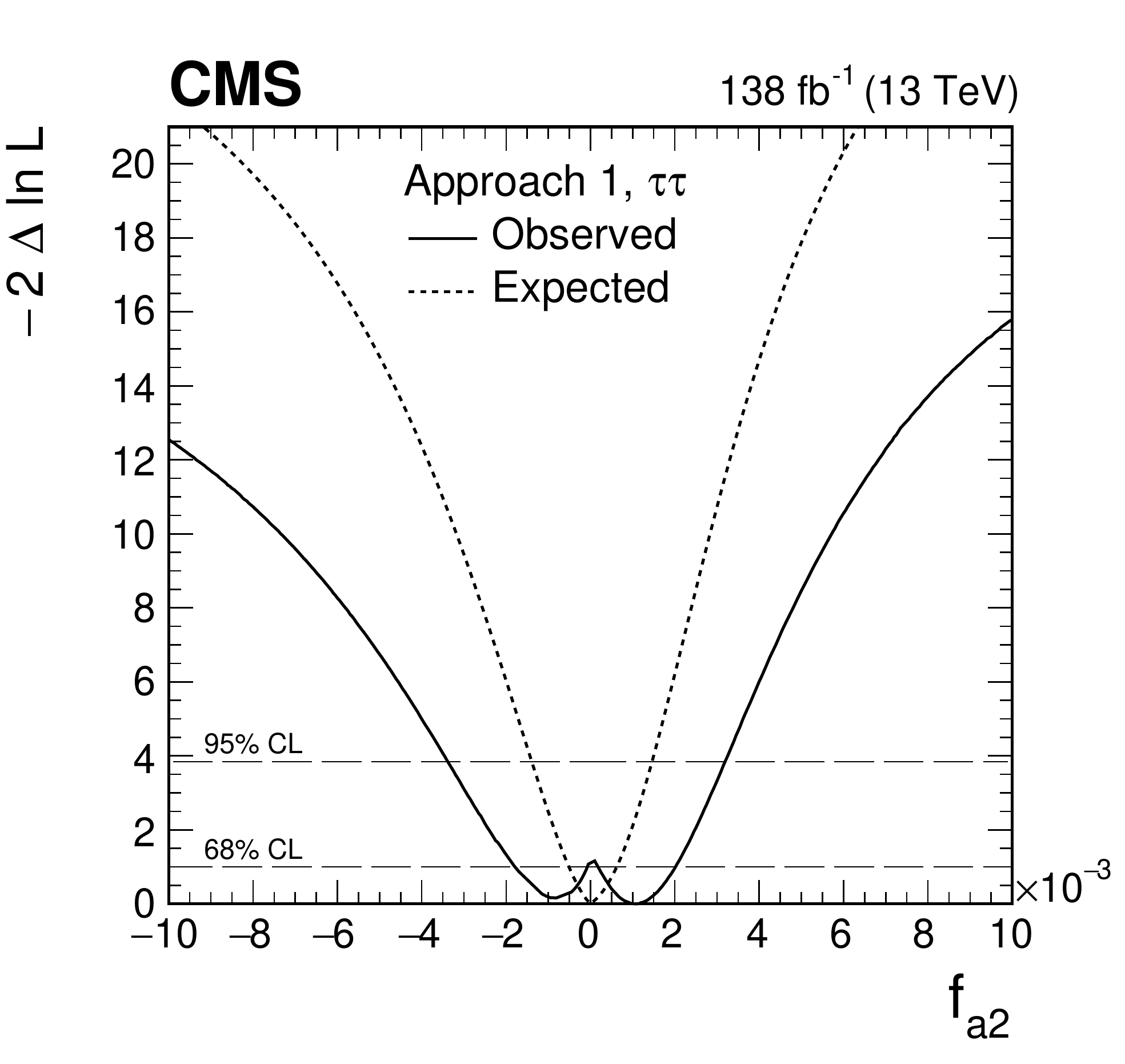}
\includegraphics[width=0.45\textwidth]{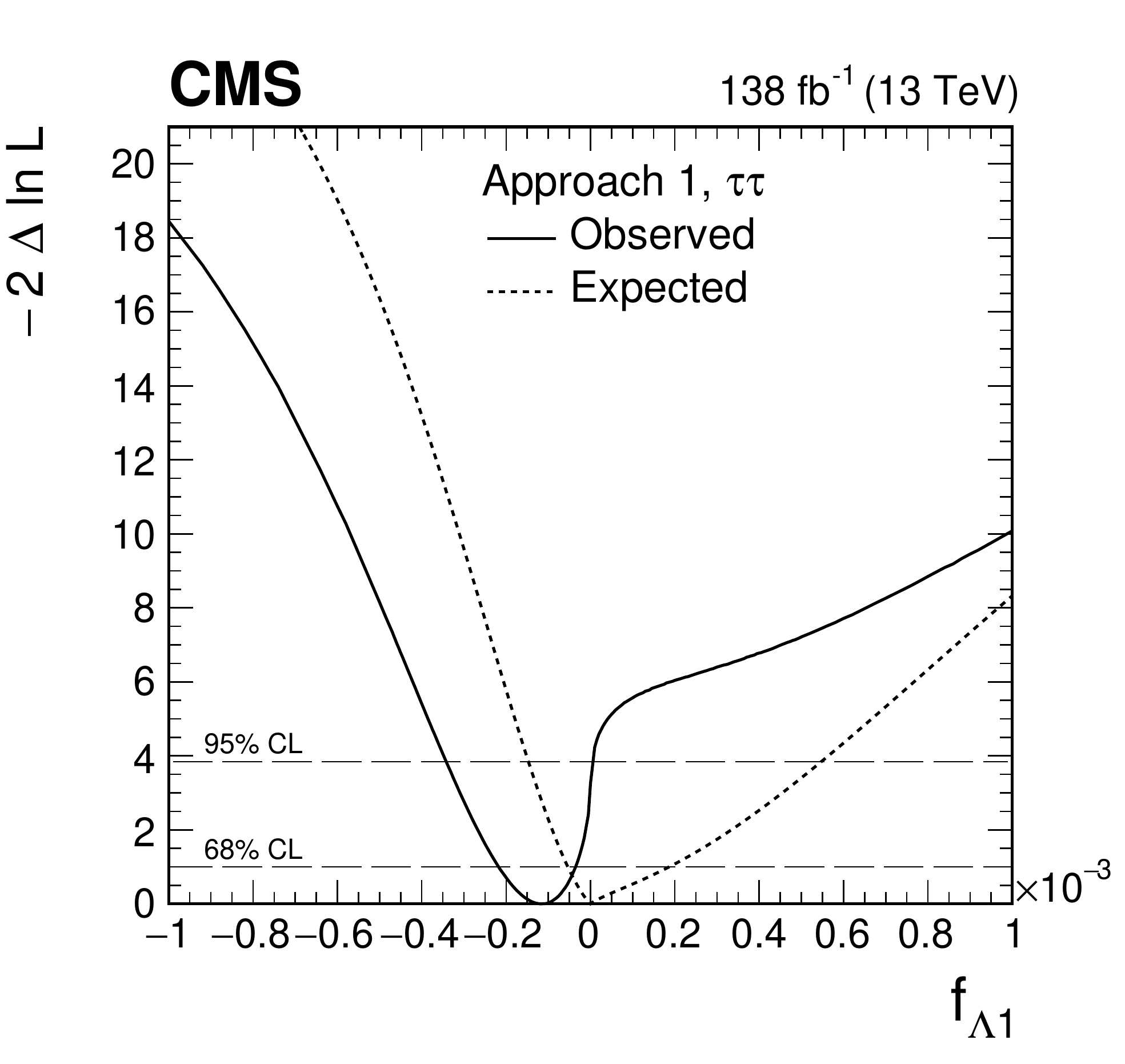}
\includegraphics[width=0.45\textwidth]{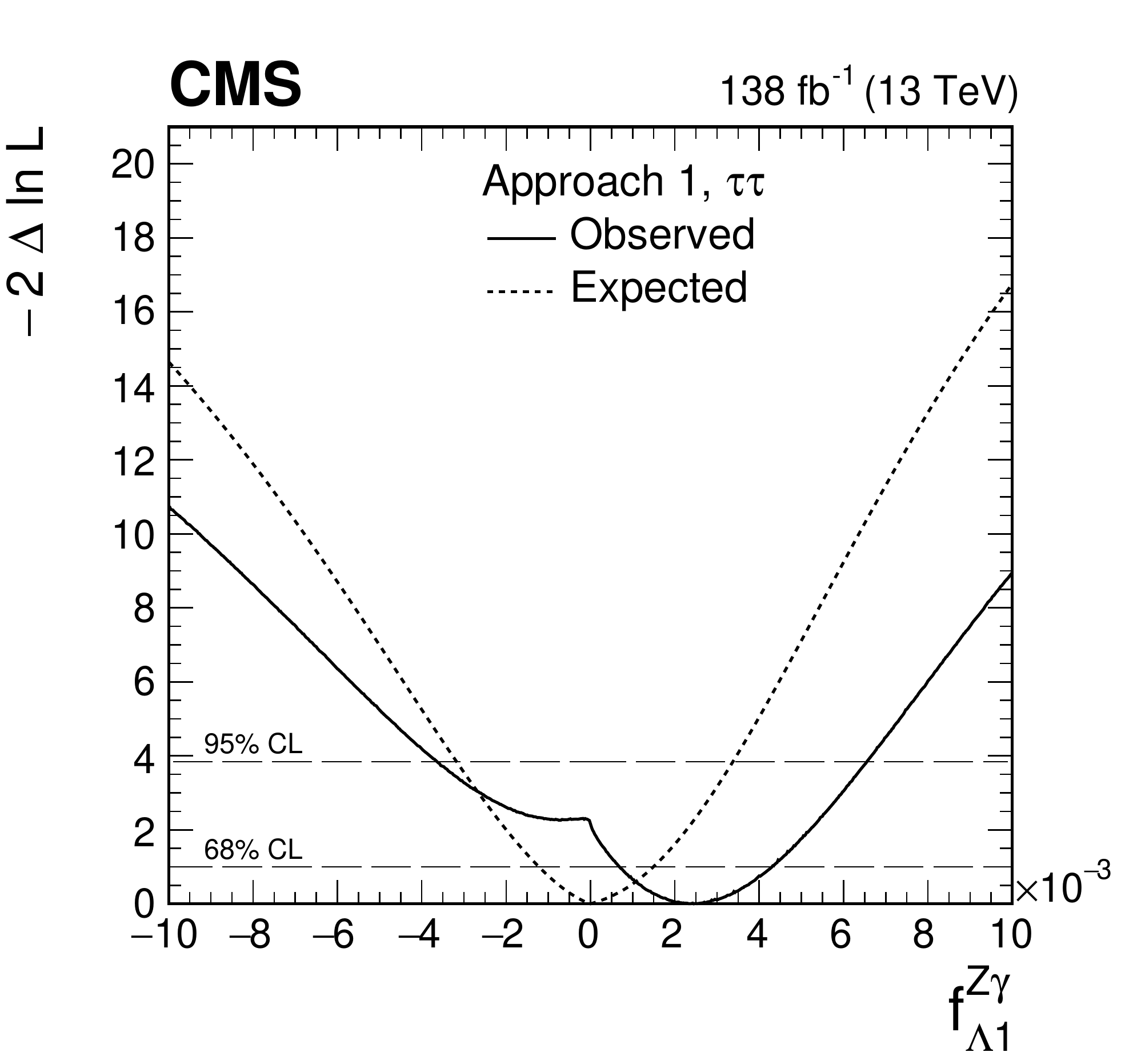}
\caption
{
Observed (solid) and expected (dashed) likelihood scans of
\fathree (upper left),
\fatwo (upper right),
\fL (lower left), and
\fLzg (lower right) in Approach 1 (${a_i^{\PW\PW}=a_i^{\PZ\PZ}}$).
\label{fig:fa3}
}
\end{figure*}

\begin{figure}[!htb]
\centering
\includegraphics[width=0.45\textwidth]{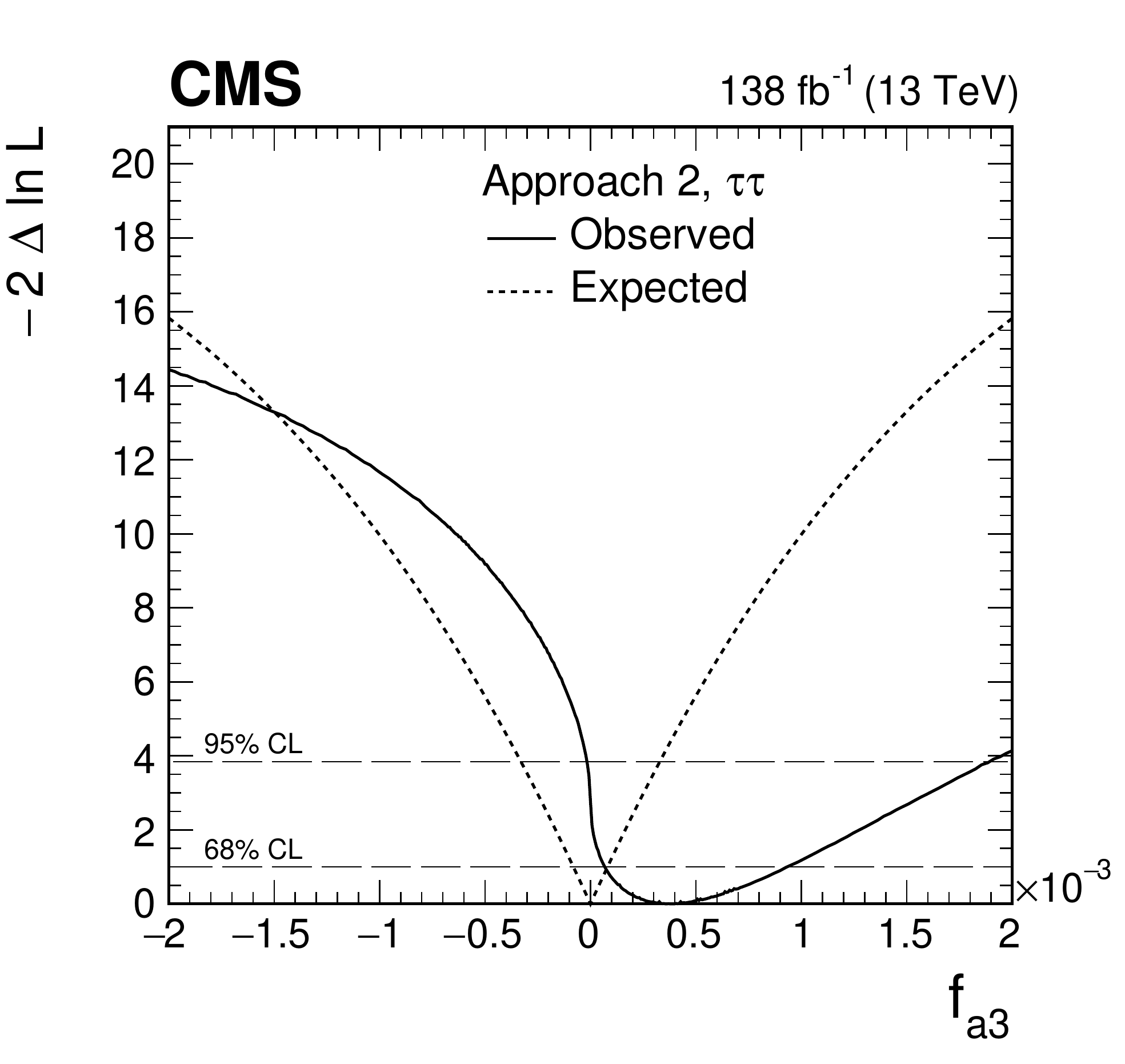}
\caption
{
Observed (solid) and expected (dashed) likelihood scans of
\fathree in Approach 2.
\label{fig:fa3_reweighted}
}
\end{figure}

An example of a pre-fit distribution in the most sensitive VBF category
for the $\tauh\tauh$ channel is shown in Fig.~\ref{fig:plot_3d_mt_tt_HVV}.
The results of the likelihood scans for Approaches 1 and 2 (defined in Sec.~\ref{sec:pheno}) are listed in Table~\ref{tab:summary_spin0} and shown in Figs.~\ref{fig:fa3} and \ref{fig:fa3_reweighted}, respectively.
In each fit, the values of the other anomalous coupling parameters are set to zero, with the exception of
the fit to the $CP$-odd parameter \fathree, which is extracted with \fggh left unconstrained.
The signal strength parameters \muqqh and \muggh are also profiled for all measurements. 
The best fit values of these parameters are consistent with unity.

\begin{table*}[h]
{\centering
\topcaption{
Allowed 68\% (central values with uncertainties) and 95\%~\CL (in square brackets)
intervals on anomalous \HVV coupling parameters using the \HTT decay.
Approaches 1 and 2 refer to the choice of the relationship between the $a_i^{\PW\PW}$ and $a_i^{\PZ\PZ}$ couplings,
defined in Section~\ref{sec:pheno}. For the observed \fatwo scan, there is a second region allowed at the 68\% CL away from the best fit value. We use the union symbol ($\cup$) to display the additional allowed \fatwo range in this case. 
}
\label{tab:summary_spin0}
\renewcommand{\arraystretch}{1.25}
\begin{scotch}{cccccc}
Approach & Parameter                                   &  \multicolumn{2}{c}{Observed$/(10^{-3})$} &  \multicolumn{2}{c}{Expected$/(10^{-3})$}    \\
&                                                     & 68\%~\CL &  95\%~\CL& 68\%~\CL &  95\%~\CL \\
\hline
\multirow{4}{*}{Approach 1} & \fathree & $0.28^{+0.38}_{-0.23}$ & $[-0.01,1.30]$    &   $0.00\pm0.06$ & $[-0.23,0.23]$  \\
& \fatwo  & $1.1^{+0.9}_{-0.9}$ $\cup[-1.8,-0.1]$ & $[-3.4,3.2]$    &   $0.0^{+0.6}_{-0.5}$ & $[-1.4,1.5]$  \\
& \fL & $-0.12^{+0.08}_{-0.10}$ & $[-0.34, 0.01]$    &   $0.00^{+0.19}_{-0.05}$& $[-0.15,0.55]$  \\
& \fLzg  & $2.5\pm1.8$ & $[-3.6,6.6]$    &   $0.0^{+1.5}_{-1.2}$ & $[-3.2,3.4]$  \\
\hline
Approach 2 & \fathree & $0.40^{+0.53}_{-0.33}$ & $[-0.01,1.90]$    &   $0.00\pm0.08$ & $[-0.33,0.33]$  \\
\end{scotch}\\
}
\end{table*}

The presence of two minima in the observed likelihood scan for \fatwo (and to a lesser extent \fLzg) is a result of the limited sensitivity to the sign of the interference between the $a_{1}$ and $a_{2}$
($\kappa_2^{\PZ\PGg}$) couplings, which in turn limits the sensitivity to the signs of the \fatwo
(\fLzg) parameters.

The CL intervals on \fai at 95\% and 68\% are more stringent compared to those utilizing the \PH
decay information in the \Hllll channel~\cite{CMS-HIG-19-009} because the VBF and \VH production processes are 
sensitive to higher values of $p_i^2$ appearing in Eq.~(\ref{eq:formfact-fullampl-spin0}). Therefore, 
the cross section of anomalous contributions in VBF and \VH production increases quickly with \fai.
As the cross section increases with respect to \fai at different rates for production and decay,
relatively small values of \fai correspond to a substantial anomalous contribution to the production cross section.
This leads to the plateau in the $-2\ln{\cal L}$ distributions for larger values of \fai in Fig.~\ref{fig:fa3}.
By using the cross section ratios for VBF production in the \fai definition in Eq.~(\ref{eq:fa_definitions}),
the appearance of the plateau and the narrow exclusion range would change.
The \fai constraints in Ref.~\cite{CMS-HIG-19-009} also utilize the VBF and \VH production information,
but the number of  reconstructed \Hllll events in these production modes is still low compared to this analysis. 

\section{Combination of the results with other decay channels}
\label{sec:combination}

The results of the anomalous coupling measurements presented in the previous sections can be further improved
by combining with other \PH production and decay channels.
The precision of the anomalous \HVV and \Hgg coupling measurements is improved by combining the \HTT and \Hllll decay channels, where we consider \PH production via VBF, \VH, and \ggH.
We additionally constrain the anomalous \Htt couplings by combining the $\Pg\Pg\PH\to\PGt\PGt/4\ell$ and $\PQt\PAQt\PH/\PQt\PH\to\PGg\PGg/4\ell$ channels.
     
For all combinations, each \PH decay channel treats anomalous couplings in \PH production processes in the likelihood
in a consistent manner.  
As with the \HTT only fits, in the likelihood fit for a given parameter the values of the other anomalous couplings are set to zero
with the exception of the fits to \fathree and \fggh, and the signal strength parameters are profiled in the combined likelihood fit.
The number of signal strength parameters in the combined fit can be reduced by using a relationship between the production cross section ratios.
For example, there are in principle four signal strength parameters for the combination of the \HTT and \Hllll channels
($\mu_{\PQq\PQq\PH}^{\PGt\PGt}$,  $\mu_{\Pg\Pg\PH}^{\PGt\PGt}$, $\mu_{\PQq\PQq\PH}^{\PZ\PZ}$,  $\mu_{\Pg\Pg\PH}^{\PZ\PZ}$).
However, one degree of freedom is removed because the ratio between the $\Pg\Pg\PH$ and VBF+$\V\PH$ cross sections is the same in both channels, 
$\mu_{\PQq\PQq\PH}^{\PGt\PGt} / \mu_{\Pg\Pg\PH}^{\PGt\PGt} = \mu_{\PQq\PQq\PH}^{\PZ\PZ} / \mu_{\Pg\Pg\PH}^{\PZ\PZ}$.
Therefore, we can parametrize the combined fit with three signal strength parameters $\mu_{\PQq\PQq\PH}$, $\mu_{\Pg\Pg\PH}$, and $\eta_{\PGt}$,
where $\eta_{\PGt}$ stands for the relative strength of the \PH coupling to the $\PGt$ leptons.
For the combination with the \ttH and $\PQt\PH$ results using the \Hllll and \Hgamgam channels,
the signal strengths $\mu_{\PQt\PAQt\PH}^{\PZ\PZ}$ and $\mu_{\PQt\PAQt\PH}^{\PGg\PGg}$ are
not related for the \fCP measurement because they could differ by the loop involved in the \Hgamgam decay.
In the EFT approach, the fully-resolved loop parametrization following Ref.~\cite{Davis:2021tiv} is used to correlate them.
All common systematic uncertainties are treated as being correlated between the channels in the combined likelihood fit.

The measurements of anomalous \Hgg and \HVV couplings using the \MELA method are
combined with the results using the on-shell \Hllll decay~\cite{CMS-HIG-19-009}. 
In the \Hllll analysis, anomalous \HVV couplings can affect both production (VBF+\VH) 
and decay ($\PH\to\V\V\to 4\ell$) processes. Information from both processes is taken into account
in the analysis. 
The combination improves the limits on the anomalous coupling parameters typically by about 20\%--50\%.

\begin{figure*}[!htb]
\centering
\includegraphics[width=0.45\textwidth]{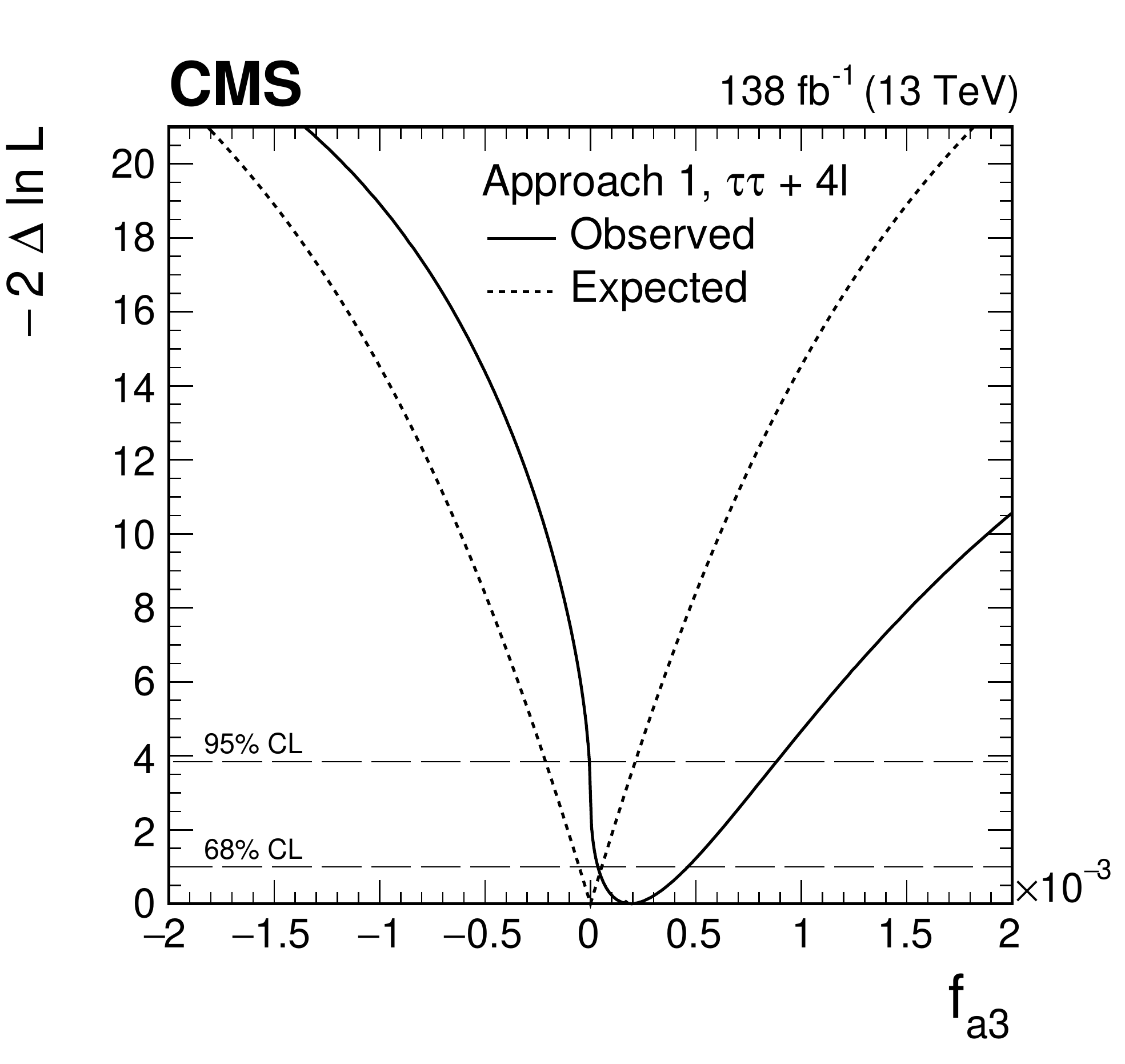}
\includegraphics[width=0.45\textwidth]{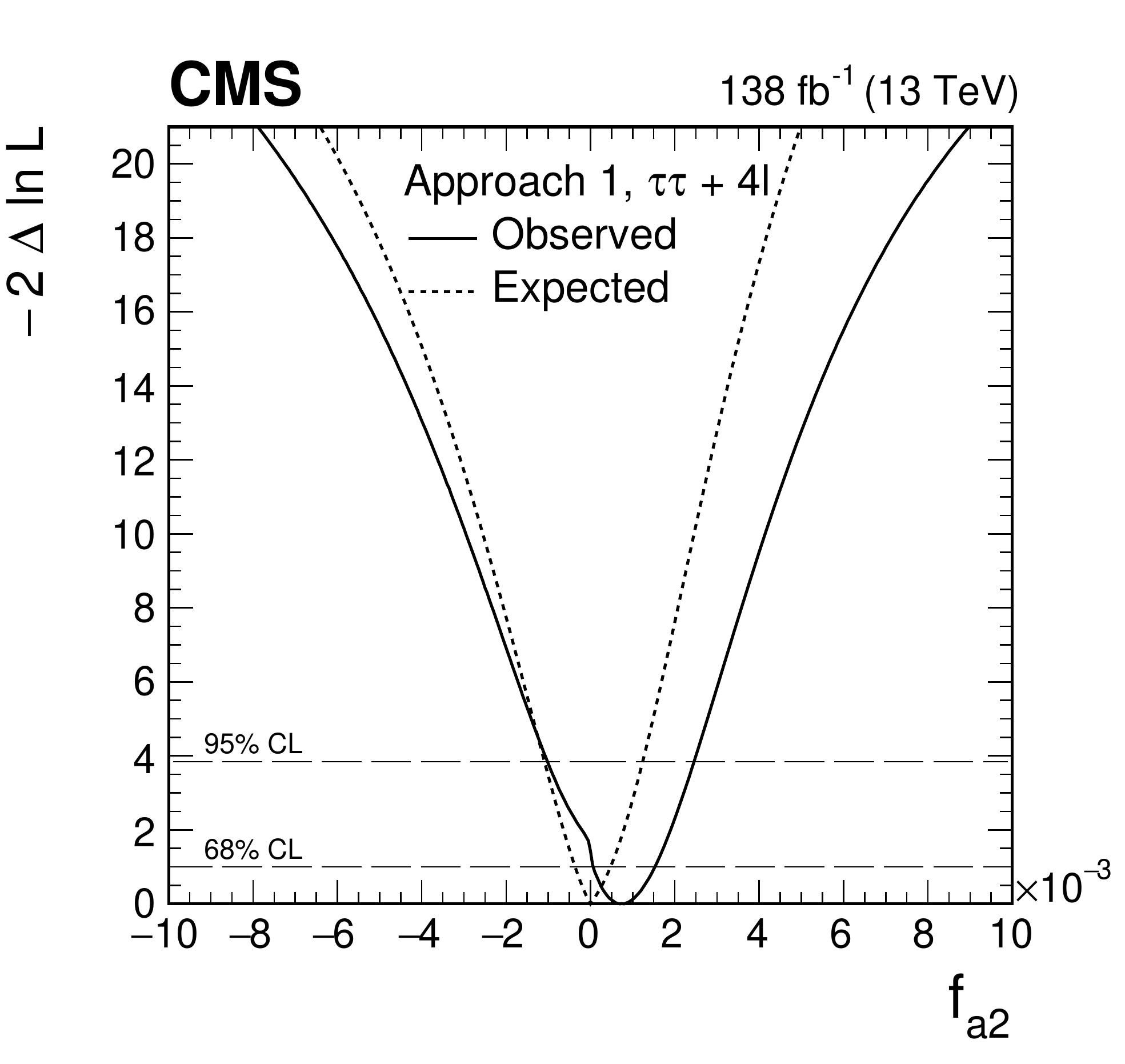}
\includegraphics[width=0.45\textwidth]{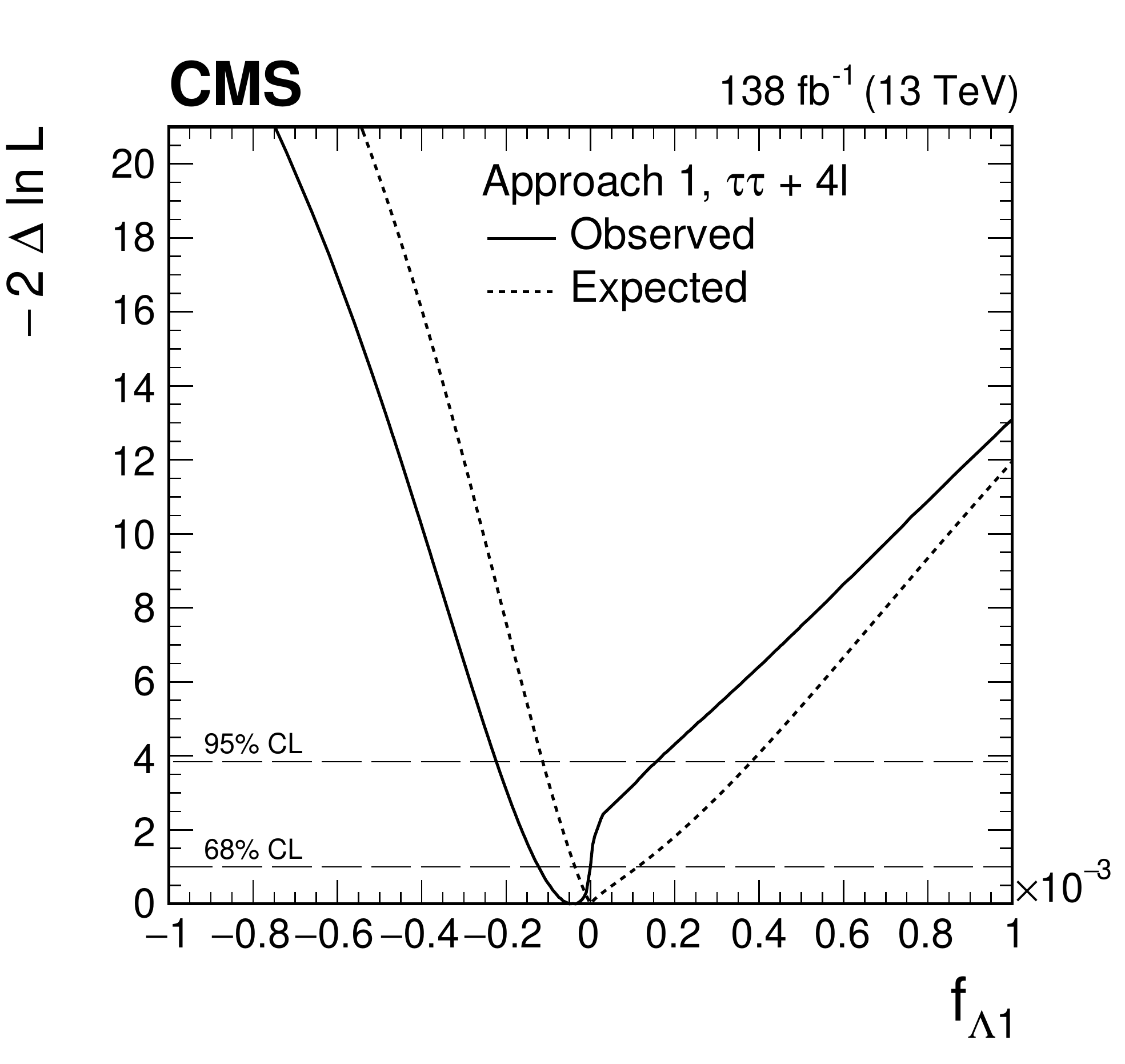}
\includegraphics[width=0.45\textwidth]{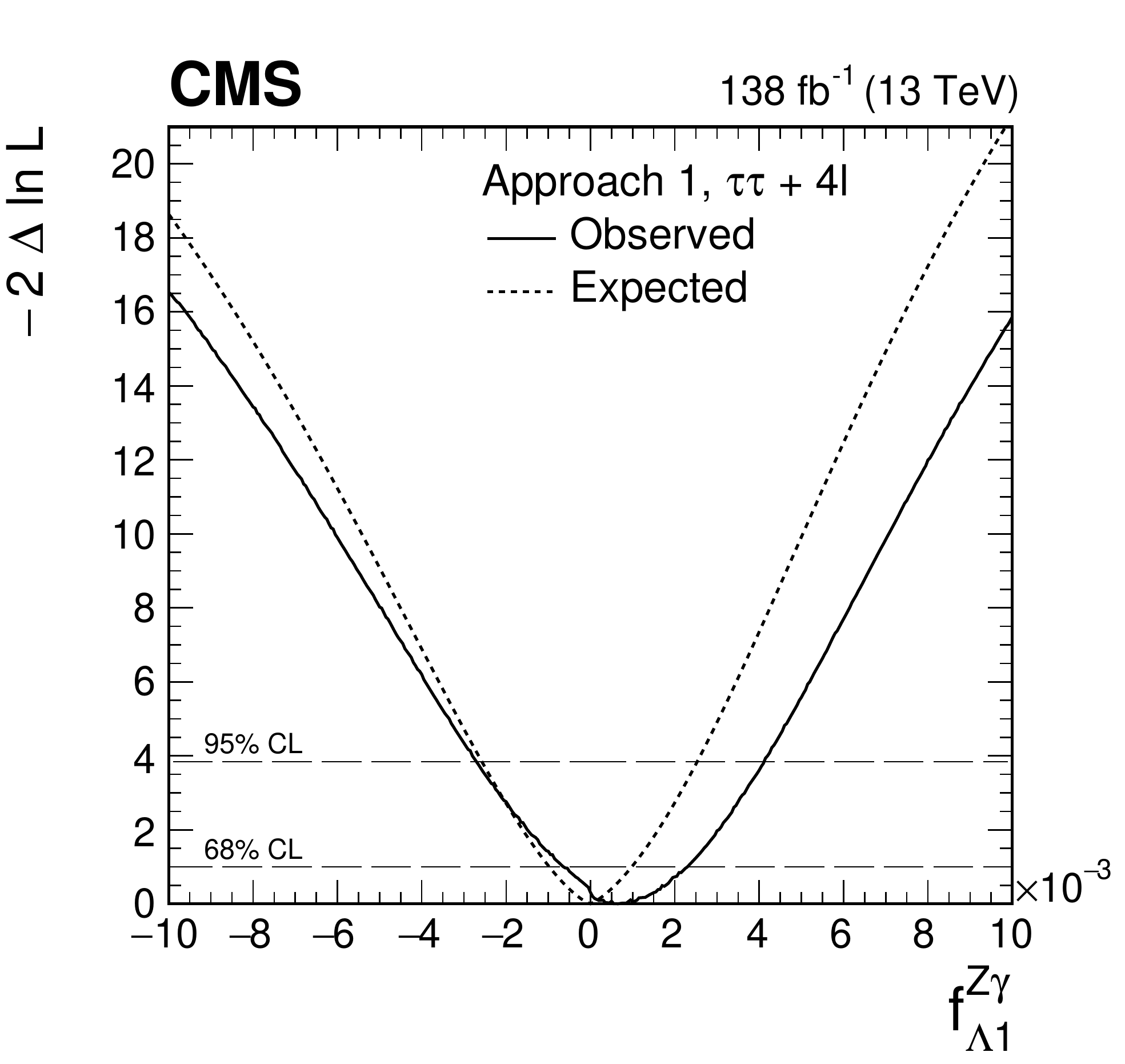}
\caption
{
Observed (solid) and expected (dashed) likelihood scans of
\fathree (upper left),
\fatwo (upper right),
\fL (lower left), and
\fLzg (lower right) in Approach 1 ($a_i^{\PW\PW}$ = $a_i^{\PZ\PZ}$) obtained with the
combination of results using the \HTT and \Hllll~\cite{CMS-HIG-19-009} decay channels.
\label{fig:combination_HVV1}
}
\end{figure*}

\begin{figure*}[!htb]
\centering
\includegraphics[width=0.45\textwidth]{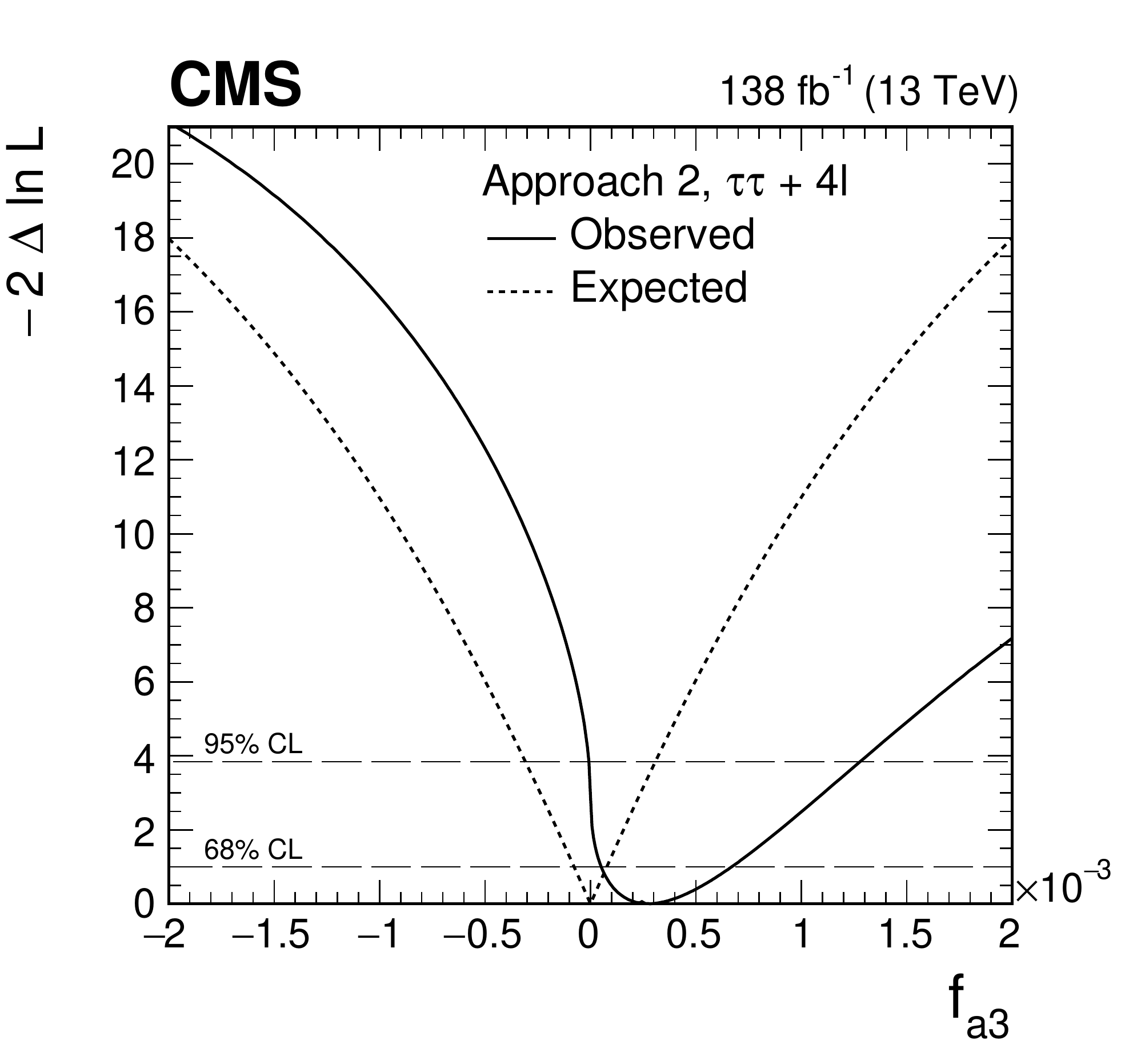}
\caption
{
Observed (solid) and expected (dashed) likelihood scans of
\fathree in Approach 2 (defined in Section~\ref{sec:pheno}) obtained with the combination of results using the \HTT and \Hllll~\cite{CMS-HIG-19-009} decay channels.
}
\label{fig:combination_HVV2}
\end{figure*}

The combined likelihood scans for the \HVV anomalous coupling measurements are shown in Figs.~\ref{fig:combination_HVV1}--\ref{fig:combination_HVV2}, and the allowed 68 and 95\%~\CL intervals are listed in Table~\ref{tab:combination_HVV}.
The \HTT channel results mainly constrain small values of
\fai where the \PH production information is the dominant factor, whereas the \Hllll analysis
provides major constraints at large values of \fai based on the decay information.

\begin{table*}[ht!]
\centering
\topcaption{
Allowed 68\% (central values with uncertainties) and 95\%~\CL (in square brackets)
intervals on anomalous \HVV coupling parameters using the \HTT and \Hllll~\cite{CMS-HIG-19-009} decay channels,
using two approaches described in Sec.~\ref{sec:pheno} that define the relationship between the
$a_i^{\PW\PW}$ and $a_i^{\PZ\PZ}$ couplings.
}
\renewcommand{\arraystretch}{1.25}
\begin{scotch}{cccccc}
Approach & Parameter                                   &  \multicolumn{2}{c}{Observed$/(10^{-3})$} &  \multicolumn{2}{c}{Expected$/(10^{-3})$}    \\
&                                                     & 68\%~\CL &  95\%~\CL& 68\%~\CL &  95\%~\CL \\
\hline
\multirow{4}{*}{Approach 1} & \fathree & $0.20^{+0.26}_{-0.16}$ & $[-0.01,0.88]$    &   $0.00\pm0.05$ & $[-0.21,0.21]$  \\
& \fatwo  & $0.7^{+0.8}_{-0.6}$ & $[-1.0,2.5] $    &   $0.0^{+0.5}_{-0.4}$ & $[-1.1,1.2]$  \\
& \fL & $-0.04^{+0.04}_{-0.08}$ & $[-0.22,0.16]$    &   $0.00^{+0.11}_{-0.04}$ & $[-0.11,0.38]$  \\
& \fLzg  & $0.7^{+1.6}_{-1.3}$ & $[-2.7,4.1]$    &   $0.0^{+1.0}_{-1.0}$ & $[-2.6,2.5]$  \\
\hline
Approach 2 & \fathree & $0.28^{+0.39}_{-0.23}$ & $[-0.01,1.28]$    &   $0.00\pm0.08$ & $[-0.30,0.30]$  \\
\end{scotch}
\label{tab:combination_HVV}
\end{table*}

\begin{figure*}[!htb]
\centering
\includegraphics[width=0.45\textwidth]{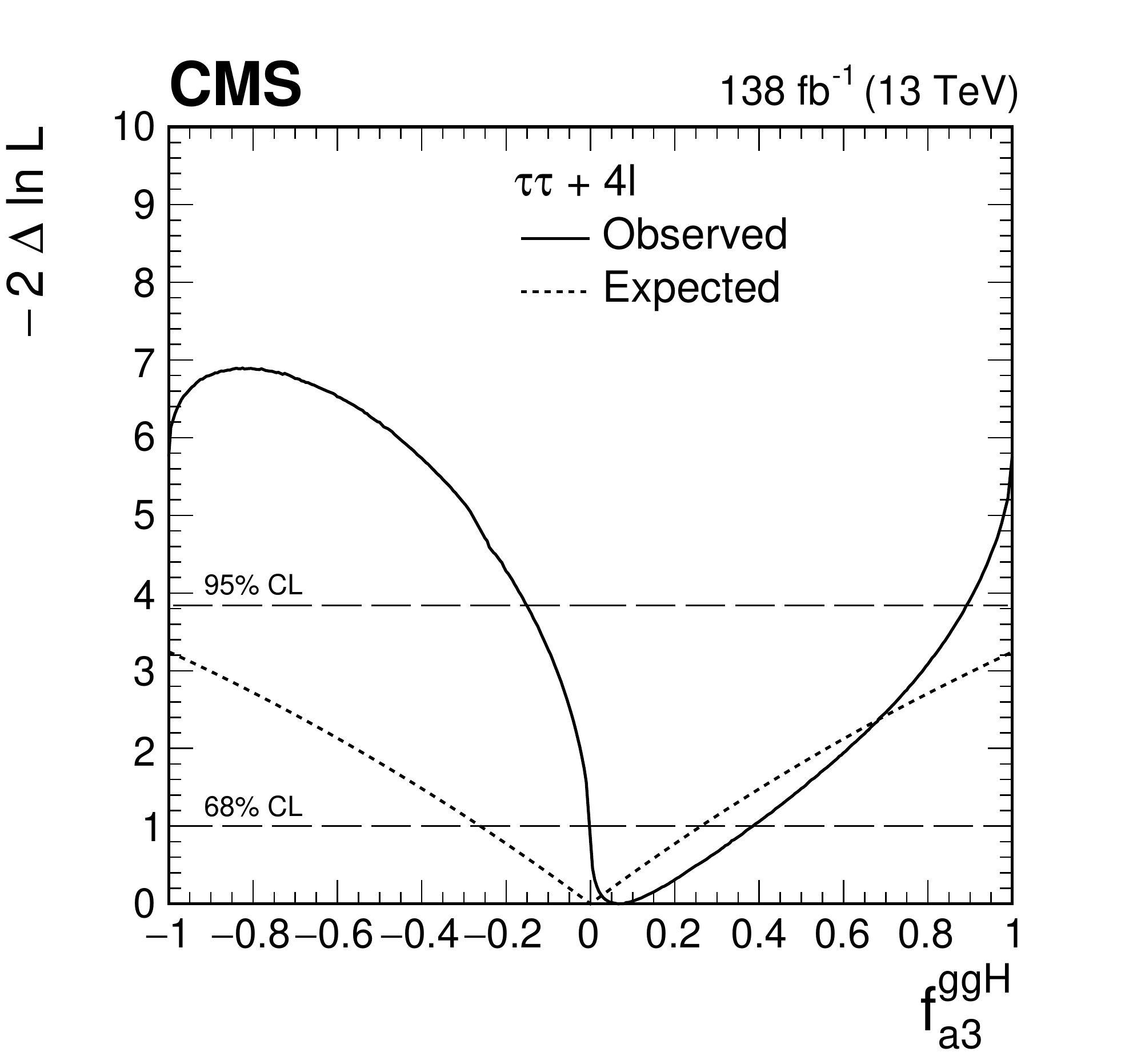}
\includegraphics[width=0.45\textwidth]{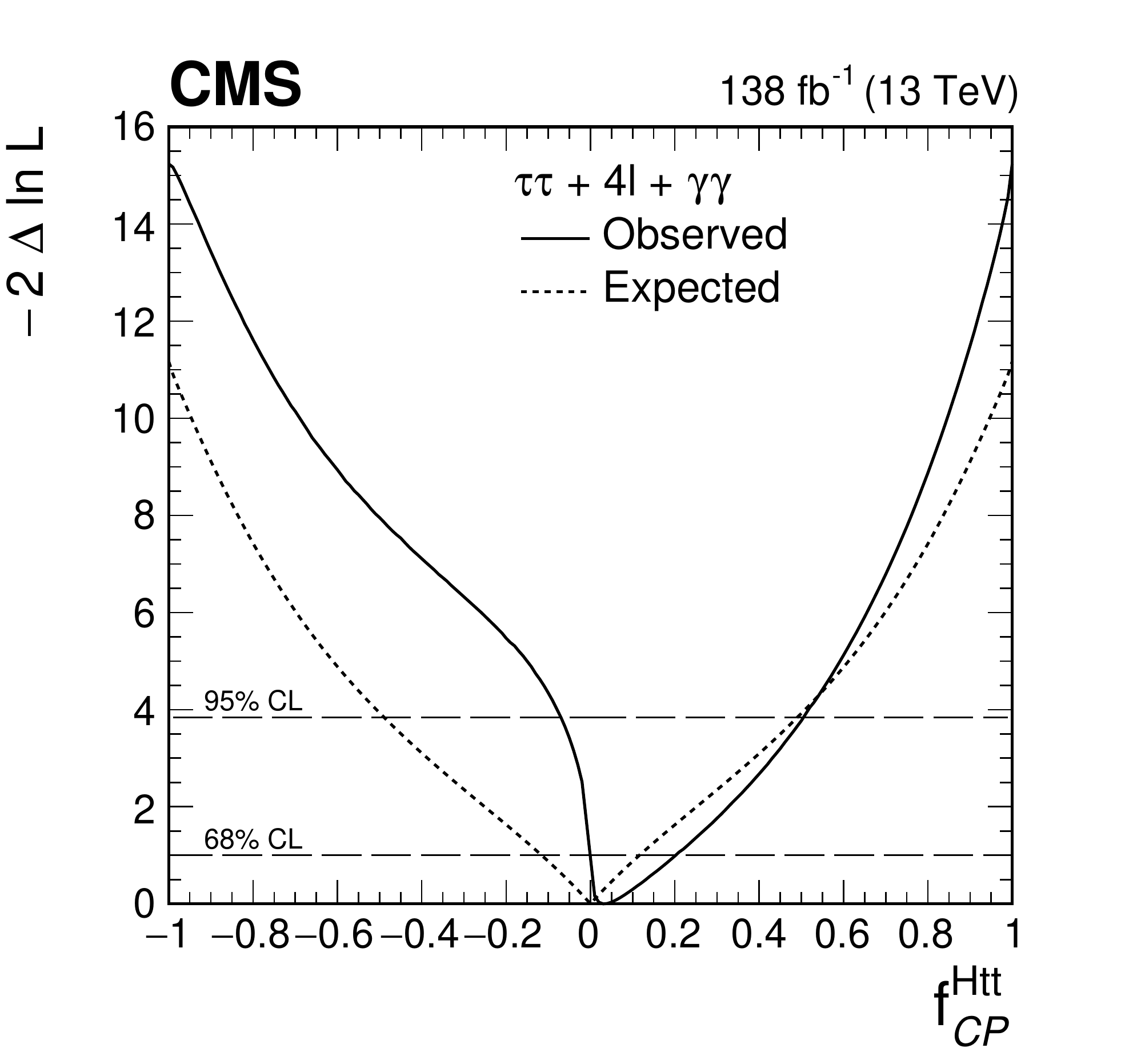}
\caption
{
Left: the observed (solid) and expected (dashed) likelihood scans of
\fggh obtained with the combination of results using the \HTT and \Hllll~\cite{CMS-HIG-19-009} decay channels.
Right: The observed (solid) and expected (dashed) likelihood scans of \fCP obtained with the combination of results using the \HTT, \Hllll~\cite{CMS-HIG-19-009}, and \Hgamgam~\cite{Chatrchyan:2020htt} decay channels. } 
\label{fig:combination_ggH}
\end{figure*}

The combined likelihood scans for the \Hgg anomalous coupling measurements are shown in Fig.~\ref{fig:combination_ggH},
and the allowed 68 and 95\%~\CL intervals are listed in Table~\ref{tab:combination_ggH}.
The \HTT channel is more sensitive to \fggh than the \Hllll channel is, but there is a significant improvement from including both channels in the combination. 
Previous measurements by the CMS and ATLAS Collaborations~\cite{CMS-HIG-19-009,ATLAS:2021pkb} were only able to differentiate between the $CP$-even and $CP$-odd scenarios with a significance slightly less than 1 standard deviation. 
With the current measurement, the pure $CP$-odd scenario is excluded with a observed (expected) significance of 2.4 standard deviations (1.8 standard deviations),
which is cross-checked with pseudoexperiments. 

\begin{table*}[h]
\centering
\topcaption{
Allowed 68\% (central values with uncertainties) and 95\%~\CL (in square brackets)
intervals on \fggh, from the combination of the \HTT and \Hllll~\cite{CMS-HIG-19-009} decay channels, and \fCP, from the combination of the \HTT, \Hllll~\cite{CMS-HIG-19-009}, and \Hgamgam~\cite{Chatrchyan:2020htt} decay channels.
}
\renewcommand{\arraystretch}{1.25}
\begin{scotch}{cccccc}
 Parameter              &  \multicolumn{2}{c}{Observed} &  \multicolumn{2}{c}{Expected}    \\
                        & 68\%~\CL &  95\%~\CL& 68\%~\CL &  95\%~\CL \\
\hline
\fggh & $0.07^{+0.32}_{-0.07}$ & $[-0.15,0.89]$    &   $0.00\pm0.26$ & \NA  \\
\fCP & $0.03^{+0.17}_{-0.03}$ & $[-0.07,0.51]$    &   $0.00\pm0.12$ & $[-0.49,0.49]$  \\
\end{scotch}
\label{tab:combination_ggH}
\end{table*}

Constraints on anomalous \Htt couplings are obtained through the combination of the \Hgg results with measurements of the \ttH and $\PQt\PH$ processes in the \Hllll~\cite{CMS-HIG-19-009} and \Hgamgam~\cite{Chatrchyan:2020htt} channels.
We measure the \fCP parameter by relating \fggh and \fCP as described in Eq.~(\ref{eq:fai-relationship-hgg-tth}), under the assumption of top quark dominance in the \ggH loop.
The results are presented in Fig.~\ref{fig:combination_ggH} and Table~\ref{tab:combination_ggH}. 

\begin{figure*}[!htb]
\centering
\includegraphics[width=0.45\textwidth]{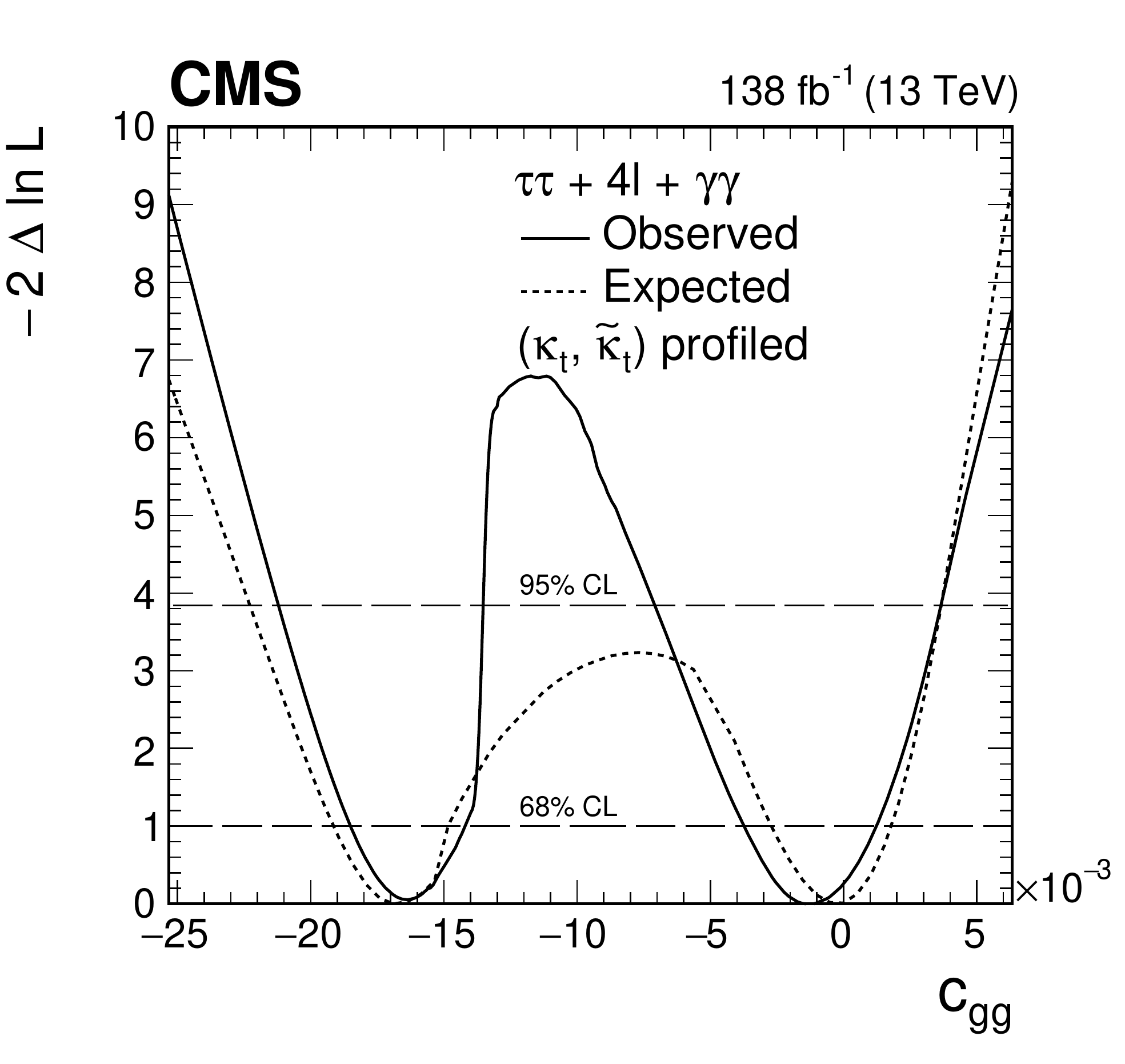}
\includegraphics[width=0.45\textwidth]{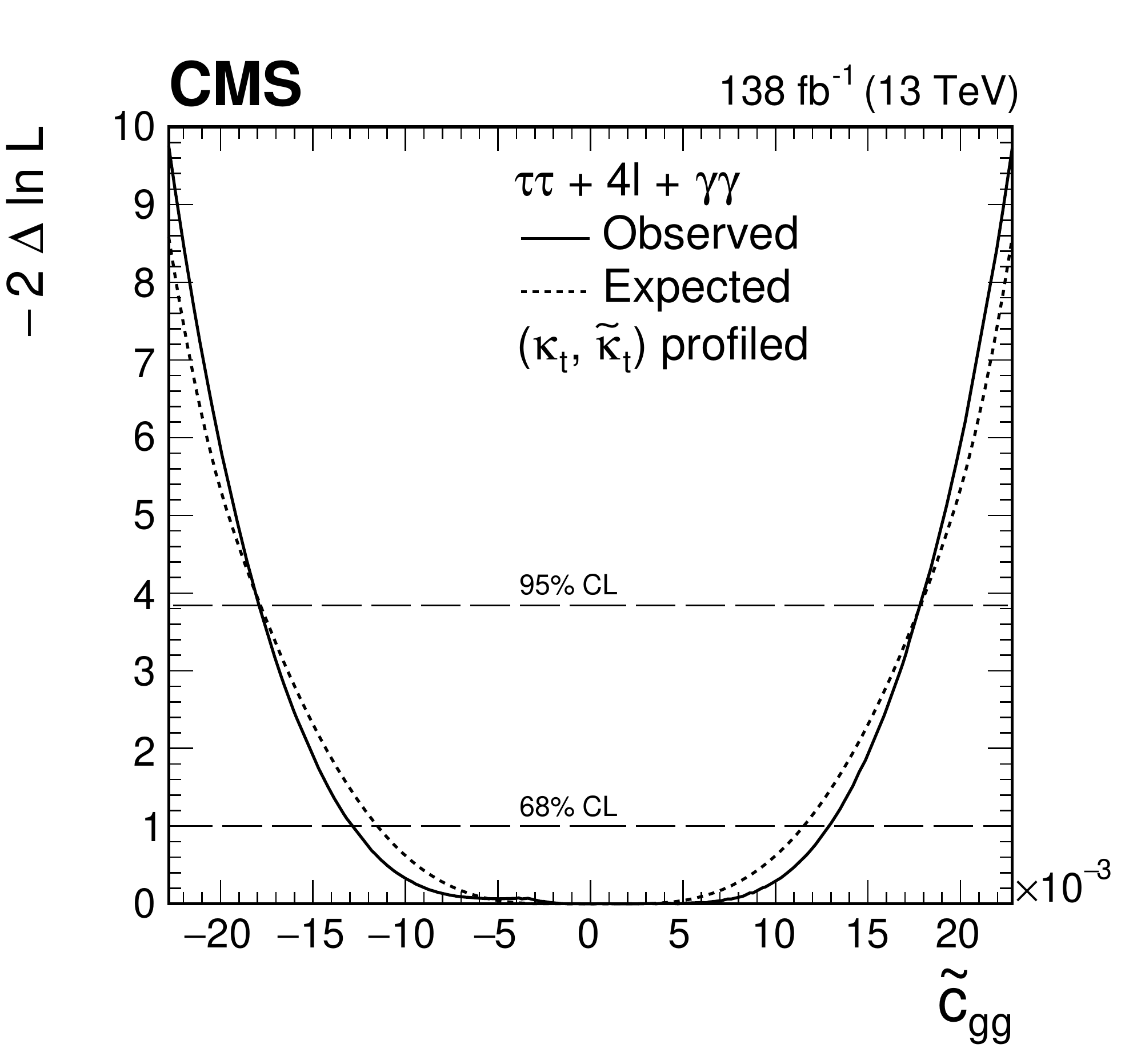}
\includegraphics[width=0.45\textwidth]{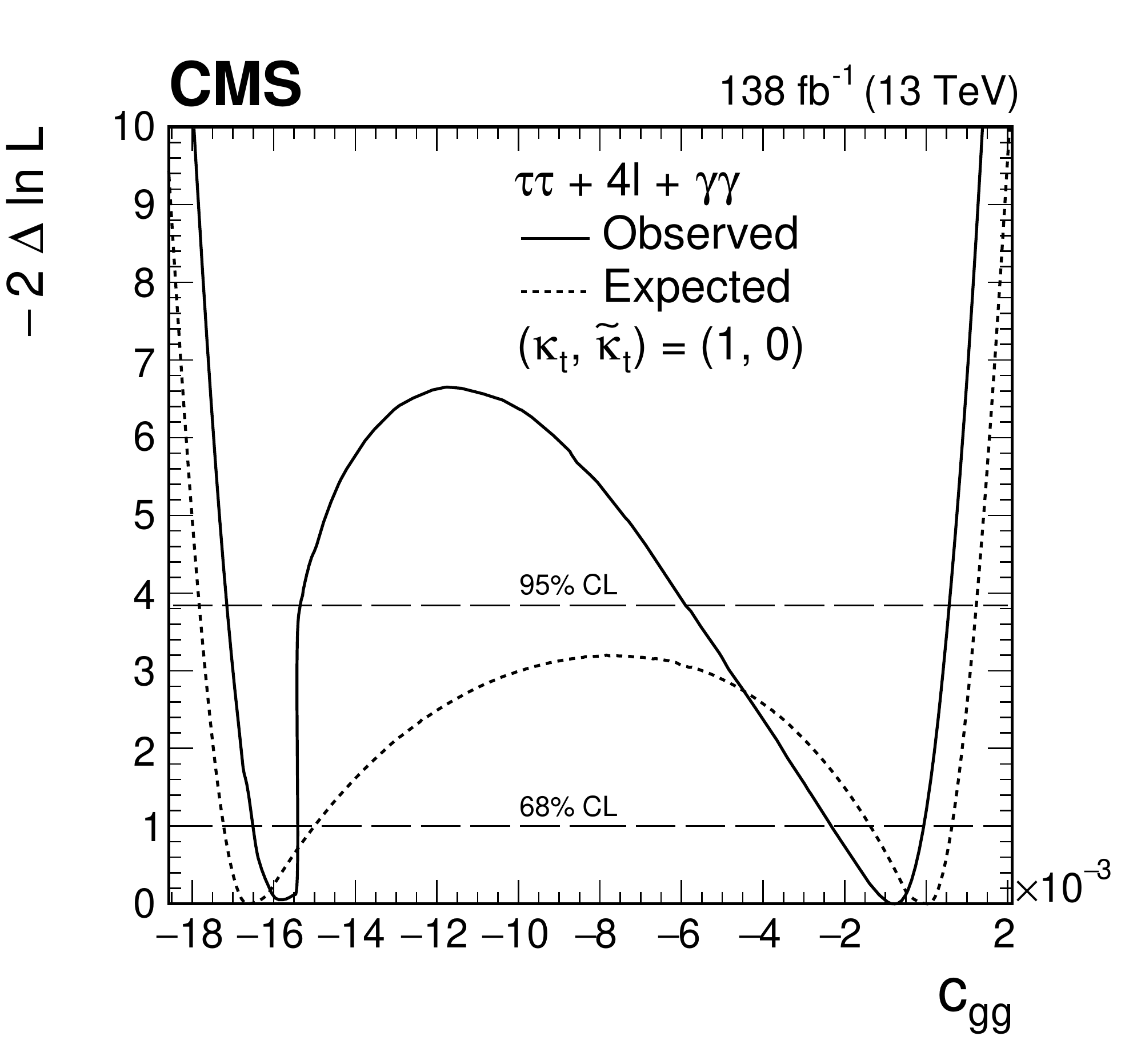}
\includegraphics[width=0.45\textwidth]{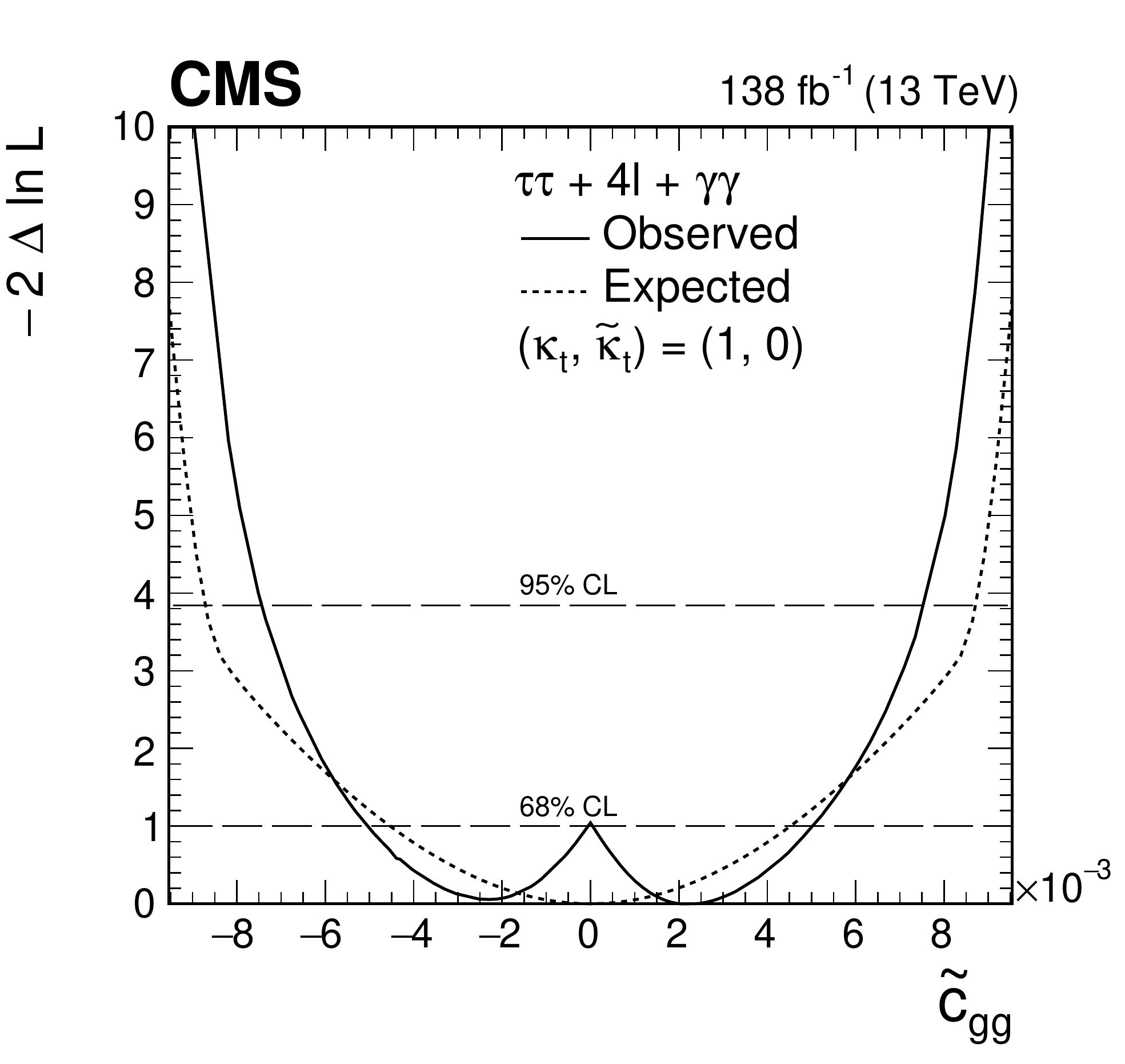}
\caption
{
Observed (solid) and expected (dashed) likelihood scans of \cgg (left) and \cggtilde (right)
with \kappat and \kappattilde profiled (upper) and fixed to SM expectation (lower)
using the \HTT, \Hllll~\cite{CMS-HIG-19-009}, and \Hgamgam~\cite{Chatrchyan:2020htt} decay channels.
\label{fig:eft_ggHttH}
}
\end{figure*}

The combination of \Hgg, \ttH, and $\PQt\PH$ results can be reinterpreted in the EFT approach as constraints on \cgg and \cggtilde.
The likelihood scans for \cgg and \cggtilde are performed with \kappat and \kappattilde 
either profiled or fixed to SM expectation ($\kappat = 1$, $\kappattilde = 0$). 
The reinterpretation is presented in Fig.~\ref{fig:eft_ggHttH} and Table~\ref{tab:eft_ggHttH}.
We note that in both \cgg scans there is a second minimum away from $\cgg=0$ due to the negative interference between the \cgg and \kappat contributions, as follows from Eq.~(\ref{eq:cgg_crosssection}).
The value of the \DeltaLL at \cgg between the two minima points is larger for the observed scan compared to the expected, due to the statistical fluctuation in the \HTT channel data described in Sec.~\ref{ssec:ggH_Results}.

\begin{table*}[h]
\centering
\topcaption{
Allowed 68\% (central values with uncertainties) and 95\%~\CL (in square brackets)
intervals on \cgg and \cggtilde using the \HTT, \Hllll~\cite{CMS-HIG-19-009}, and \Hgamgam~\cite{Chatrchyan:2020htt} decay channels. Results are presented for two scenarios: \kappat and \kappattilde profiled in the fit, and \kappat and \kappattilde fixed to the SM expectation. In instances where there is a second allowed region away from the best fit value at a given CL, we use the union symbol ($\cup$) to display the additional allowed \cggtilde/\cgg range. 
}
\renewcommand{\arraystretch}{1.25}
\cmsTable{
\begin{scotch}{ccccc}
 Parameter   &  Scenario  &  & 68\%~\CL$/(10^{-2})$ &  95\%~\CL$/(10^{-2})$ \\
\hline
\multirow{2}{*}{\cgg}       & \multirow{2}{*}{Profiled}& Observed & $-0.11^{+0.20}_{-0.26}$ $\cup[-1.85,-1.42]$ & $[-2.12,-1.35]\cup[-0.71,0.36]$ \\
                                      &                          & Expected & $0.00^{+0.18}_{-0.27}$ $\cup[-1.91,-1.48]$  & $[-2.23,0.37]$  \\
\multirow{2}{*}{\cggtilde} & \multirow{2}{*}{Profiled}& Observed & $0.00\pm1.29$                      & $[-1.79,1.79]$  \\   
                                      &                          & Expected & $0.00\pm1.15$                               & $[-1.78,1.78]$  \\
\multirow{2}{*}{\cgg}       & \multirow{2}{*}{Fixed}   & Observed & $-0.08^{+0.07}_{-0.15}$ $\cup[-1.65,-1.54]$ & $[-1.71,-1.54]\cup[-0.59,0.05]$ \\
                                      &                          & Expected & $0.00^{+0.06}_{-0.14}$ $\cup[-1.73,-1.50]$  & $[-1.78,0.12]$  \\
\multirow{2}{*}{\cggtilde} & \multirow{2}{*}{Fixed}   & Observed & $0.22^{+0.28}_{-0.22}$ $\cup[-0.50,0.00]$   & $[-0.74,0.75]$  \\
                                      &                          & Expected & $0.00\pm0.45$ & $[-0.87,0.87]$  \\
\end{scotch}
}
\label{tab:eft_ggHttH}
\end{table*}

\clearpage

\section{Summary}
\label{sec:Summary}

A study is presented of anomalous interactions of the Higgs boson (\PH) with vector bosons, including
$CP$ violation, using its associated production with two hadronic jets in gluon fusion (\ggH), vector boson
fusion (VBF), and associated production with a vector boson, and a subsequent decay to a pair
of \PGt leptons. Constraints have been set on the $CP$-violating effects in \ggH production in terms of
the effective cross section ratio \fggh, or equivalently the effective mixing angle \alphacp, 
using matrix element techniques. The \ggH production analysis results in the most stringent limits
on $CP$ violation in \ggH production to date.
In the VBF production analysis, constraints on the $CP$-violating parameter
\fathree and on the $CP$-conserving parameters
\fatwo, \fL, and
\fLzg have been set using matrix element techniques.
Further constraints were obtained in the combination of the \HTT, \Hllll, and \Hgamgam channels. 
The combination improves the limits on the anomalous coupling parameters typically by about 20\%--50\%.
The analysis excludes the pure $CP$-odd scenario of the Higgs coupling to gluons with a significance of 2.4 standard deviations. 

\begin{acknowledgments}
We congratulate our colleagues in the CERN accelerator departments for the excellent performance of the LHC and thank the technical and administrative staffs at CERN and at other CMS institutes for their contributions to the success of the CMS effort. In addition, we gratefully acknowledge the computing centres and personnel of the Worldwide LHC Computing Grid and other centres for delivering so effectively the computing infrastructure essential to our analyses. Finally, we acknowledge the enduring support for the construction and operation of the LHC, the CMS detector, and the supporting computing infrastructure provided by the following funding agencies: BMBWF and FWF (Austria); FNRS and FWO (Belgium); CNPq, CAPES, FAPERJ, FAPERGS, and FAPESP (Brazil); MES and BNSF (Bulgaria); CERN; CAS, MoST, and NSFC (China); MINCIENCIAS (Colombia); MSES and CSF (Croatia); RIF (Cyprus); SENESCYT (Ecuador); MoER, ERC PUT and ERDF (Estonia); Academy of Finland, MEC, and HIP (Finland); CEA and CNRS/IN2P3 (France); BMBF, DFG, and HGF (Germany); GSRI (Greece); NKFIH (Hungary); DAE and DST (India); IPM (Iran); SFI (Ireland); INFN (Italy); MSIP and NRF (Republic of Korea); MES (Latvia); LAS (Lithuania); MOE and UM (Malaysia); BUAP, CINVESTAV, CONACYT, LNS, SEP, and UASLP-FAI (Mexico); MOS (Montenegro); MBIE (New Zealand); PAEC (Pakistan); MES and NSC (Poland); FCT (Portugal); MESTD (Serbia); MCIN/AEI and PCTI (Spain); MOSTR (Sri Lanka); Swiss Funding Agencies (Switzerland); MST (Taipei); MHESI and NSTDA (Thailand); TUBITAK and TENMAK (Turkey); NASU (Ukraine); STFC (United Kingdom); DOE and NSF (USA).

\hyphenation{Rachada-pisek} Individuals have received support from the Marie-Curie programme and the European Research Council and Horizon 2020 Grant, contract Nos.\ 675440, 724704, 752730, 758316, 765710, 824093, 884104, and COST Action CA16108 (European Union); the Leventis Foundation; the Alfred P.\ Sloan Foundation; the Alexander von Humboldt Foundation; the Belgian Federal Science Policy Office; the Fonds pour la Formation \`a la Recherche dans l'Industrie et dans l'Agriculture (FRIA-Belgium); the Agentschap voor Innovatie door Wetenschap en Technologie (IWT-Belgium); the F.R.S.-FNRS and FWO (Belgium) under the ``Excellence of Science -- EOS" -- be.h project n.\ 30820817; the Beijing Municipal Science \& Technology Commission, No. Z191100007219010; the Ministry of Education, Youth and Sports (MEYS) of the Czech Republic; the Hellenic Foundation for Research and Innovation (HFRI), Project Number 2288 (Greece); the Deutsche Forschungsgemeinschaft (DFG), under Germany's Excellence Strategy -- EXC 2121 ``Quantum Universe" -- 390833306, and under project number 400140256 - GRK2497; the Hungarian Academy of Sciences, the New National Excellence Program - \'UNKP, the NKFIH research grants K 124845, K 124850, K 128713, K 128786, K 129058, K 131991, K 133046, K 138136, K 143460, K 143477, 2020-2.2.1-ED-2021-00181, and TKP2021-NKTA-64 (Hungary); the Council of Science and Industrial Research, India; the Latvian Council of Science; the Ministry of Education and Science, project no. 2022/WK/14, and the National Science Center, contracts Opus 2021/41/B/ST2/01369 and 2021/43/B/ST2/01552 (Poland); the Funda\c{c}\~ao para a Ci\^encia e a Tecnologia, grant CEECIND/01334/2018 (Portugal); the National Priorities Research Program by Qatar National Research Fund; MCIN/AEI/10.13039/501100011033, ERDF ``a way of making Europe", and the Programa Estatal de Fomento de la Investigaci{\'o}n Cient{\'i}fica y T{\'e}cnica de Excelencia Mar\'{\i}a de Maeztu, grant MDM-2017-0765 and Programa Severo Ochoa del Principado de Asturias (Spain); the Chulalongkorn Academic into Its 2nd Century Project Advancement Project, and the National Science, Research and Innovation Fund via the Program Management Unit for Human Resources \& Institutional Development, Research and Innovation, grant B05F650021 (Thailand); the Kavli Foundation; the Nvidia Corporation; the SuperMicro Corporation; the Welch Foundation, contract C-1845; and the Weston Havens Foundation (USA).
\end{acknowledgments}

\bibliography{auto_generated} 

\cleardoublepage \appendix\section{The CMS Collaboration \label{app:collab}}\begin{sloppypar}\hyphenpenalty=5000\widowpenalty=500\clubpenalty=5000
\cmsinstitute{Yerevan Physics Institute, Yerevan, Armenia}
{\tolerance=6000
A.~Tumasyan\cmsAuthorMark{1}\cmsorcid{0009-0000-0684-6742}
\par}
\cmsinstitute{Institut f\"{u}r Hochenergiephysik, Vienna, Austria}
{\tolerance=6000
W.~Adam\cmsorcid{0000-0001-9099-4341}, J.W.~Andrejkovic, T.~Bergauer\cmsorcid{0000-0002-5786-0293}, S.~Chatterjee\cmsorcid{0000-0003-2660-0349}, K.~Damanakis\cmsorcid{0000-0001-5389-2872}, M.~Dragicevic\cmsorcid{0000-0003-1967-6783}, A.~Escalante~Del~Valle\cmsorcid{0000-0002-9702-6359}, R.~Fr\"{u}hwirth\cmsAuthorMark{2}\cmsorcid{0000-0002-0054-3369}, M.~Jeitler\cmsAuthorMark{2}\cmsorcid{0000-0002-5141-9560}, N.~Krammer\cmsorcid{0000-0002-0548-0985}, L.~Lechner\cmsorcid{0000-0002-3065-1141}, D.~Liko\cmsorcid{0000-0002-3380-473X}, I.~Mikulec\cmsorcid{0000-0003-0385-2746}, P.~Paulitsch, F.M.~Pitters, J.~Schieck\cmsAuthorMark{2}\cmsorcid{0000-0002-1058-8093}, R.~Sch\"{o}fbeck\cmsorcid{0000-0002-2332-8784}, D.~Schwarz\cmsorcid{0000-0002-3821-7331}, S.~Templ\cmsorcid{0000-0003-3137-5692}, W.~Waltenberger\cmsorcid{0000-0002-6215-7228}, C.-E.~Wulz\cmsAuthorMark{2}\cmsorcid{0000-0001-9226-5812}
% ...
\par}
\vskip\cmsinstskip
\dag:~Deceased\\
$^{1}$Also at Yerevan State University, Yerevan, Armenia\\
$^{2}$Also at TU Wien, Vienna, Austria\\
% ...
\end{sloppypar}
\end{document}